\documentclass[energies,article,accept,moreauthors,pdftex,12pt,a4paper]{mdpi} 
\lastpage{x}
\doinum{10.3390/------}
\pubvolume{8}
\pubyear{2015}
\history{Received: 21 August 2014 / Accepted: 20 January 2015 / Published: 2015}
\pdfoutput=1

\usepackage{graphicx}
\usepackage{subcaption}
\usepackage{float}
\usepackage{soul}
\usepackage{caption}

\newcommand{\fat}[1]{\mathbf{#1}}
\newcommand{\bdm}{\begin{displaymath}}
\newcommand{\edm}{\end{displaymath}}
\newcommand{\beq}{\begin{equation}}
\newcommand{\eeq}{\end{equation}}
\newcommand{\beqa}{\begin{eqnarray}}
\newcommand{\eeqa}{\end{eqnarray}}

\newcommand{\bmat}[1]{\( \begin{array}{#1}}
\newcommand{\emat}{\end{array}\)}

\newcommand{\integrate}[2]{\int\limits^{#2}_{#1} \hspace{-0.15cm}}

\Title{Towards a Simplified Dynamic Wake Model Using POD Analysis}

\Author{David Bastine *, Bj\"orn Witha, Matthias W\"achter and Joachim Peinke}

\address[1]{%
 Institute of Physics, ForWind, University of Oldenburg, Ammerl\"ander Herrstr. 136, 26129 Oldenburg, Germany; E-Mails:bjoern.witha@forwind.de (B.W.); matthias.waechter@forwind.de (M.W.); peinke@uni-oldenburg.de (J.P.)\vspace{-12pt}}

\corres{E-Mail: david.bastine@uni-oldenburg.de; \linebreak Tel.: +49-441-798-5054. \\
\vspace{12pt} Academic Editor: Frede Blaabjerg \vspace{6pt}}

\abstract{We apply a modified proper orthogonal decomposition (POD) to large eddy simulation data
of a wind turbine wake in a turbulent atmospheric boundary layer. 
The turbine is modeled as an actuator disk.
Our analysis mainly focuses on the pragmatic identification of spatial modes, which
yields a low order description of the wake flow. This reduction to a few degrees of freedom is a crucial first
step for the development of simplified dynamic wake models based on modal decompositions.
It is shown that only a few modes are necessary to capture the
basic dynamical aspects of quantities that are relevant to a turbine in the wake flow.
Furthermore, we show that the importance of the individual modes depends
on the relevant quantity chosen. Therefore, the optimal choice of modes for a possible model could in principle
depend on the application of interest.
We additionally present a possible interpretation of the extracted modes by relating them
to the specific properties of the wake. For example, the first mode is related to the horizontal large-scale
movement.
}
\keyword{wake; proper orthogonal decomposition (POD); wake model; meandering; large eddy simulations (LES); actuator disk; loads; dynamic wake; reduced order model}
\begin{document}
%
 

\section{Introduction}

Due to economical and technological benefits, wind turbines are often arranged 
in clusters that can contain up to hundreds of turbines.
One of the key issues of these wind farms is the wake effect. A~wind turbine in the wake of
another turbine experiences a strongly altered inflow, resulting in reduced power production and
higher dynamic loads acting on the turbine. In order to minimize these effects, a detailed
understanding and efficient modeling of wakes is essential. This is of particular importance in the
planning phase of a farm, where an optimized layout can result in a much higher energy production~\cite{Barthelmie2010} and longer lifetimes of the turbines. Furthermore, good wake models are an
important tool for an efficient wind farm control \cite{Goit2014,Storey2014}.
It has, for example, been shown that wake deflection through yawing of 
the turbines can mitigate the wake effect \cite{Medici14563,Jimenez2010,Fleming2014}.

The hydrodynamic equations describing the dynamics of a wind turbine wake and the flow through~a wind farm are well known. However, solving these equations in acceptable time is currently
impossible due to the many scales involved in turbulent flows. A very successful modeling tool for
turbulent flows is large eddy simulation (LES), which solves the hydrodynamic equations
on the large scales of the flow, but models the influence of the smaller scales by
sub-grid-scale turbulence models (e.g.,~\cite{Pope2000}). Combined~with simplified turbine models, such as actuator disk
and actuator line, LESs have been increasingly applied in wind energy research
\cite{Jimenez2007,Calaf2010,Porte-Agel2011,Goit2014,Storey2014,Witha2014}.
Unfortunately, they are still too time consuming for many practical applications, such as the
optimization of a wind farm layout or wind farm control.

Therefore, steady-state wake models, such as \cite{Jensen1983, AINSLIE1988, Larsen1988, Frandsen2003, Frandsen2006},
are still the state-of-the-art for many purposes. These models can, for example, use empirical descriptions 
of the velocity deficit, such as \cite{Jensen1983}, or mean field solutions to strongly approximated versions
of the fluid dynamical equations, such as \cite{AINSLIE1988}. More~recently, mean field
wakes stemming from computational fluid dynamic simulations have been used in the context of layout optimization
 with respect to power \cite{Schmidt2014}. Instead of modeling the velocity deficit itself, Frandsen~\cite{Frandsen2003} uses an effective turbulence intensity to model the fatigue loads in a wind park.
However, all of the models neglect the dynamical aspects of the wake, which are commonly expected
to be relevant for the loads and power production of a turbine in a wake. 

The first step to incorporate the dynamical aspects of the wake is to take the meandering movement of
the velocity deficit into account. Usually, this is done under the assumption that the meandering is mainly
caused by the large-scale dynamics of the atmospheric boundary layer (ABL) 
\cite{Larsen2007, Larsen2008, Trujillo2011, Espana2012}. While~this is a 
very promising approach \cite{Thomsen2005,Thomsen2007}, these models neglect, for example, the dynamically changing
shape of the velocity deficit, which is also expected to be relevant, particularly for the loads on the turbine.
Moreover, it is not fully clear whether the passive tracer assumption
always yields a good description of the meandering \cite{Andersen2013}.
Thus, there is still a need for improved simplified dynamic wake models that yield a good description
of the inflow conditions of a wind turbine, but avoid too complex numerical~simulations.

An alternative approach to simplified modeling, used in fluid dynamics, is to decompose
the flow into a superposition of spatial modes \cite{BERKOOZ1993,Citriniti2000, Jung2004, Iqbal2007, Tutkun2008}
and to develop reduced order models for specific flow problems~\cite{BERKOOZ1993, Delville1999, Bergmann2005, Cordier2013}. Andersen \textit{et al.} \cite{Andersen2014,Andersen2013}
applied the proper orthogonal decomposition (POD) to LES data of a wake in an infinitely long row of turbines modeled with an
actuator line approach. They~found clearly structured POD modes, which could be a first step toward a reduced order model of wakes in a
long row of turbines. Additionally, their results indicate that the low-frequency dynamics of the wake
are not only caused by the large-scale dynamics of the ABL and can, thus, not be treated 
separately from the wake itself.

This work is also motivated by the idea of reduced order modeling based on modal decompositions
of the wake flow. Such reduced models could play an important role in the layout optimization and
controlling of wind farms. They could therefore provide a useful alternative to the
kinematic approach followed in dynamic wake meandering models
\cite{Larsen2007, Larsen2008}. The main goal of this paper is to
identify spatial modes that yield a low dimensional description of the velocity field, while
still preserving important aspects of the wake, relevant to a sequential turbine.
A strong dimensional reduction of the flow is a crucial step 
for the development of low order models, since it facilitates
the application of the huge toolbox of advanced modeling methods of nonlinear 
dynamical systems \cite{Kantz2004,Argyris2015,kleinhans2007, Friedrich2011, Milan2014}. 

For this purpose, we apply the POD to LES data of a wind turbine wake.
In contrast to \linebreak Andersen \textit{et~al.} \cite{Andersen2014,Andersen2013}, we analyze the wake of a single turbine modeled
by an actuator disk and use a turbulent ABL as the inflow condition. Additionally,
we apply threshold criteria to focus on the wake structure.
After this preprocessing procedure, we estimate POD modes and use them to reconstruct the
original wake flow. 
We propose to assess the quality of such reconstructions
based on the potential impact on a turbine in the wake instead of using
the turbulent kinetic energy.
The results in \cite{bastine2014} gave the first hint that relevant quantities of the wake 
can be described well, even though only a small part of the turbulent kinetic energy is recovered.
Here, we investigate three additional quantities, which are related to the power, thrust and yaw load on 
a virtual turbine in the wake and perform a detailed comparison. This comparison naturally
leads to the idea of selecting modes depending on the desired application of a possible model.
A similar approach has also been suggested by Saranyasoontorn and Manuel
\cite{Saranyasoontorn2005, Saranyasoontorn2006} in the case of an ABL inflow.
Furthermore, we interpret the role of the different modes.
This enables us to relate our reduced order modeling approach to
features of other dynamic wake models.

In Section \ref{sec:lesdata} of this work, we describe the LES data used for our analysis.
Section \ref{sec:rom} introduces the concept of modal decompositions and motivates our main
goal of dimensional reduction in a more formal way. This is followed by a brief description of
the POD theory (Section \ref{sec:podtheory}). Section \ref{sec:preprocessing} describes
the preprocessing, which we apply before estimating the corresponding
POD modes in Section \ref{sec:podmodes}. Next, we use the extracted modes
to investigate approximations of the original velocity field depending on the number of modes
used for reconstruction (Section \ref{sec:podrecon}).
In Section \ref{sec:alternative}, we introduce the general concept of assessing the quality of POD reconstructions
through measures related to a turbine in the wake flow.
We define four simple measures (e.g., the energy flux through
a disk) and compare the corresponding results in Section \ref{sec:results}.
In Section \ref{sec:interpretation}, we further investigate the role of the individual POD modes and try
to relate them to the specific properties of the wake. Finally, the conclusions drawn from our
results are presented in Section \ref{sec:conclusions}.

\section{LES Data}
\label{sec:lesdata}

The LES data that will be analyzed in this work have been
generated by performing simulations with the parallelized LES Model PALM \cite{Raasch2001, Maronga2015},
which has been widely used for studies of the atmospheric boundary layer.
PALM solves the filtered, incompressible, non-hydrostatic Navier--Stokes
equations under the Boussinesq approximation. The sub-grid-scale
turbulence parameterization is based on the 1.5-order scheme of
Deardorff \cite{Deardorff1980}. The lateral boundary conditions are periodic, and no-slip conditions
are used at the lower boundary. At the top of the domain, a Dirichlet boundary condition
with a constant geostrophic wind of $u_g$ = $10$ ms$^{-1}$ is used.
For further details on the PALM model, we refer the reader to~\cite{Raasch2001}.

The wind turbine is modeled by a standard, uniformly-loaded actuator disk.
This approach yields very similar far wake characteristics as more complex
turbine parameterizations, such as the actuator line method \cite{Witha2014, Wu2011}.
The simulated wind turbine has a rotor diameter of $D=100$ m and a hub height of $160$~m.
A constant thrust coefficient of $C_T = 0.75$ was applied.
     
\begin{figure}[H]
\centering
\begin{subfigure}[t]{.24\textwidth}
 \centering
 \includegraphics[width=.98\linewidth]{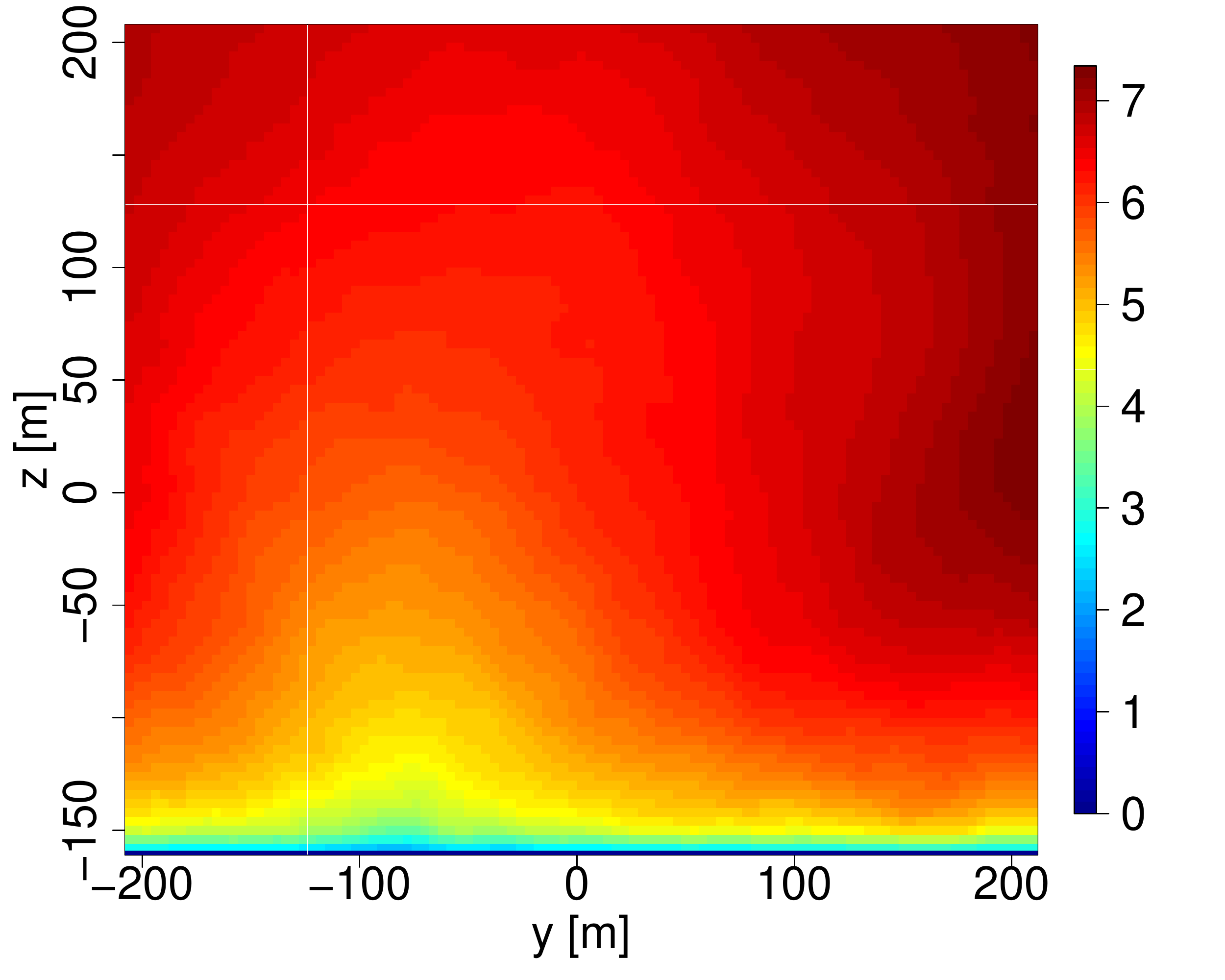}
 \subcaption{}
 \label{fig:q0mean}
\end{subfigure}%
\begin{subfigure}[t]{.24\textwidth}
 \centering
 \includegraphics[width=.98\linewidth]{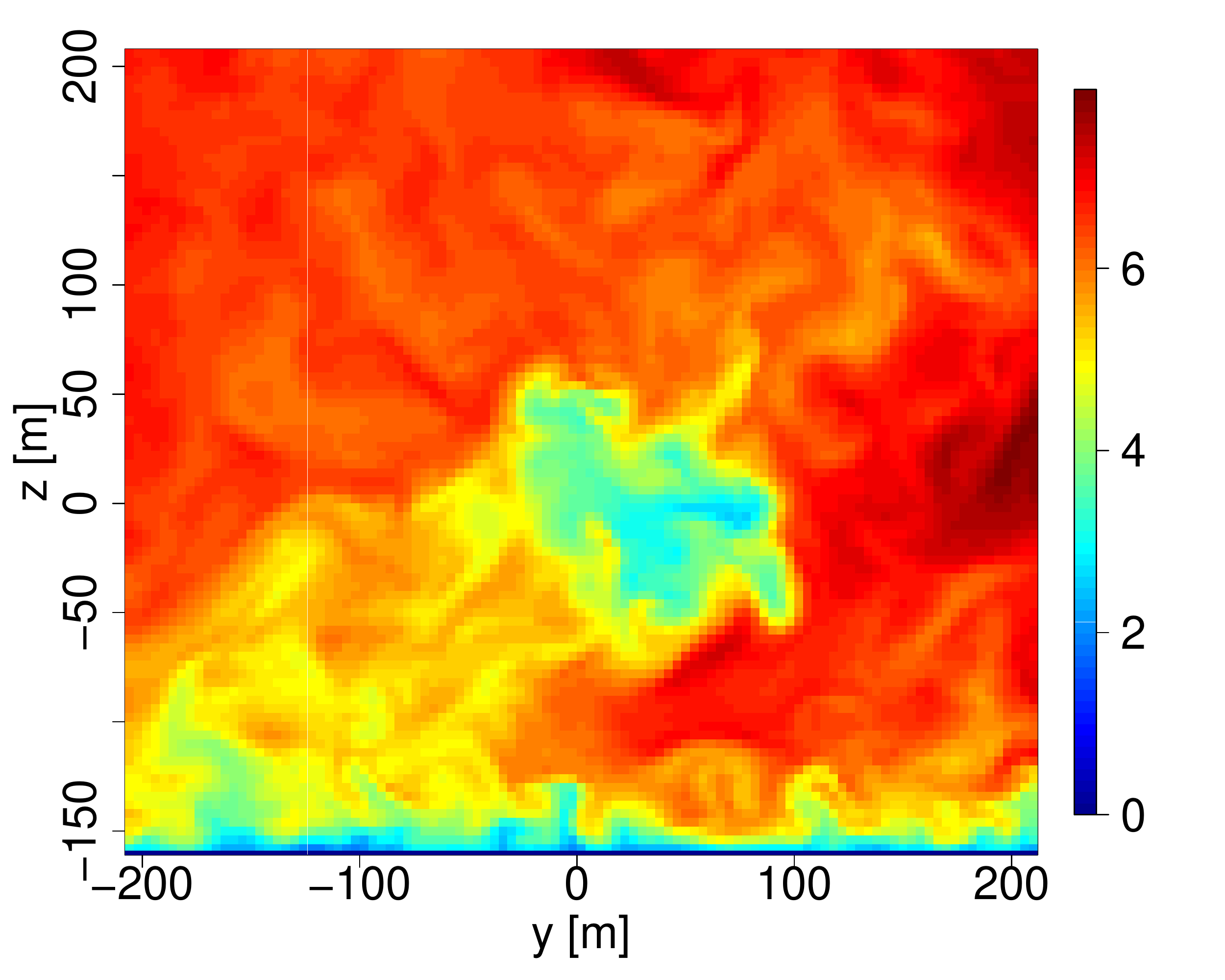}
 \subcaption{}
 \label{fig:qsnap02}
\end{subfigure}
\begin{subfigure}[t]{.24\textwidth}
 \centering
 \includegraphics[width=.98\linewidth]{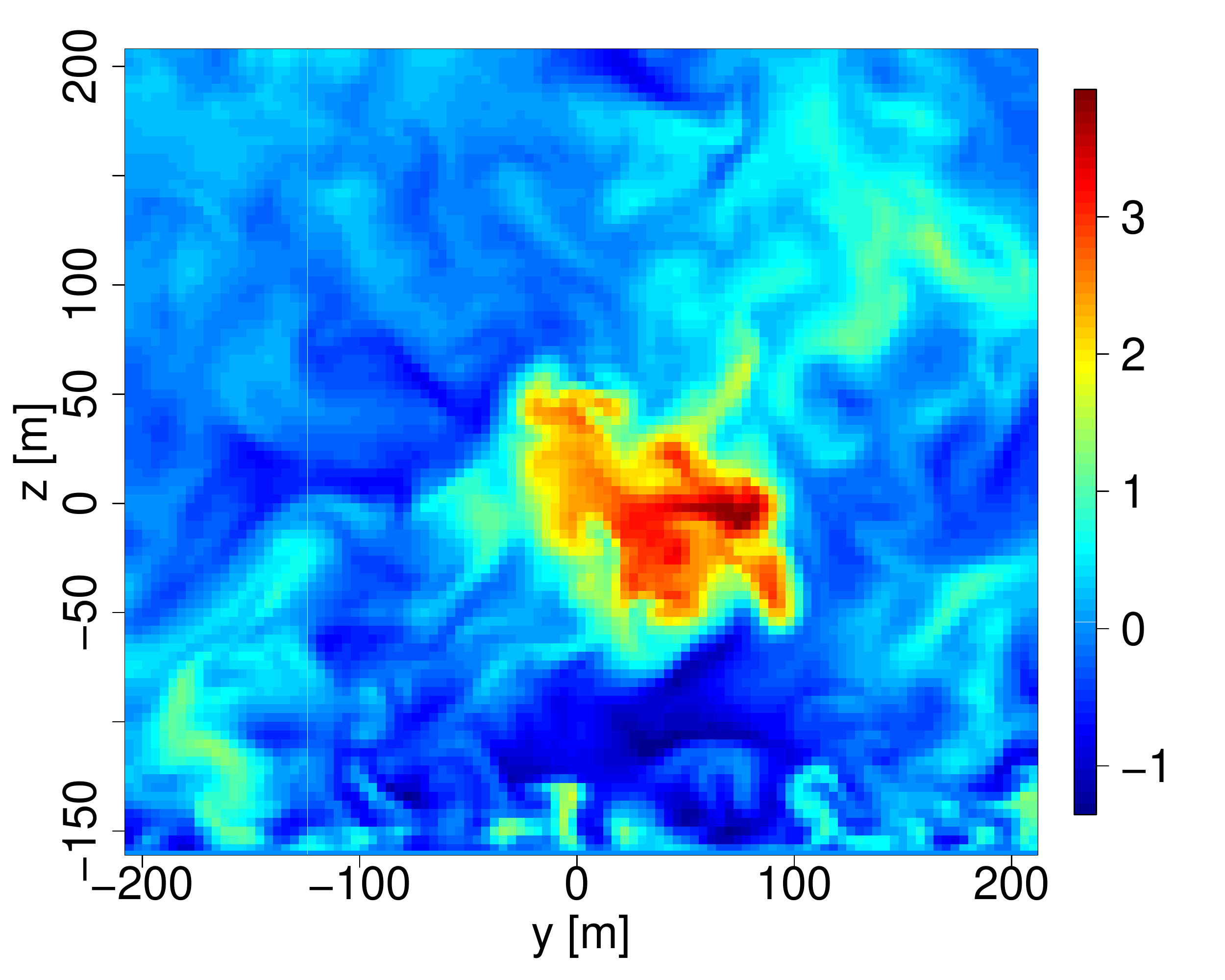}
 \subcaption{}
 \label{fig:qsnap0b}
\end{subfigure}%
\begin{subfigure}[t]{.24\textwidth}
 \centering
 \includegraphics[width=.98\linewidth]{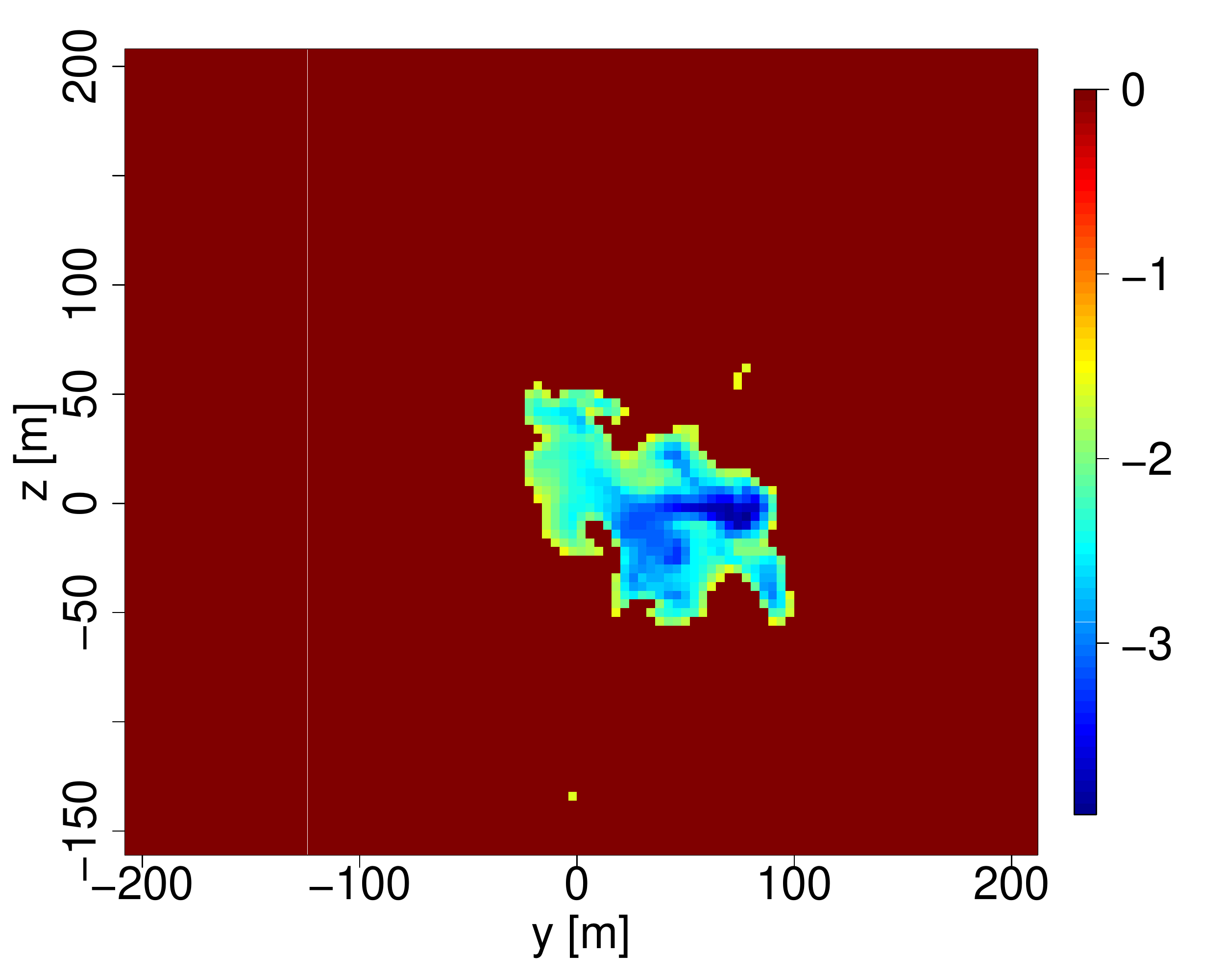}
 \subcaption{}
 \label{fig:qsnap0c}
\end{subfigure}
\vspace {-12pt}
\caption{(\textbf{a}) Time-averaged velocity field without turbine; (\textbf{b}) wake $u(y,z)$ at distance $x=4D$ and $t=500$~s; (\textbf{c}) wake after subtracting the time-averaged
velocity field without the turbine; (\textbf{d}) wake after applying the threshold.}
\label{Fig:q0mean}
\end{figure}

A stationary and fully-turbulent neutral ABL is generated in a precursor simulation without the wind
turbine using periodic inflow conditions.
An aerodynamic roughness length of $z_0$ = $0.05$ m is prescribed. The Coriolis force is neglected, so that
the mean flow is aligned with the \textit{x}-axis in all heights.
After 12 h, a stationary inflow is achieved. In the main
simulation with the wind turbine, non-cyclic inflow conditions are applied.
The final horizontally and time-averaged velocity and temperature
profiles of the precursor simulation are used as the fixed inflow profile.
Furthermore, the instantaneous fields of the precursor simulation are
periodically mapped onto the domain of the main simulation. The
turbulence is continuously recycled at every time step by picking up the
turbulent fluctuations in a~\textit{yz}-plane downstream of the
inflow and imposing them onto the fixed mean inflow profile. The distance
between this \textit{yz}-plane and the inflow boundary is chosen large enough that all turbulent
scales are preserved.
More details on this recycling method can be found in \cite{Witha2014, Kataoka2002}. 
A notable property of this ABL flow is the existence of large-scale coherent regions of high or low velocity.
These regions have the form of stripes aligned with the stream-wise velocity direction and originate
from the surface layer. They~are a known phenomenon, called streaks, and occur in neutral and stable atmospheres 
(see, e.g., \cite{Witha2014b, Moeng1994}).
We~expect the low and high velocity regions in the mean field of the flow without turbine
(Figure \ref{fig:q0mean}) to be a fingerprint of such structures. 

\begin{figure}[H]
\centering
\begin{subfigure}[t]{.52\textwidth}
 \centering
 \includegraphics[width=.90\linewidth]{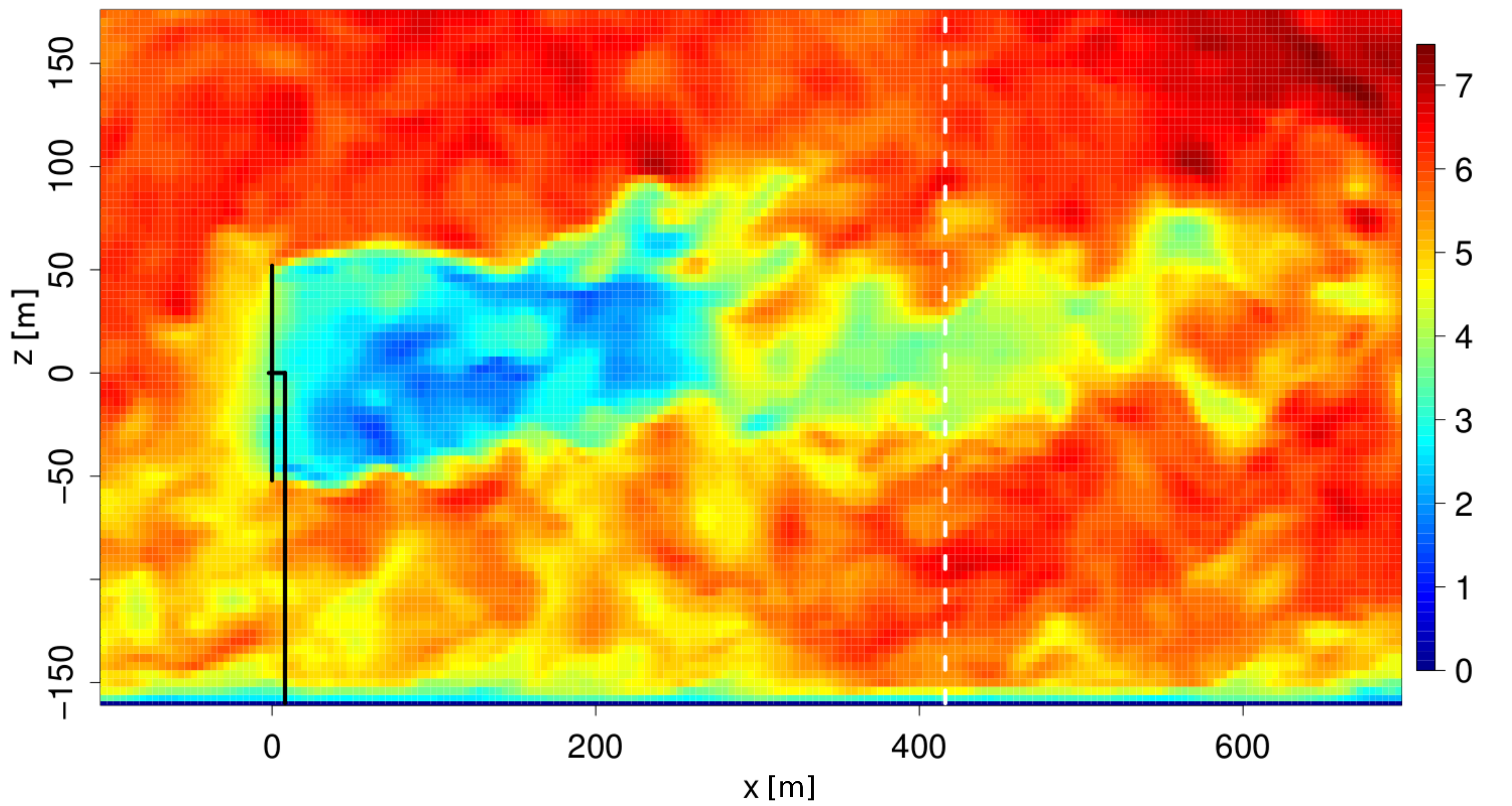}
 \subcaption{}
 \label{fig:les-xz}
\end{subfigure}%
\begin{subfigure}[t]{.38\textwidth}
 \centering
 \includegraphics[width=.90\linewidth]{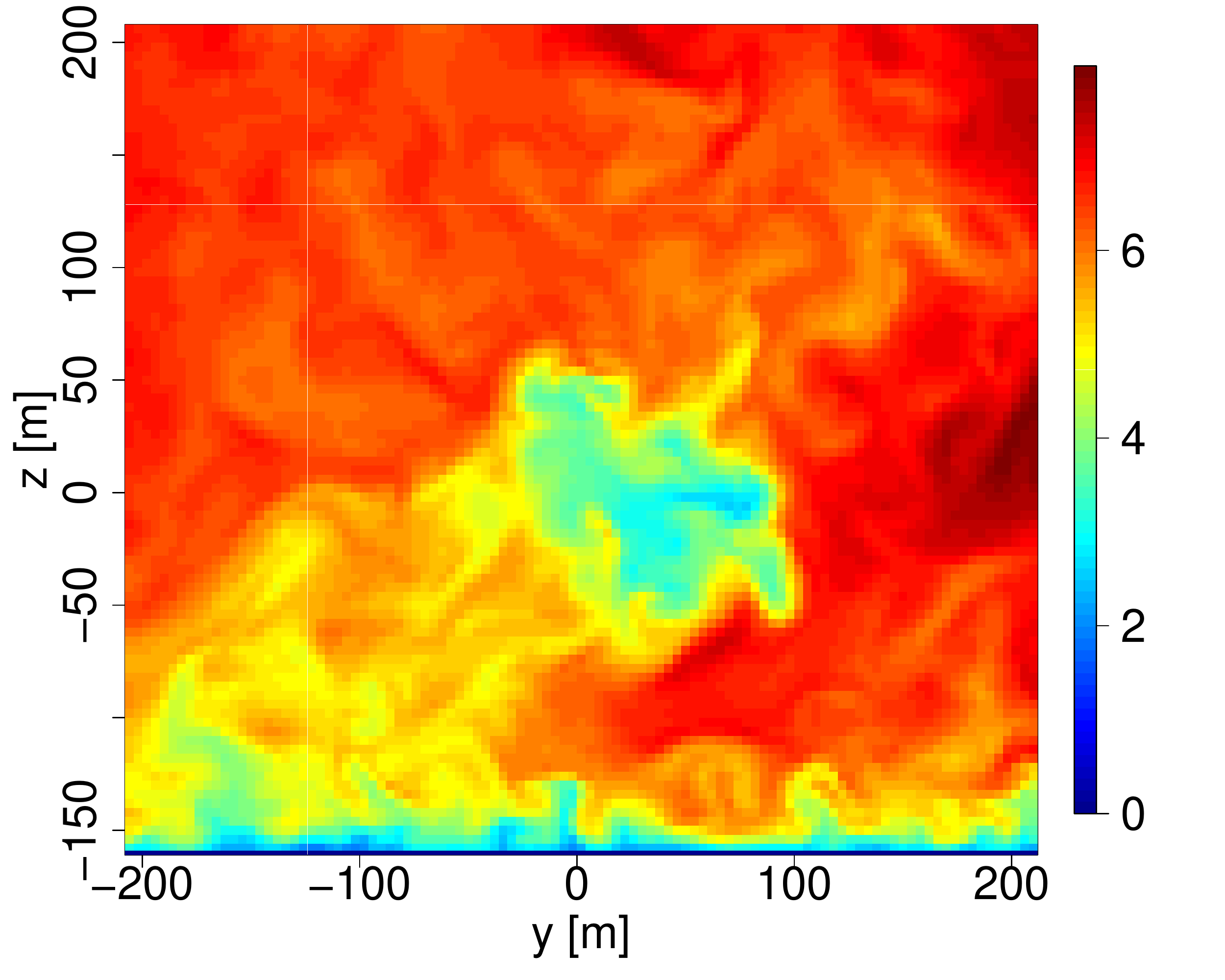}
 \subcaption{}
 \label{fig:qsnap0a}
\end{subfigure}\vspace{-12pt}
\caption{Snapshots of the analyzed LES data. (\textbf{a}) $xz$-plane ($y = 0$ m);
(\textbf{b}) $yz$-plane ($x~=~4~D$). The color contours denote the main velocity component $u$ in ms$^{-1}$.
The white dashed line marks the $yz$-plane at $x = 4 D$, which was used for the POD analysis.}
\label{Fig:les-xz}
\end{figure}

The domain size of the main simulation is about $8$ km $\times$ $2$ km $\times$ $0.5$ km with a grid size of $4$ m.
With this large domain, we ensure that the wake is virtually unaffected by the boundaries. 
As shown in Figure~\ref{Fig:les-xz}, we define the $x$-direction aligned with the main flow, $y$ as
the lateral and $z$ as the vertical coordinate. The turbine is placed in the center of the domain in
$y$-direction and 2500~m downstream of the inflow.
Hereafter, the origin of the coordinate system will be located at the hub of the~turbine.

The generated data are written out at a rate of $2$ Hz with a time series length of 
11,750 s after excluding the first $500$ s of the simulation.
For our analysis, we focus on data in a $yz$-plane perpendicular to the main flow,
which is placed $4 D
$ downstream of the turbine (the white dashed line in Figure \ref{fig:les-xz}). At this distance, we expect
only minor differences between the standard actuator disk model used here and more sophisticated
 actuator models \cite{Wu2011, Witha2014}. Furthermore, we confine our analysis to the main
velocity component yielding a spatio-temporal data field $u(y,z,t)$.
A snapshot of this data field for $t = 500$ s 
can be seen in Figure \ref{fig:qsnap0a}. It shows a pronounced wake downstream of the turbine with a strong reduction
of the flow velocity.

\vspace {-12pt}
\section{Decomposing the Velocity Field}\vspace{-12pt}
\label{sec:podanalysis}
\subsection{Modal Decompositions and Reduced Order Modeling}
\label{sec:rom}

To develop a simplified dynamic wake model based on a modal decomposition, two essential steps are necessary. 
The first step is to find a reduced order description of the flow given by a superposition 
of spatial modes $\phi_j(y,z)$. Thus, we define:
\begin{equation}
 u^{(N)}(y,z,t)=\langle u(y,z,t) \rangle_t + \sum\limits_{j=1}^{N} {a}_j(t) {\phi}_j(y,z)~
 \label{eq:decomposition}
\end{equation}
where $\langle ... \rangle_t$ denotes the time average and $N$ the order of the reduced description $u^{(N)}$ 
in terms of the number of modes taken into account.
For orthogonal modes, as used in this work, the $a_j(t)$ are given~through:
\begin{equation}
a_j(t)=(\phi_j|u'):=\integrate{}{} dy dz~ \phi_j^{*}(y,z) u'(y,z,t)~
\label{eq:scalarprod}
\end{equation}
where $u'(y,z,t):=u(y,z,t)-\langle u(y,z,t) \rangle_t$ denotes the fluctuating part of the field.

The second step to develop a dynamic wake model is to model the time-dependent weighting coefficients $a_j(t)$.
Such a model could then be given by an $N$-dimensional dynamical system in the form of:
\begin{equation}
 \dot{a}_j=f_j(a_1,..a_N)~,\mbox{with}~j=1,...,N~
 \label{eq:rom}
\end{equation}

For this temporal modeling, the order $N$ of the reduced system Equation (\ref{eq:rom}) is crucial.
In the case of large $N$, estimating such models from data becomes principally very difficult, and often, huge amounts
of data are needed. If $N$ is chosen too small, important information of the velocity field is lost.
Thus, the success of the temporal modeling of the $a_j(t)$ strongly depends on the reduced order description chosen
in the first step. Hence, we aim for identifying modes that yield the necessary dimensional reduction, while preserving 
important aspects of the flow. 

A useful tool to obtain spatial modes and a corresponding decomposition in the spirit of
Equation (\ref{eq:decomposition}) is the POD, which is described in the next section
(Section \ref{sec:podtheory}). In this work, we will not focus on the standard POD modes of the
velocity field $u$, but choose the $\phi_j$ in a
slightly different manner, as will be explained in Section \ref{sec:preprocessing}.

\subsection{POD Theory}
\label{sec:podtheory}

Subsequently, we briefly illustrate the mathematical definition and important properties of the POD.
Details can be found, e.g., in \cite{BERKOOZ1993}.

As its name suggests, the proper orthogonal decomposition (POD) is
a decomposition in the spirit of Equation (\ref{eq:decomposition}). The modes $\phi_j$ 
are then called POD modes and are the optimal modes with
respect to the turbulent kinetic energy of the flow. More exactly, they are defined
as the orthogonal functions that minimize the mean squared error:
\begin{equation}
\langle \| u'(y,z,t)-\sum\limits_{j=1}^{N}a_j(t) \phi_j(y,z) \|_2^2 \rangle_t~,~\mbox{with}~ a_j(t)=(\phi_j|u')~
\label{eq:optimize}
\end{equation}

Note that, in our case and in the rest of this work, only the stream-wise turbulent kinetic energy is taken into account.

Using a variational approach, one can show that the described minimization problem is equivalent to the eigenvalue
equation of the covariance operator.
\begin{equation}
\integrate{}{} dy'dz'~ \langle u'(y,z,t)u'^{*}(y',z',t) \rangle_{t}~ \phi_j(y',z')=\lambda_j \phi_j(y,z) ~\mbox{with}~ \lambda_1>\lambda_2>...
\label{eq:evproblem}
\end{equation}

This equivalence is the main reason why the POD is a simple, but powerful tool. 
Instead of treating a~complex optimization problem, we can now simply solve an eigenvalue problem of an
Hermitian operator. In practice, Equation (\ref{eq:evproblem}) is
often approximated by the eigenvalue problem of the discretized covariance matrix
$C_{ij}=\langle u_i(t)u_j^{*}(t) \rangle_{t}$, where $u_k$ is the value of $u$ at the k
-th grid
point.

We can conclude from the Hilbert--Schmidt theory that the modes are orthogonal with $(\phi_i|\phi_j)=\delta_{ij}$
and that the covariance operator is diagonal in the POD basis, yielding a linearly decorrelated description:
\begin{equation}
 \langle a_i(t)^{*}a_j(t)\rangle_t=\lambda_{i} \delta_{ij}~
\end{equation}
\noindent
Due to Parseval's relation:
\begin{equation}
 \langle \| u'(y,z,t) \|^2_2 \rangle_t:=\langle \integrate{}{} dy dz~ |u'(y,z,t)|^2 \rangle_t = \sum\limits_{j=1}^{\infty}\langle|a_j|^2\rangle_t~~
 \mbox{and}~~\langle|a_j(t)|^2\rangle_t=\lambda_j~
 \label{eq:parseval}
\end{equation}
the $\mbox{j}^{\mbox{th}}$ eigenvalue can be interpreted as a measure for the kinetic energy contained in 
the $\mbox{j}^{\mbox{th}}$ mode yielding for the mean squared error of a 
reconstruction: 
\begin{equation}
\langle \| u'-{u'}^{(N)} \|_2^2 \rangle_t =\sum\limits_{j=N+1}^{\infty}\lambda_j=
\langle \| u' \|_2^2- \| {u'}^{(N)} \|_2^2 \rangle_t~
\label{eq:rerror}
\end{equation}
where ${u'}^{(N)}=\sum\limits_{j=1}^{N}a_j(t) \phi_j(y,z)$.
For many applications, the method of snapshots (see, e.g., \cite{BERKOOZ1993}) is used instead of directly solving Equation (\ref{eq:evproblem}). Here,
we solve the direct problem, since the number of time steps used for the analysis is of the same order as the number of grid points. Therefore, the method of snapshots is not more
efficient.

We could now apply the POD method to the data field $u(y,z,t)$.
The results based on this standard decomposition are discussed in Appendix \ref{app:classicpod}. 
In the next section, we introduce a slightly different approach, which is used throughout the rest of this work.

\subsection{Preprocessing and Modified Decomposition}
\label{sec:preprocessing}

In this work, we aim to describe the wake in the turbulent ABL without describing the dynamics of the ABL itself. Thus,
in the following, we introduce two preprocessing steps that separate the wake structure
from the ABL flow. Note that despite this separation, the extracted wake structure
still contains information about the response of the wake to the ABL dynamics.

As the first preprocessing step, the time-averaged velocity field of the flow without turbine (Figure~\ref{fig:q0mean})
is subtracted from
$u(y,z,t)$. This leads to a slightly better separation from the ABL structures (Figure~\ref{fig:qsnap0b}).
Second, we extract the deficit by using a (temporally local) relative threshold. 
This means that we set all values smaller than 40\% of the current deficit maximum to zero
(Figure~\ref{fig:qsnap0c}). A similar procedure has also been applied in \cite{Espana2011}. 
The results presented in the following are relatively robust against the choice of the threshold. Similar
results are obtained for choosing 20\%--50\%. In the following, the preprocessed field will be called $\tilde{u}(y,z,t)$,
with the corresponding POD modes $\tilde{\phi_j}$.

In Section \ref{sec:podrecon}, we apply the obtained modes $\tilde{\phi_j}$ to reconstruct the original velocity field
$u(y,z,t)$ (not $\tilde{u}$). 
Thus, Equation (\ref{eq:decomposition}) becomes:
\begin{equation}
 u^{(N)}(y,z,t):=\langle u(y,z,t) \rangle_t + \sum\limits_{j=1}^{N}{a}_j(t) \tilde{\phi}_j(y,z)~,~\mbox{with}~{a}_j:=(\tilde{\phi}_j|u')~
 \label{eq:podreconalt}
\end{equation}
This way, we do not investigate a standard proper orthogonal decomposition of $u$, but a modified
decomposition of $u$ using the POD modes of the preprocessed field $\tilde{u}$. In Appendix \ref{app:classicpod},
we briefly point out the advantages of this alternative approach, which probably stem from the fact that
the modes $\tilde{\phi_j}$ focus solely on the wake structure. To model the complete inflow of a turbine in the
wake, a model, such as~\cite{Mann1998}, could be added to model the ambient field.

It is important to note that similar results as presented in this work
can be obtained when using $\tilde{\phi}_j$ for a reconstruction
of the corresponding preprocessed field $\tilde{u}$. 

\subsection{POD Modes of the Preprocessed Field}
\label{sec:podmodes}

\begin{figure}[H]
\begin{center}
\begin{subfigure}[H]{.32\textwidth}
 \includegraphics[width=.98\linewidth]{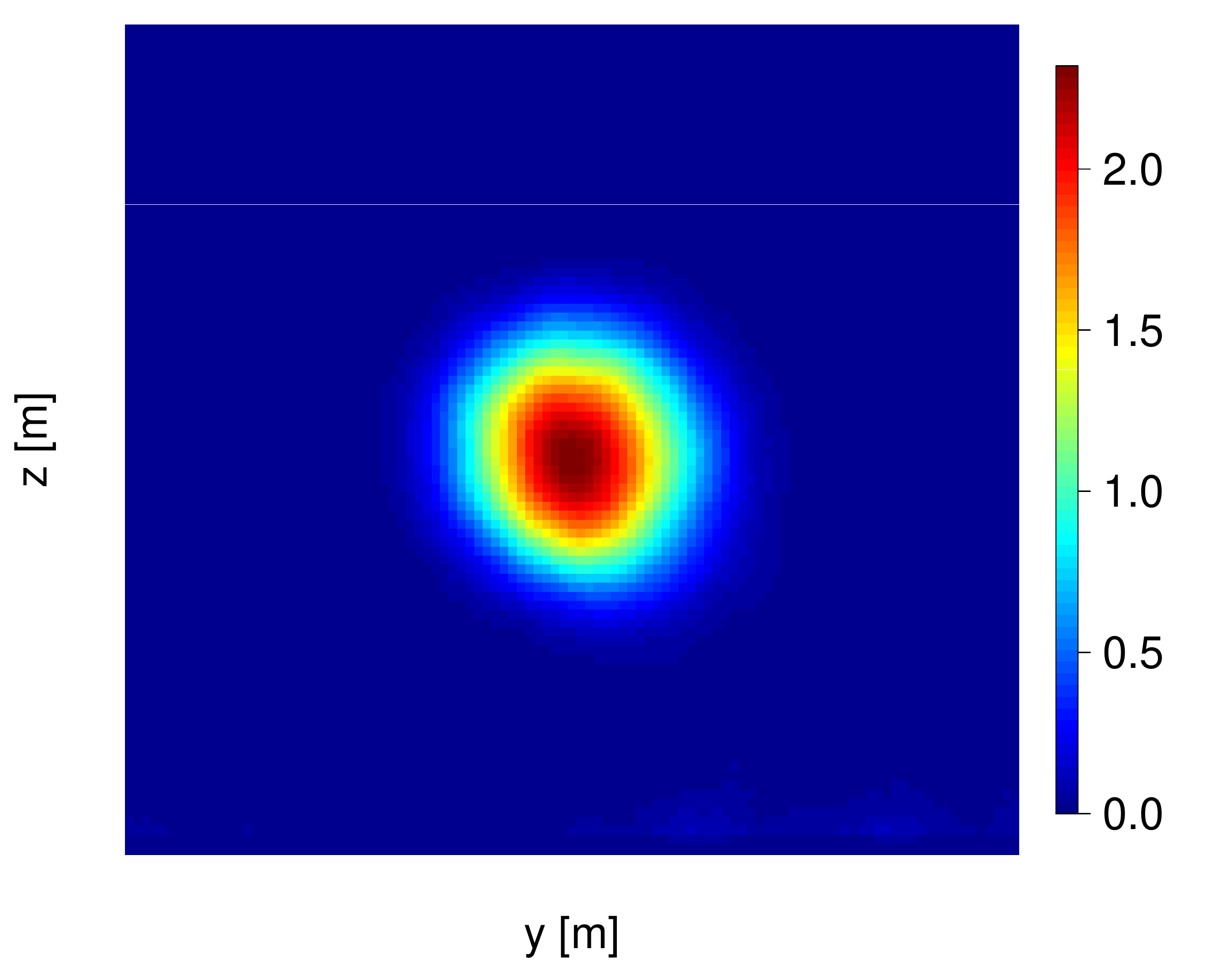}
 \subcaption{}
 \label{fig:qcutmean}
\end{subfigure}%
\vspace {-18pt}
\begin{subfigure}[H]{.32\textwidth}
 \includegraphics[width=.98\linewidth]{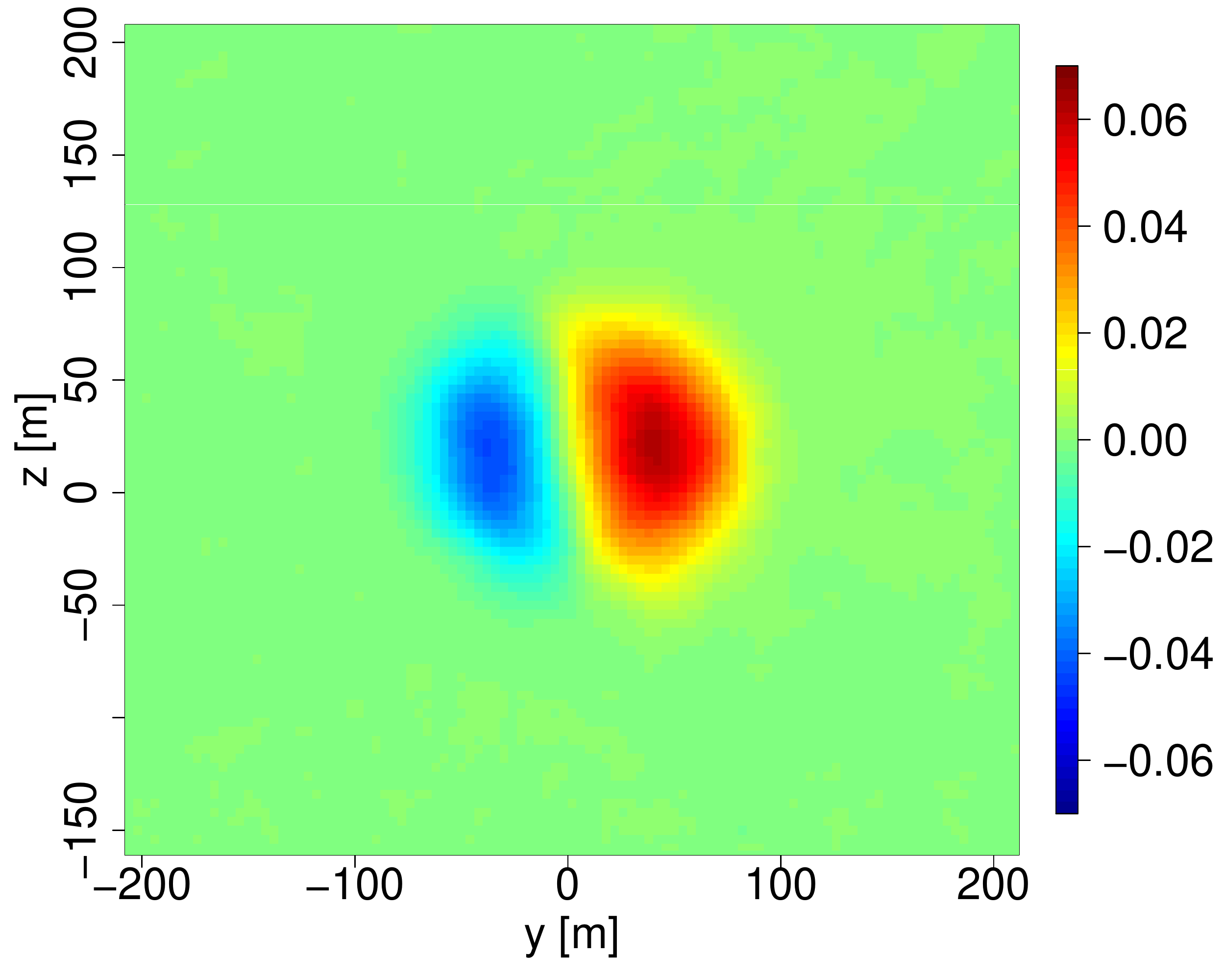}
 \subcaption{}
 \label{fig:mode1}
\end{subfigure}
\begin{subfigure}[H]{.32\textwidth}
 \includegraphics[width=.98\linewidth]{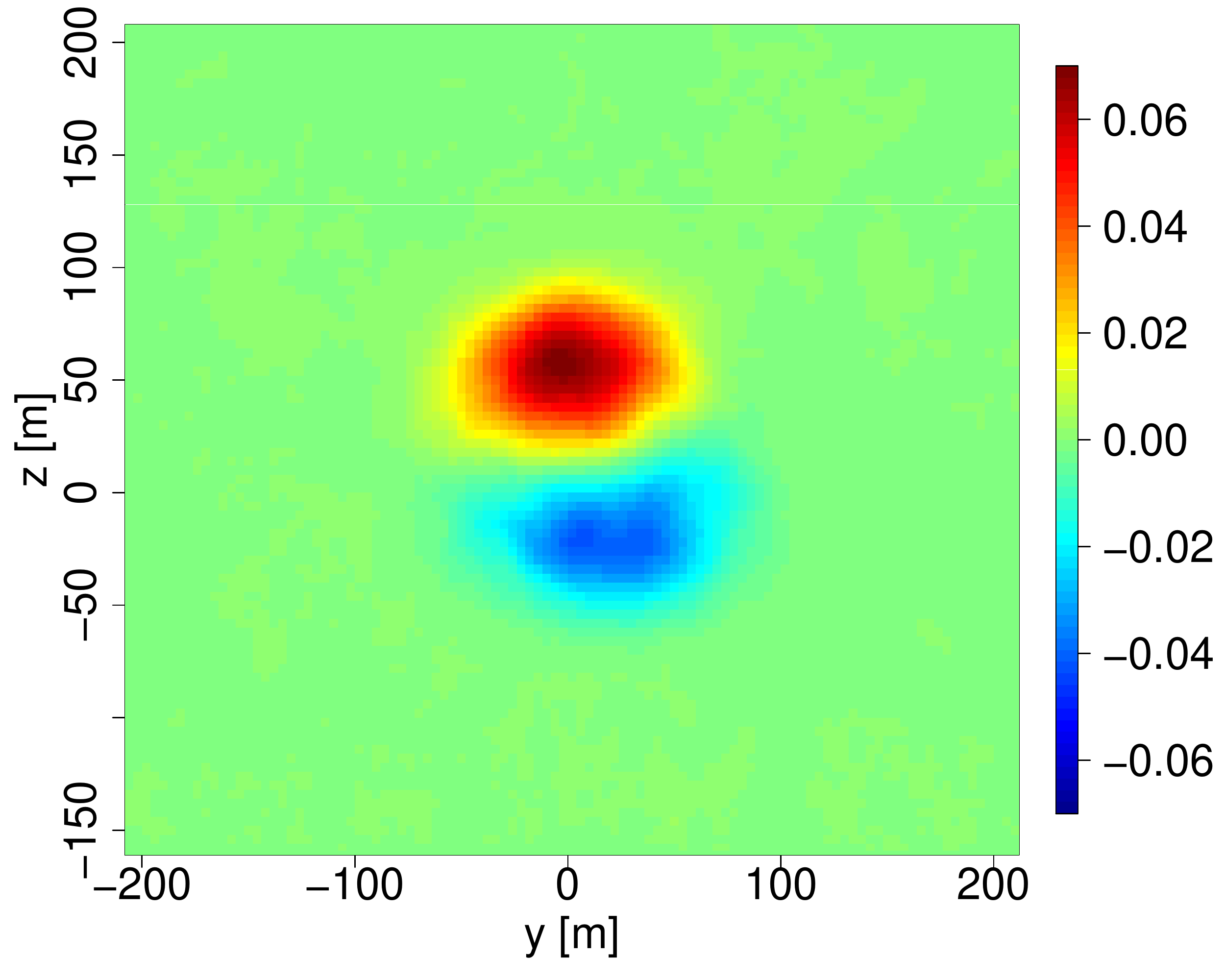}
 \subcaption{}
 \label{fig:mode2}
\end{subfigure}%
\end{center}
\begin{center}
\begin{subfigure}[H]{.32\textwidth}
 \includegraphics[width=.98\linewidth]{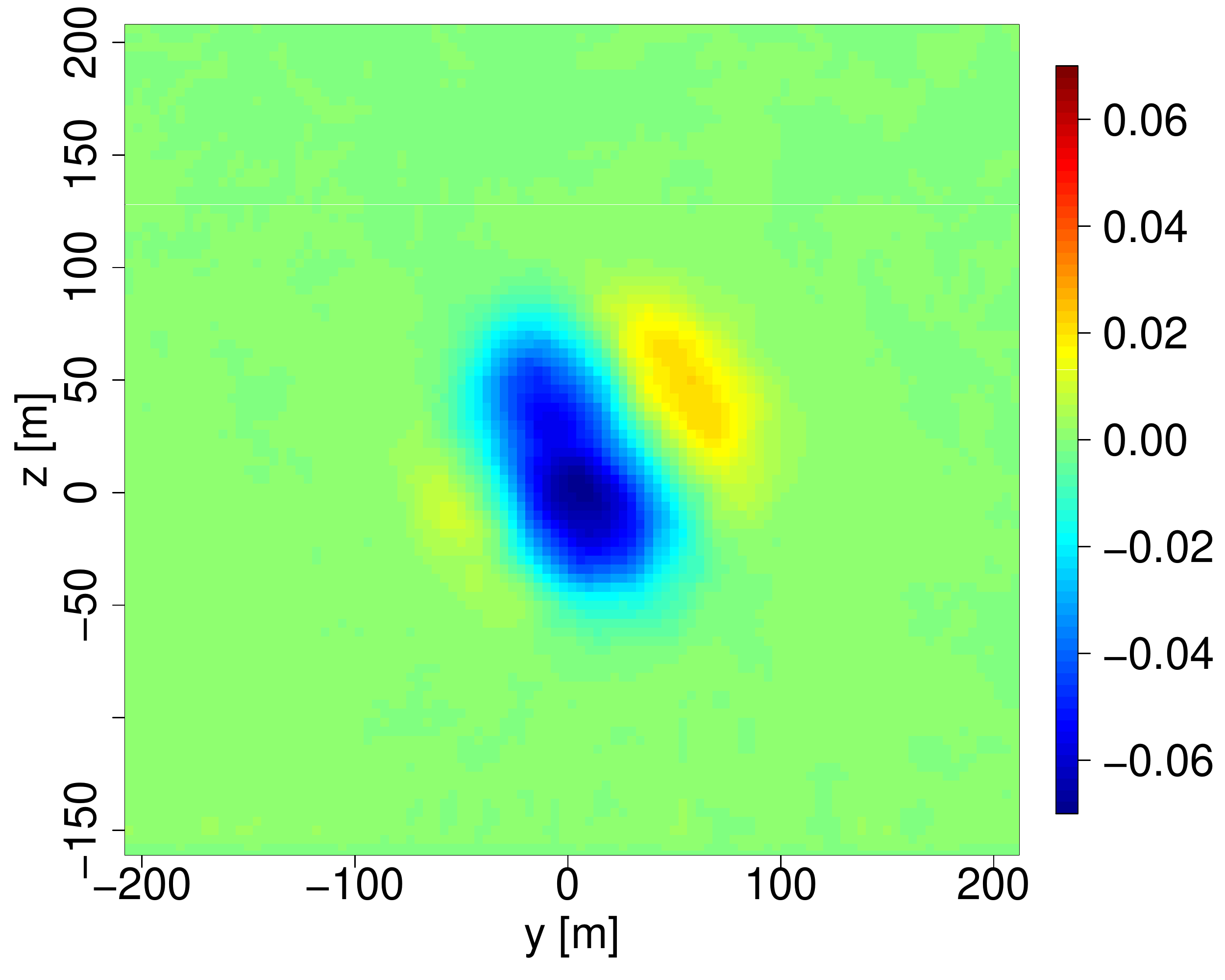}
 \subcaption{}
 \label{fig:mode3}
\end{subfigure}%
\vspace {-18pt}
\begin{subfigure}[H]{.32\textwidth}
 \includegraphics[width=.98\linewidth]{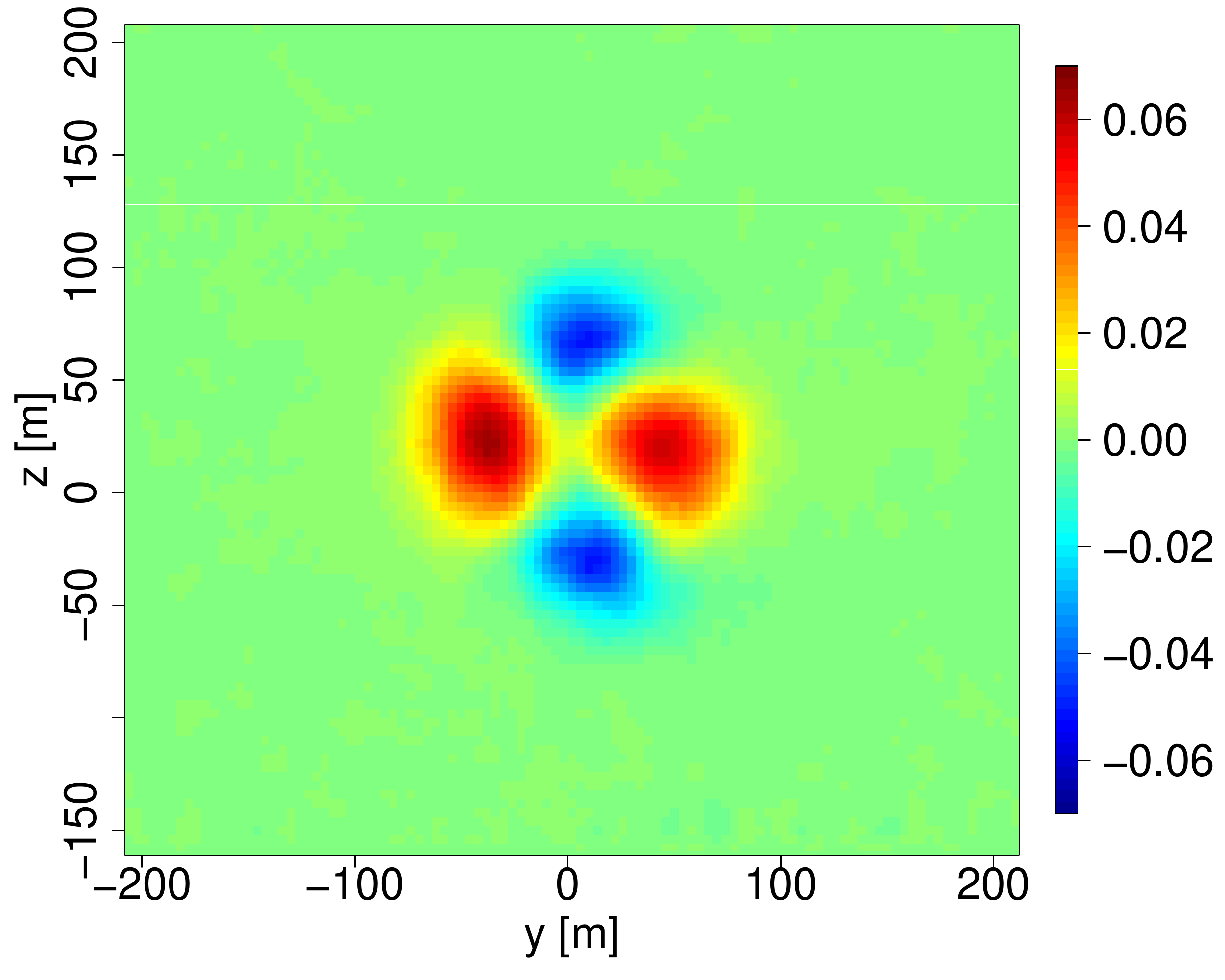}
 \subcaption{}
 \label{fig:mode4}
\end{subfigure}
\begin{subfigure}[H]{.32\textwidth}
 \includegraphics[width=.98\linewidth]{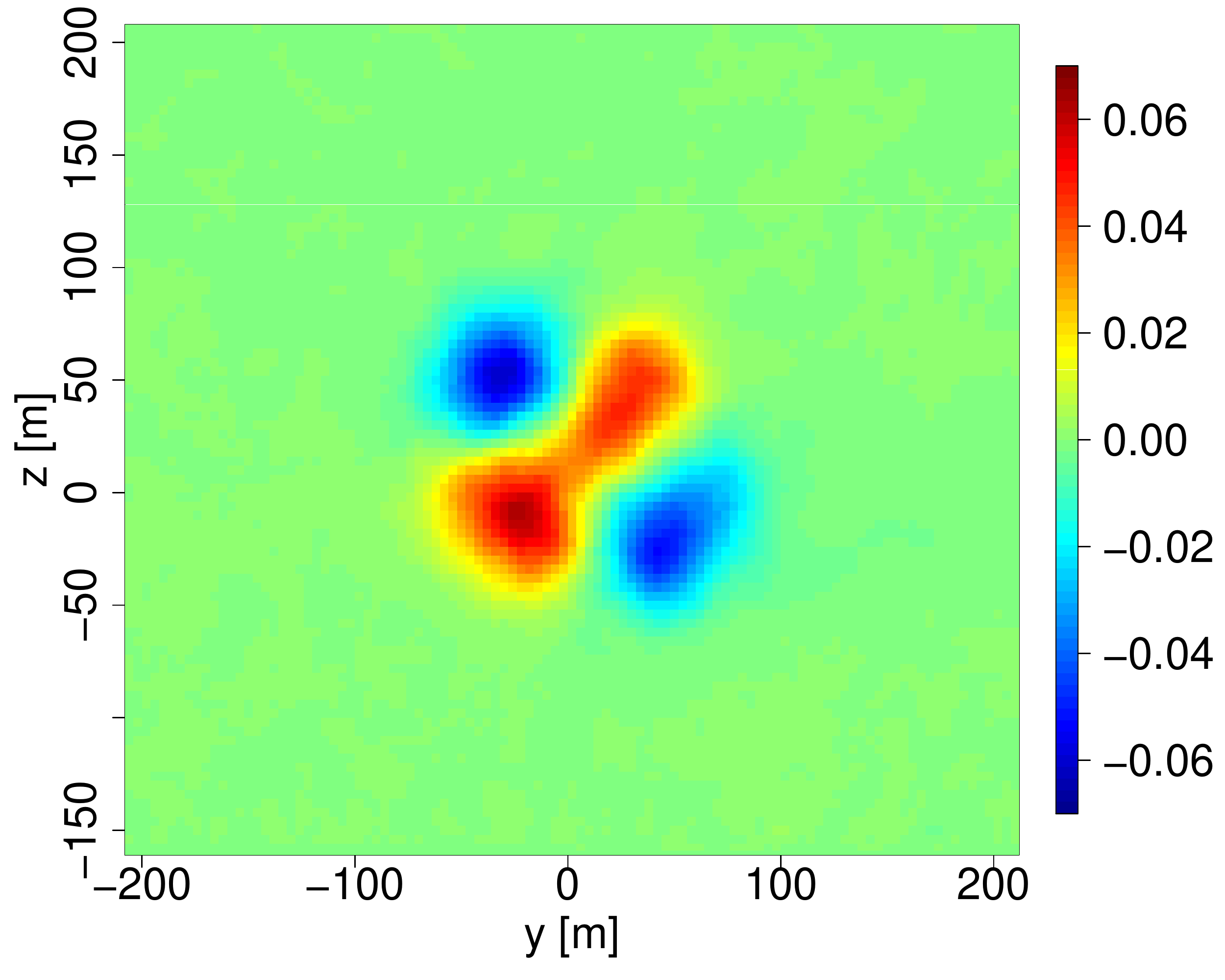}
 \subcaption{}
 \label{fig:mode5}
\end{subfigure}%
\end{center}
\begin{center}
\begin{subfigure}[H]{.32\textwidth}
 \includegraphics[width=.98\linewidth]{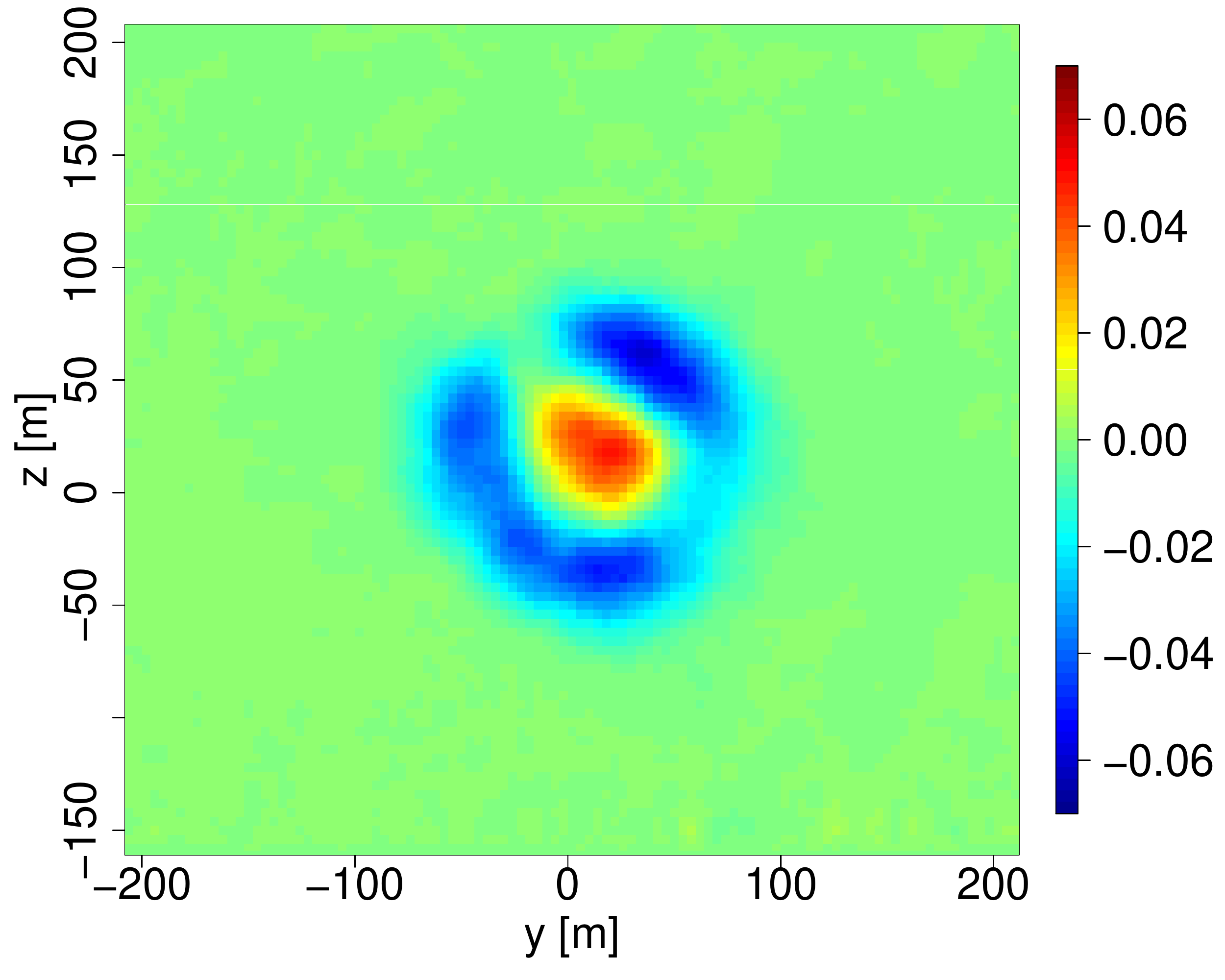}
 \subcaption{}
 \label{fig:mode6}
\end{subfigure}%
\vspace {-18pt}
\begin{subfigure}[H]{.32\textwidth}
 \includegraphics[width=.98\linewidth]{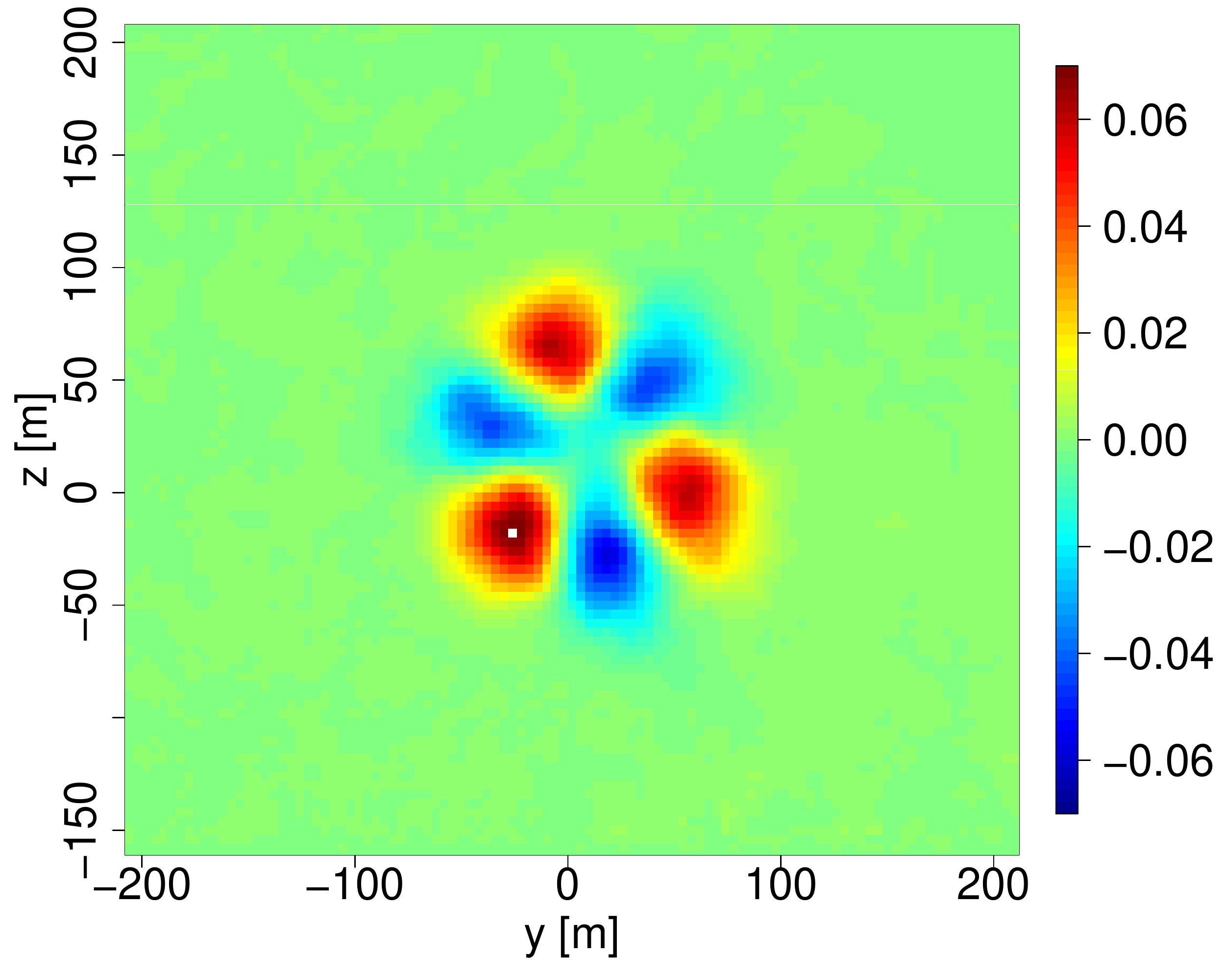}
 \subcaption{}
 \label{fig:mode7}
\end{subfigure}
\begin{subfigure}[H]{.32\textwidth}
 \includegraphics[width=.98\linewidth]{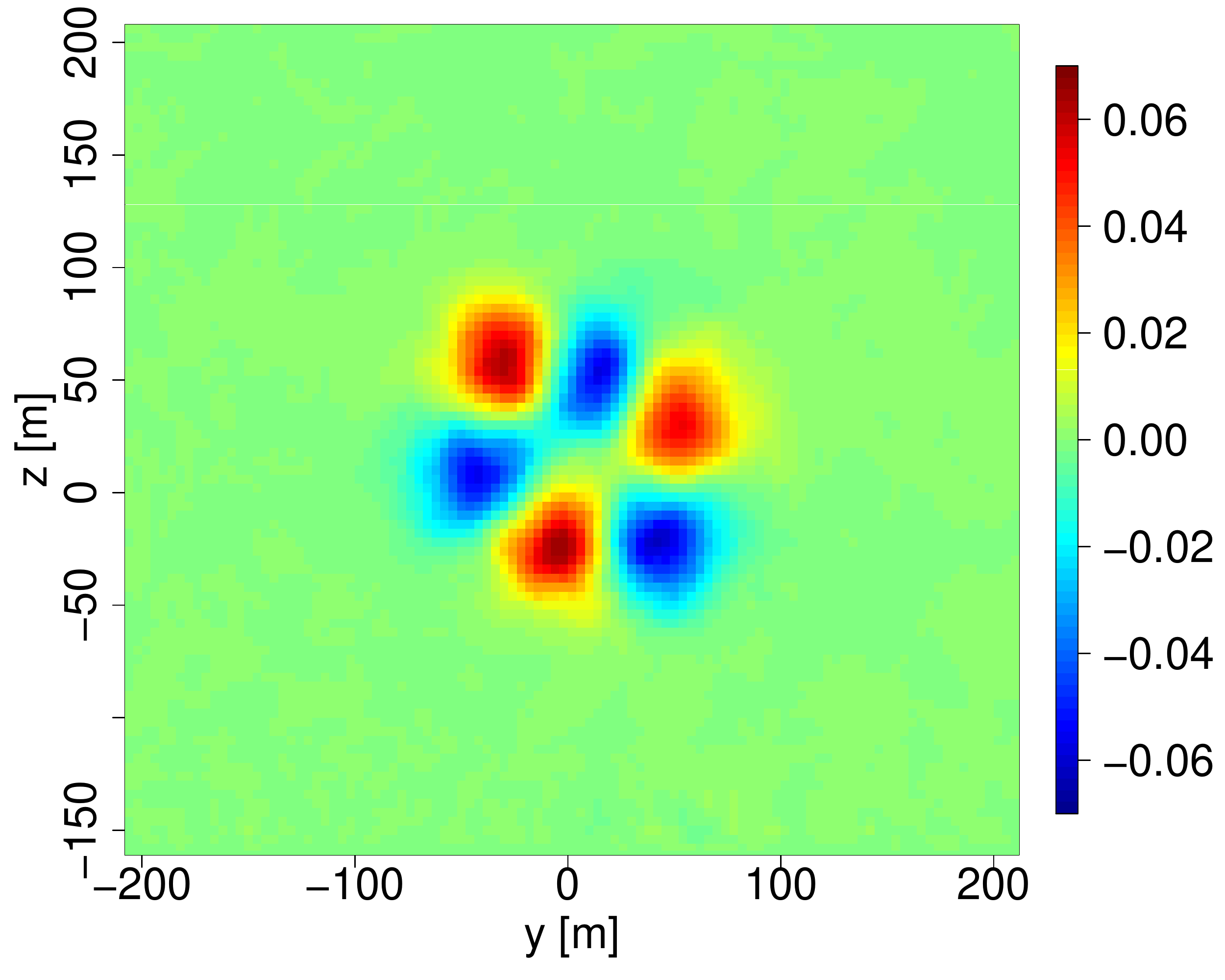}
 \subcaption{}
 \label{fig:mode8}
\end{subfigure}
\end{center}

\caption{Mean field and POD modes of the preprocessed field $\tilde{u}$. (\textbf{a}) Mean field; (\textbf{b})~Mode~1; (\textbf{c}) Mode 2; (\textbf{d}) Mode 3; (\textbf{e}) Mode 4; (\textbf{f}) Mode 5; (\textbf{g}) Mode 6; (\textbf{h}) Mode 7; (\textbf{i}) Mode 8.}
\label{Fig:qcutmean}
\end{figure}

Next, we calculate the covariance matrix corresponding to the preprocessed field $\tilde{u}$ and
numerically solve its eigenvalue problem analogous to Equation (\ref{eq:evproblem}). 
The obtained eigenvectors are the POD modes $\tilde{\phi}_j$,
which are shown in Figure \ref{Fig:qcutmean}.
The modes show clear structures with a trend from larger to smaller scales with increasing mode number.
This trend is easily explained by the fact that the modes are sorted with respect to energy
and that the kinetic energy in a turbulent flow typically decreases with scale.

\begin{figure}[H]
\centering
\begin{subfigure}[t]{.45\textwidth}
 \centering
 \includegraphics[width=.95\linewidth]{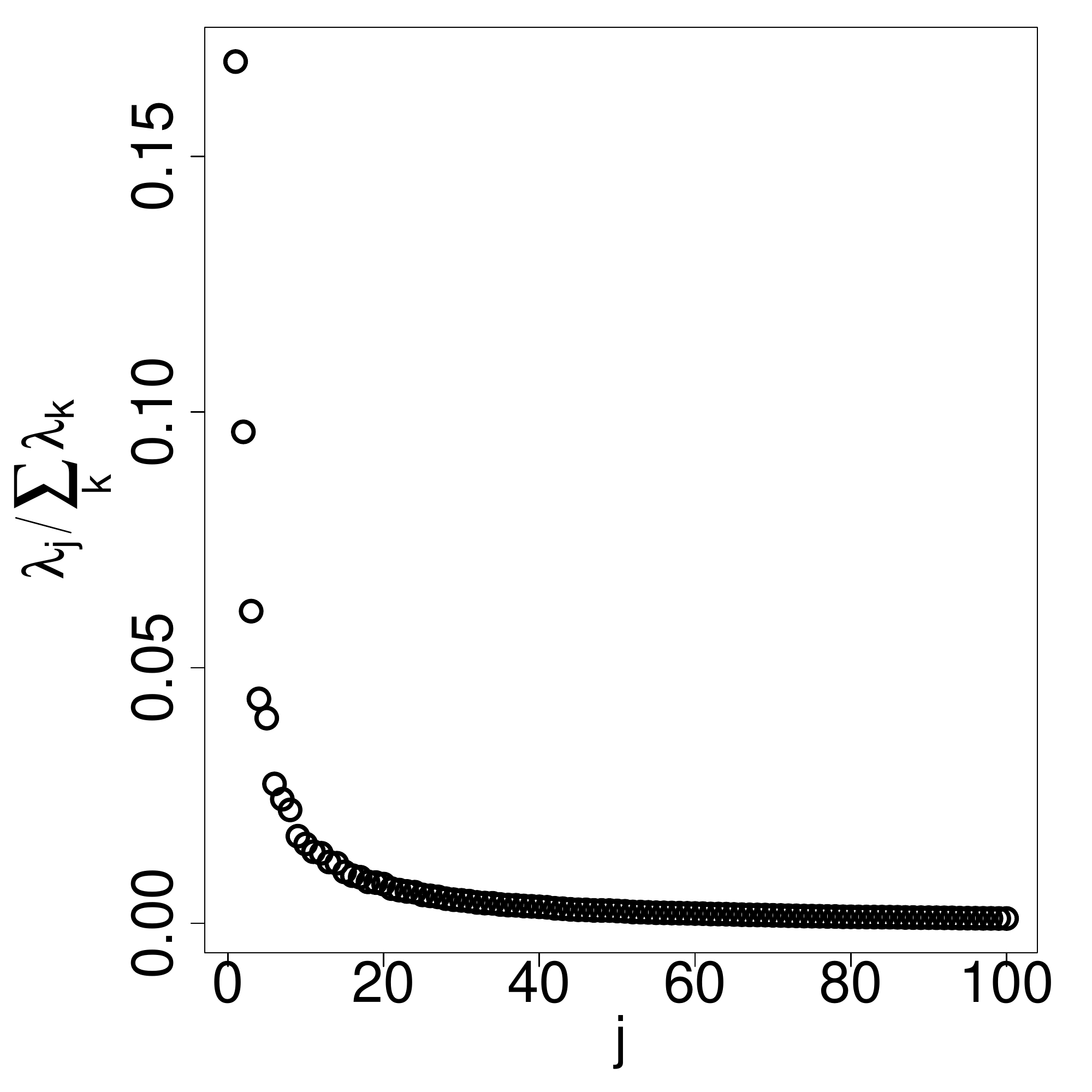}
 \subcaption{}
 \label{fig:podspectrum1}
\end{subfigure}%
\begin{subfigure}[t]{.45\textwidth}
 \centering \includegraphics[width=.95\linewidth]{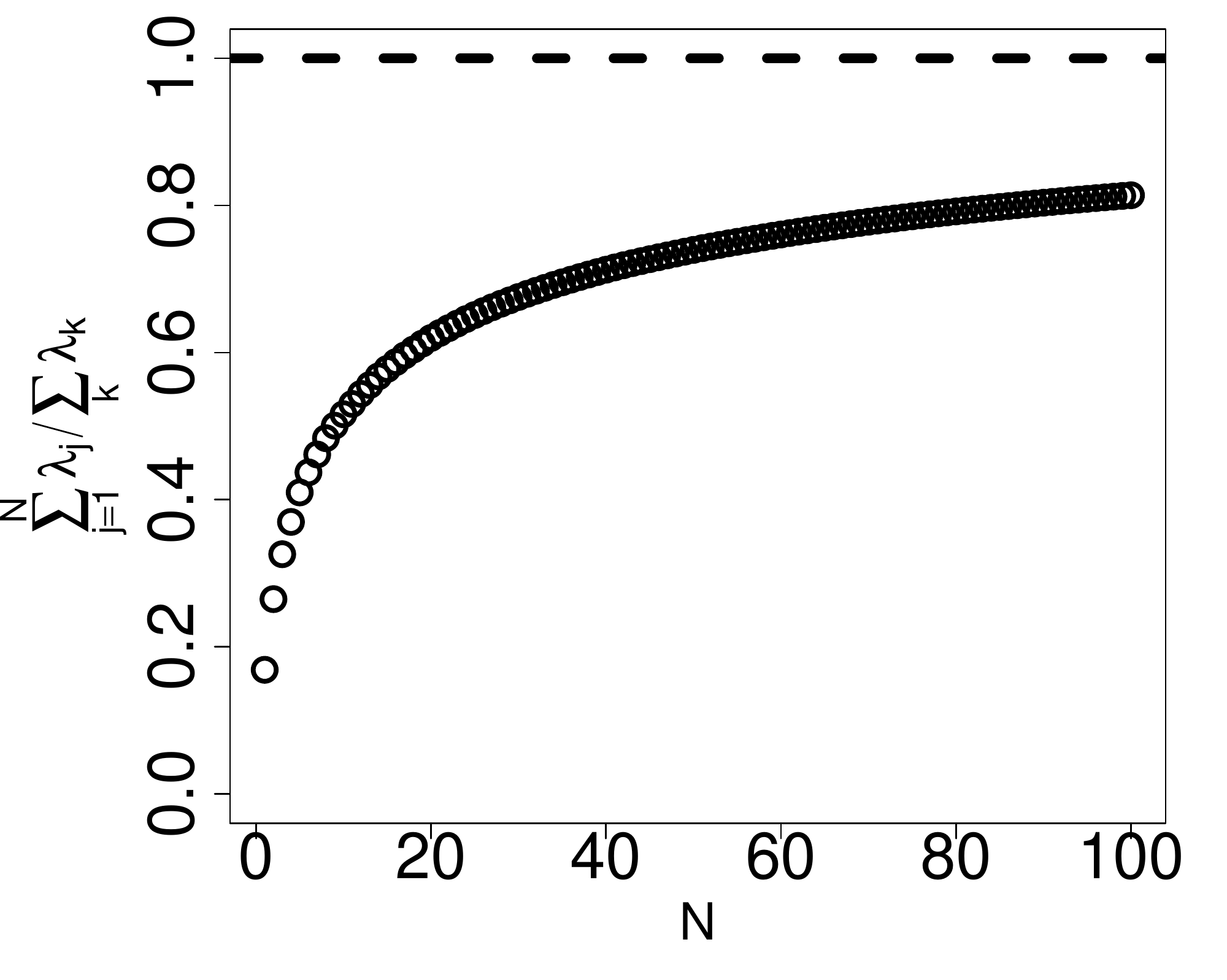}

 \subcaption{}
 \label{fig:podspectrum1c}
\end{subfigure}
\caption{(\textbf{a}) POD eigenvalues of the preprocessed field $\tilde{u}$;
(\textbf{b}) cumulative energy \textit{versus} the number of modes $N$.}
\label{Fig:podspectrum1}
\end{figure}

The modes resemble Fourier modes in the azimuthal direction. We find roughly rotationally symmetric 
Modes $3$ and $6$ and pairs showing a dipole-like structure $(1,2)$, quadrupole structure $(4,5)$ and hexapole
structure $(7,8)$.
These observations indicate an approximate statistical isotropy \cite{BERKOOZ1993}. This is a remarkable
result, since the rotational symmetry is clearly broken by the ABL.
The extracted wake structure therefore behaves approximately isotropic in the non-isotropic ABL flow. 
Obviously, the rotational symmetry is not perfect. Mode $3$, for example, also indicates the
presence of principal axes that do not coincide with the horizontal and vertical direction.
Furthermore,
the eigenmode pairs described above do not correspond to the same eigenvalue
(see Figure \ref{fig:podspectrum1}). 

It is important to note that the symmetric properties of the wake structure are revealed
through the threshold applied to the field. The original field includes more
information about the spatial interaction region between wake and ABL. Consequently, the original
field behaves less isotropic, which can also be seen in the corresponding POD modes, as 
described in Appendix \ref{app:classicpod}.

To illustrate the importance of the different modes, we show the POD spectrum and its cumulative
version in Figure \ref{Fig:podspectrum1}. According to Equation (\ref{eq:rerror}), these figures reveal
that we need around $100$ modes to recover around 80\% of the turbulent kinetic energy
of the field $\tilde{u}$.
This is not surprising, since in a turbulent flow, the energy 
is spread out over a wide range of scales. The POD basis does not solve this problem, even though
it might perform better in inhomogeneous fields than a standard Fourier basis.

The convergence behavior of the POD modes and values is discussed briefly in Appendix \ref{app:convergence}.

\subsection{Reconstructions of the Velocity Field}
\label{sec:podrecon}

\begin{figure}[H]
\begin{center}
\begin{subfigure}[H]{.32\textwidth}
 \includegraphics[width=.98\linewidth]{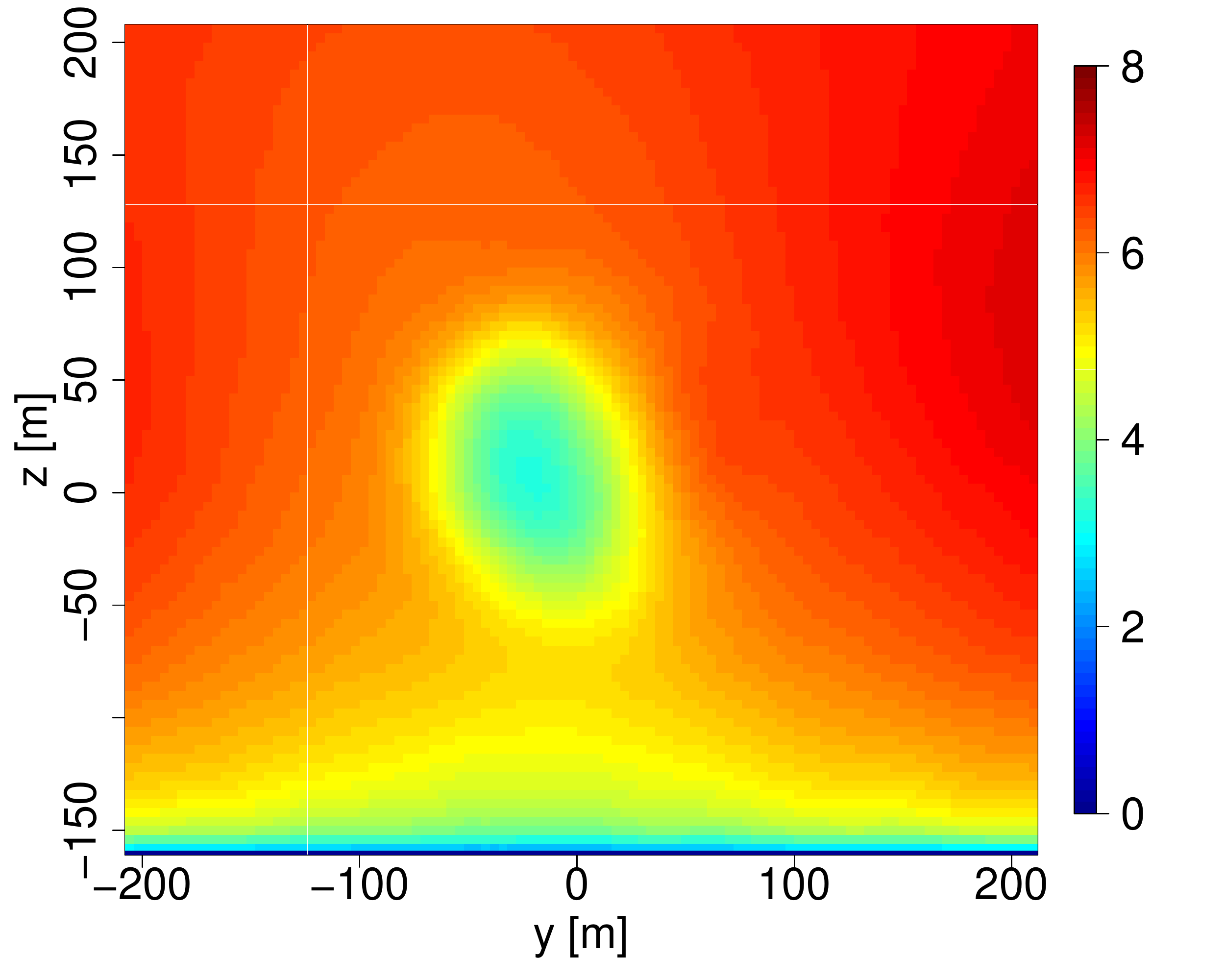}
 \subcaption{}
 \label{fig:recon2}%
 \end{subfigure}%
\vspace {-12pt}
\begin{subfigure}[H]{.32\textwidth}
 \includegraphics[width=.98\linewidth]{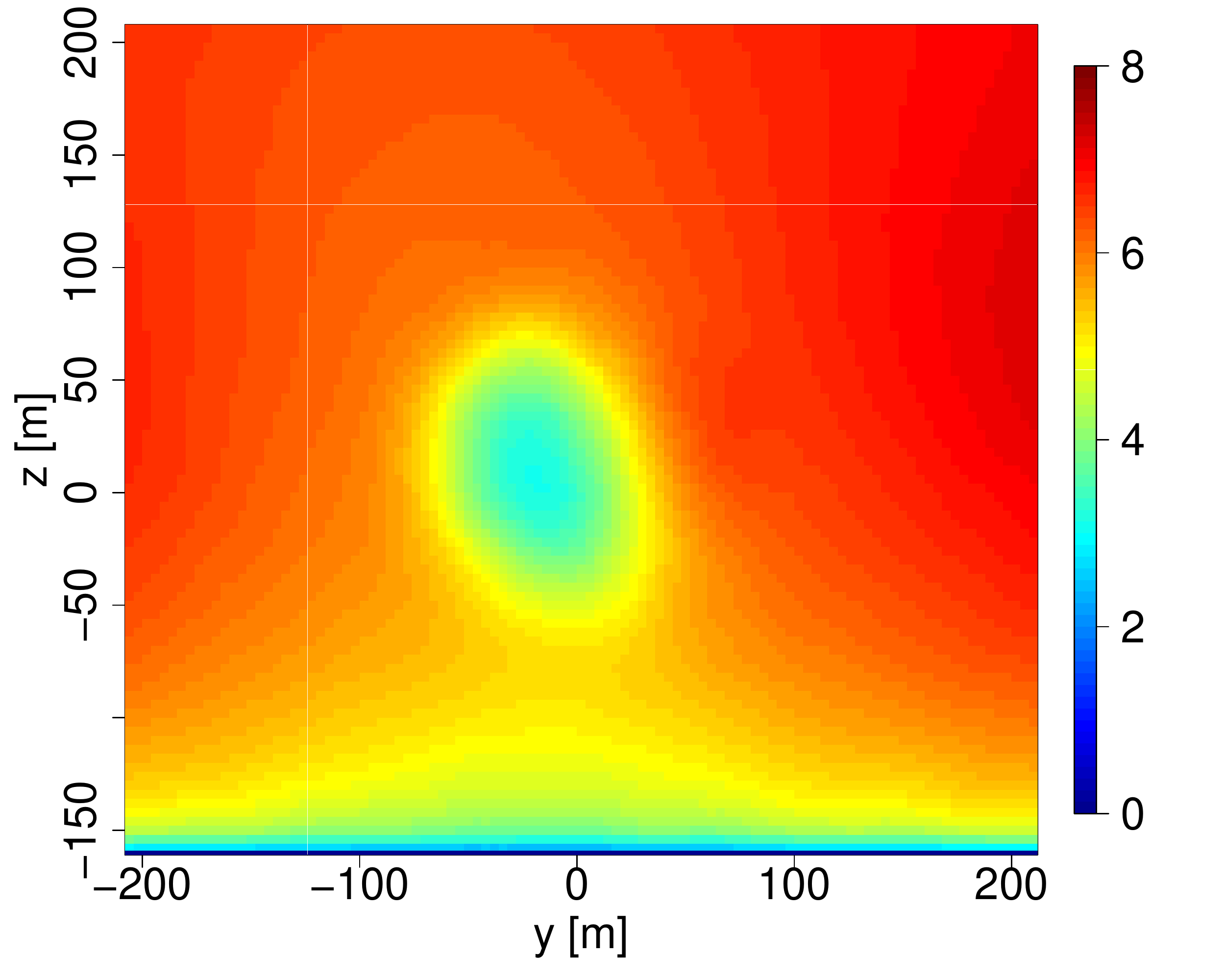}
 \subcaption{}
 \label{fig:recon3}
\end{subfigure}
\begin{subfigure}[H]{.32\textwidth}
 \includegraphics[width=.98\linewidth]{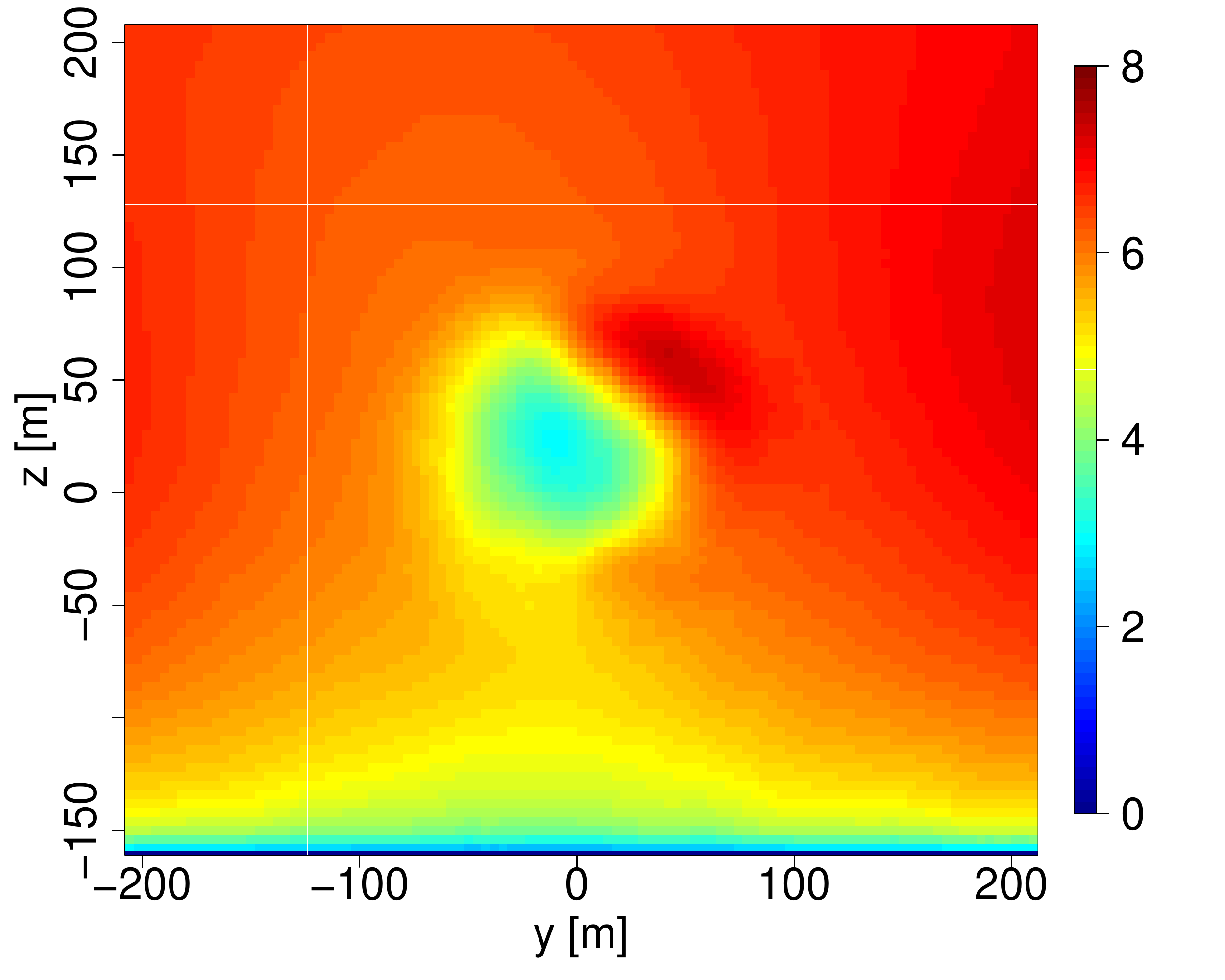}
 \subcaption{}
 \label{fig:recon6}
\end{subfigure}
\end{center}%
\begin{center}
\begin{subfigure}[H]{.32\textwidth}
 \includegraphics[width=.98\linewidth]{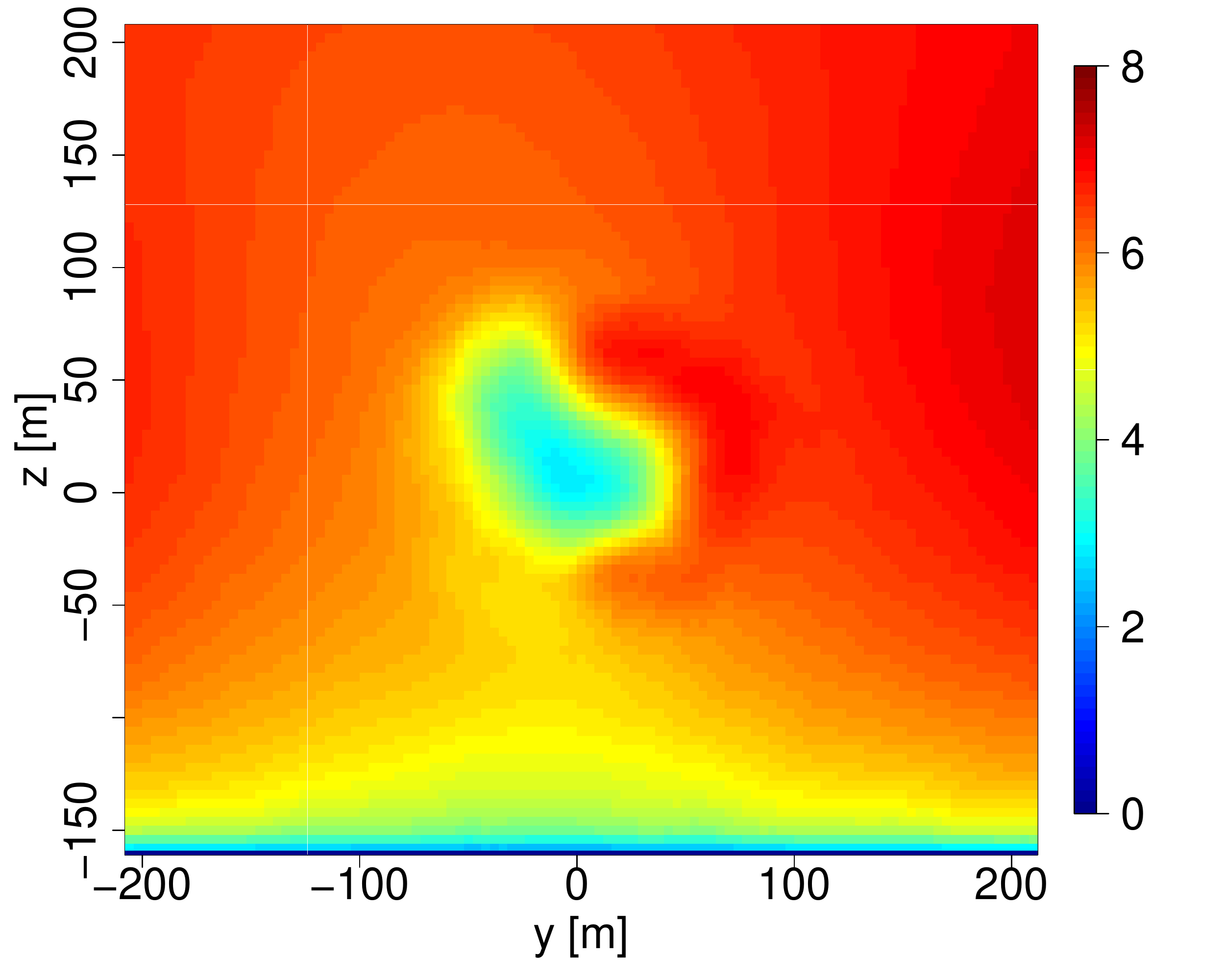}
 \subcaption{}
 \label{fig:recon10}%
 \end{subfigure}%
\vspace {-12pt}
\begin{subfigure}[H]{.32\textwidth}
 \includegraphics[width=.98\linewidth]{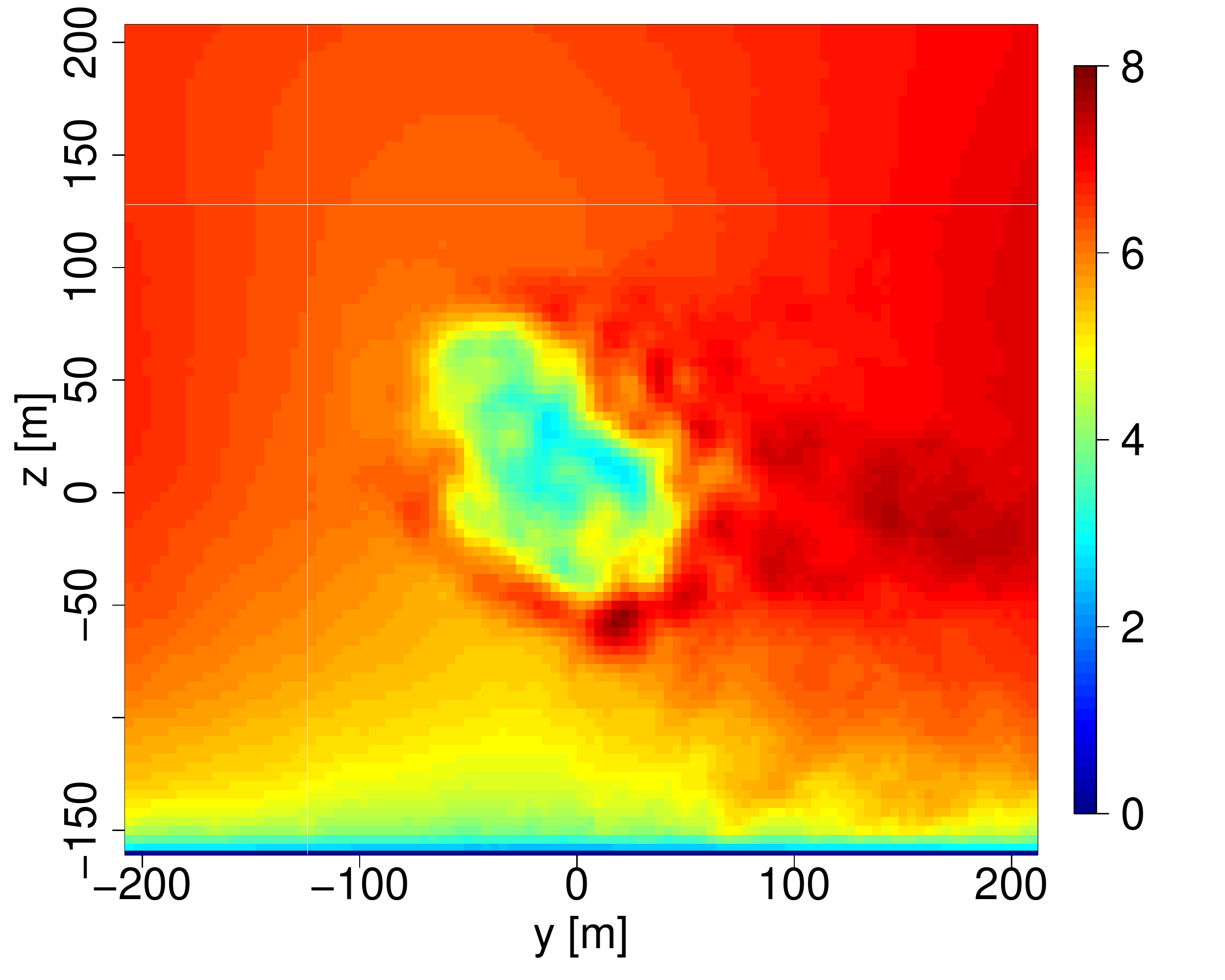}
 \subcaption{}
 \label{fig:recon100}
\end{subfigure}
\begin{subfigure}[H]{.32\textwidth}
 \includegraphics[width=.98\linewidth]{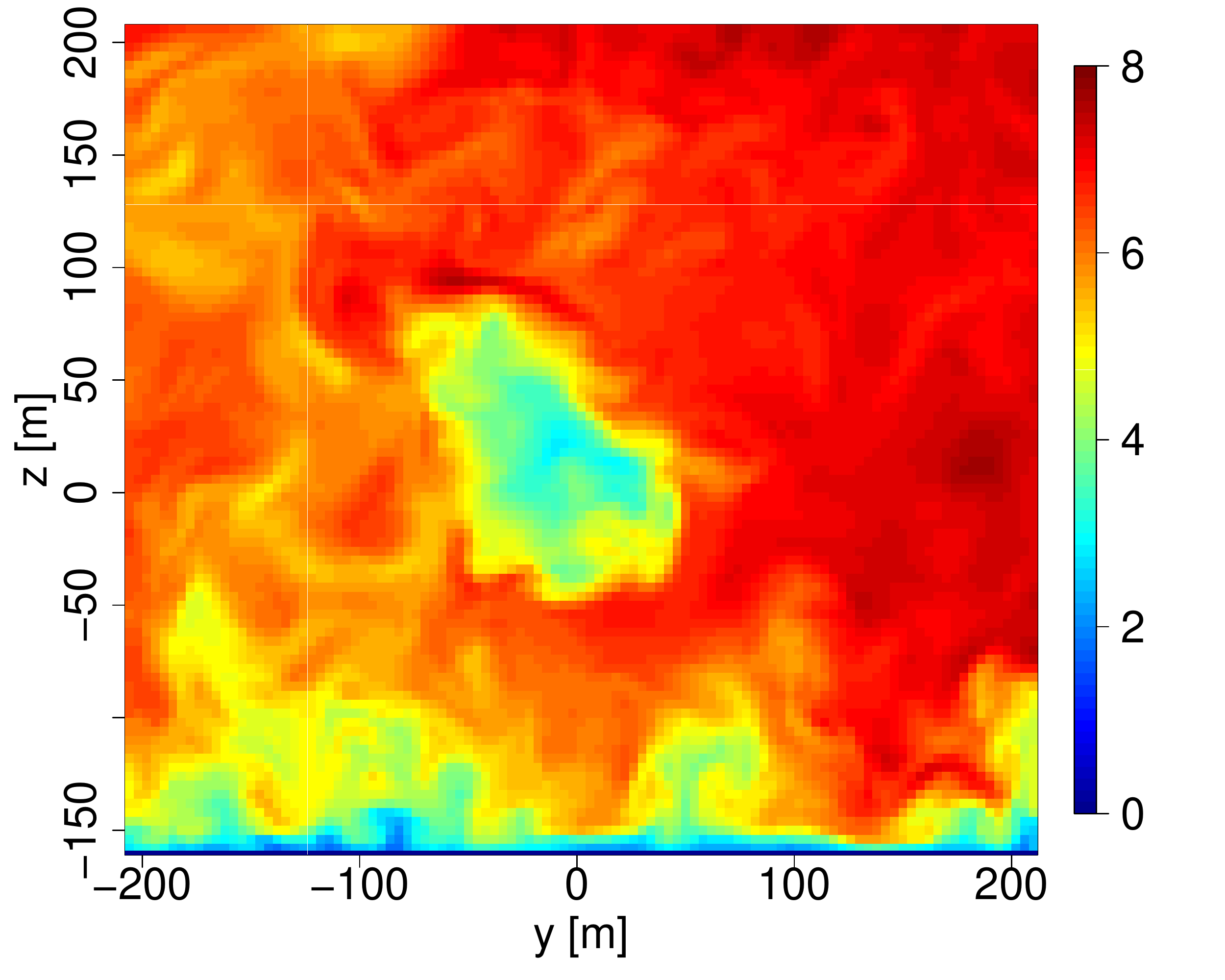}
 \subcaption{}
 \label{fig:reconfull}
\end{subfigure}
\end{center}%
\caption{Reconstructions of the velocity field $u$ following Equation \eqref{eq:podreconalt} ($t$ = 1000 s). \protect\linebreak (\textbf{a}) Two modes; (\textbf{b}) three modes; (\textbf{c}) six modes; (\textbf{d}) 10 modes; (\textbf{e}) 100 modes; (\textbf{f})~full~field.}
\label{Fig:recon2}
\end{figure}

We now use the extracted POD modes $\tilde{\phi_j}$ to approximate the original 
field $u(y,z,t)$ according to Equation (\ref{eq:podreconalt}) and vary the number of modes $N$ used for
the reconstruction.
Examining the reconstructed snapshots (Figure \ref{Fig:recon2}), we see that more modes obviously
lead to a more accurate description of the wake structure. To recover the spatial small-scale structures of the wake,
many POD modes are necessary. 
Unfortunately, this pure visualization does not offer many clues for the number of modes
necessary to obtain a useful low order description. 
The slow increase of the recovered turbulent kinetic energy with $N$, discussed
in the former section, indicates that useful reduced order
models need to contain a lot of modes. However, in the context of wake modeling, we
are also interested in other quantities describing the impact of the turbine on the wake.
Therefore, we propose the use of other quantitative measures of quality in the next section.

\section{Alternative Measures for the Quality of Reconstructions}\vspace{-18pt}
\label{sec:alternative}
\subsection{A General Quality Measure}
\label{sec:generalmeasure}

Our main interest lies in the impact of the wake on a sequential turbine. 
A measure used to draw conclusions on the quality of a reconstructed field should therefore be
related to forces or other effects on a turbine.
Such a measure could, for example, be given by a specific load or the power a turbine would
produce if it were standing in the wake. 
Thus, we define a general scalar quality measure $M(t)$ as the result
from a corresponding operator $\fat{\hat{M}}$ applied to the velocity field:
\begin{equation} 
 \fat{\hat{M}}[u(y,z,t)]=M(t)~
 \label{eq:mdef}
\end{equation}

\noindent
Usually, $M$ will depend on the field $u(y,z,t)$ confined to the virtual rotor area only (Figure
\ref{fig:qsnapcirc}) and not on the velocity in the rest of the domain.
Analogous to Equation (\ref{eq:mdef}), we define for a reconstructed field:
\begin{equation}
 M^{(N)}(t):=\fat{\hat{M}}[u^{(N)}(y,z,t)]~
 \label{eq:mdef2}
\end{equation}

\noindent
The quality of a reconstructed field can now be evaluated by comparing $M^{(N)}(t)$ and $M(t)$.
If $M^{(N)}(t)$ is a good approximation of $M(t)$, a reconstruction with $N$ modes yields
a good description of the wake with respect to the measure $M(t)$.
To quantify the difference of $M^{(N)}(t)$ and $M(t)$, we define two different errors.
The first error is given through the two-norm 
of the time series:

\begin{equation}
 \varepsilon_{\mbox{std}}(N)=\frac{\|M^{(N)}-M\|_2}{\| M \|_2}~
 \label{eq:errorstd}
\end{equation}

\noindent
which will be referred to as the {standard error} in the following.
Another possible choice is motivated by the idea that sometimes, not the mean $\langle M \rangle_t$ is important, but the fluctuations $M(t)-\langle M \rangle_t$. 
It is, e.g., often assumed that for the calculation of fatigue loads, the influence of the mean is negligible. Hence, we~define:\vspace{6pt}
\begin{equation}
 \varepsilon_{\mbox{dyn}}(N)=\frac{\|(M^{(N)}-\langle M^{(N)} \rangle_t)- (M - \langle M \rangle_t)\|_2}{\| M - \langle M \rangle_t \|_2}~
 \label{eq:errordyn}\vspace{6pt}
\end{equation}
which will be referred to as the {dynamical error} in the following.

\subsection{Introducing Alternative Measures}
\label{sec:measures}

\begin{figure}[H]
\centering
\begin{subfigure}[H]{.32\textwidth}
 \centering
 \includegraphics[width=.96\linewidth]{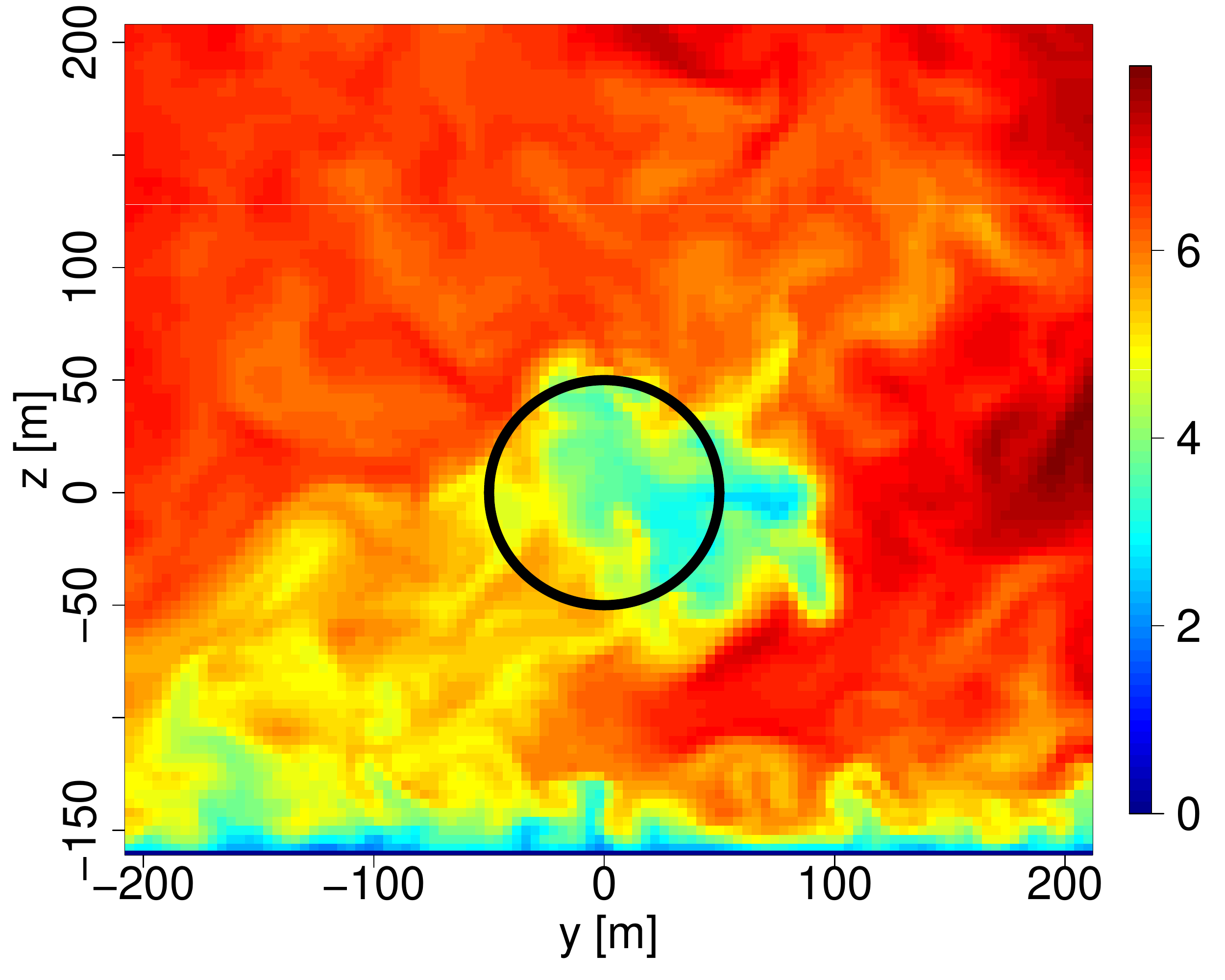}
 \subcaption{
 }
 \label{fig:qsnapcirc}
\end{subfigure}%
\begin{subfigure}[H]{.32\textwidth}
 \centering
 \includegraphics[width=.99\linewidth]{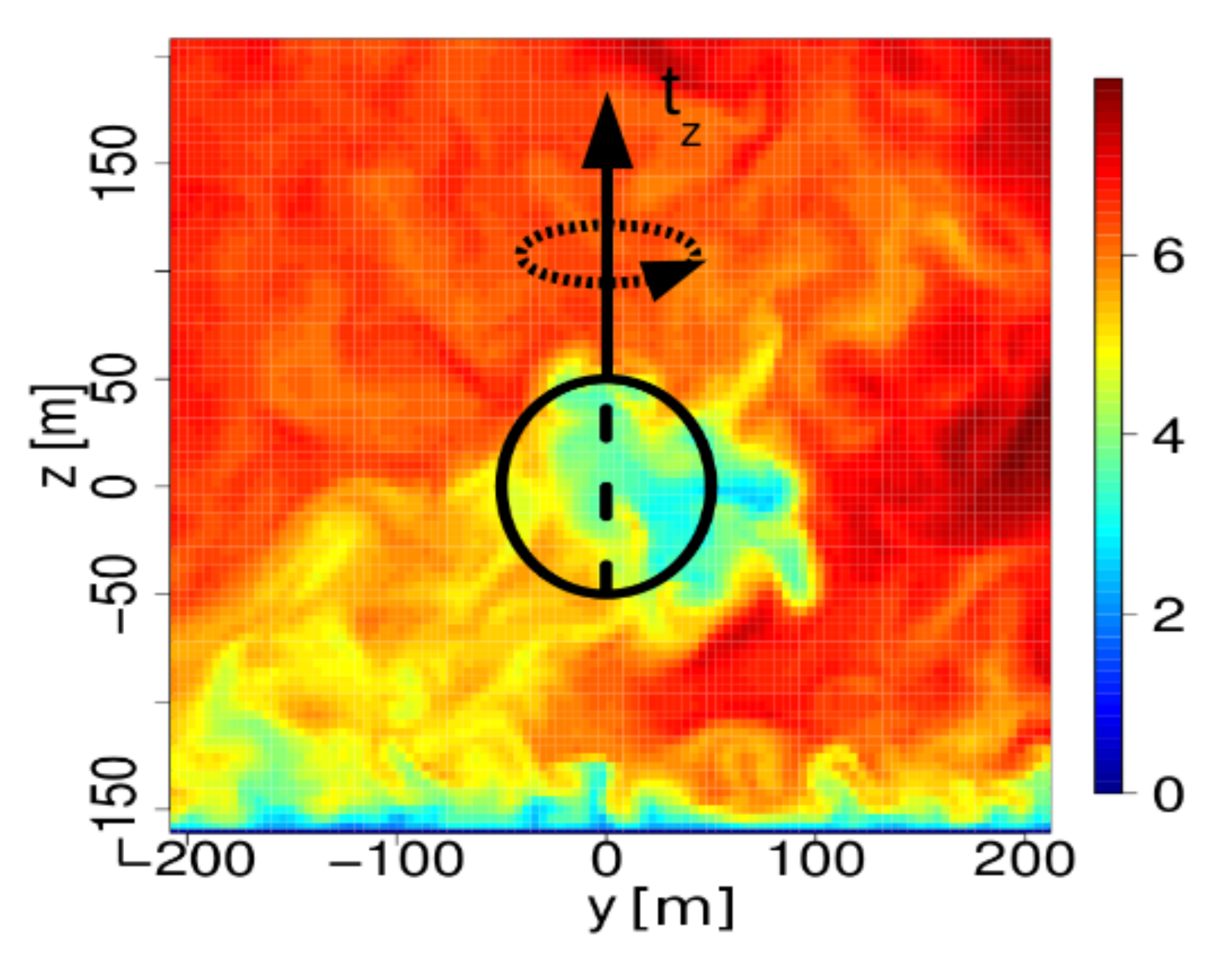}
 \subcaption{}
 \label{fig:qsnapcircyaw}
\end{subfigure}
\begin{subfigure}[H]{.32\textwidth}
 \centering
 \includegraphics[width=.96\linewidth]{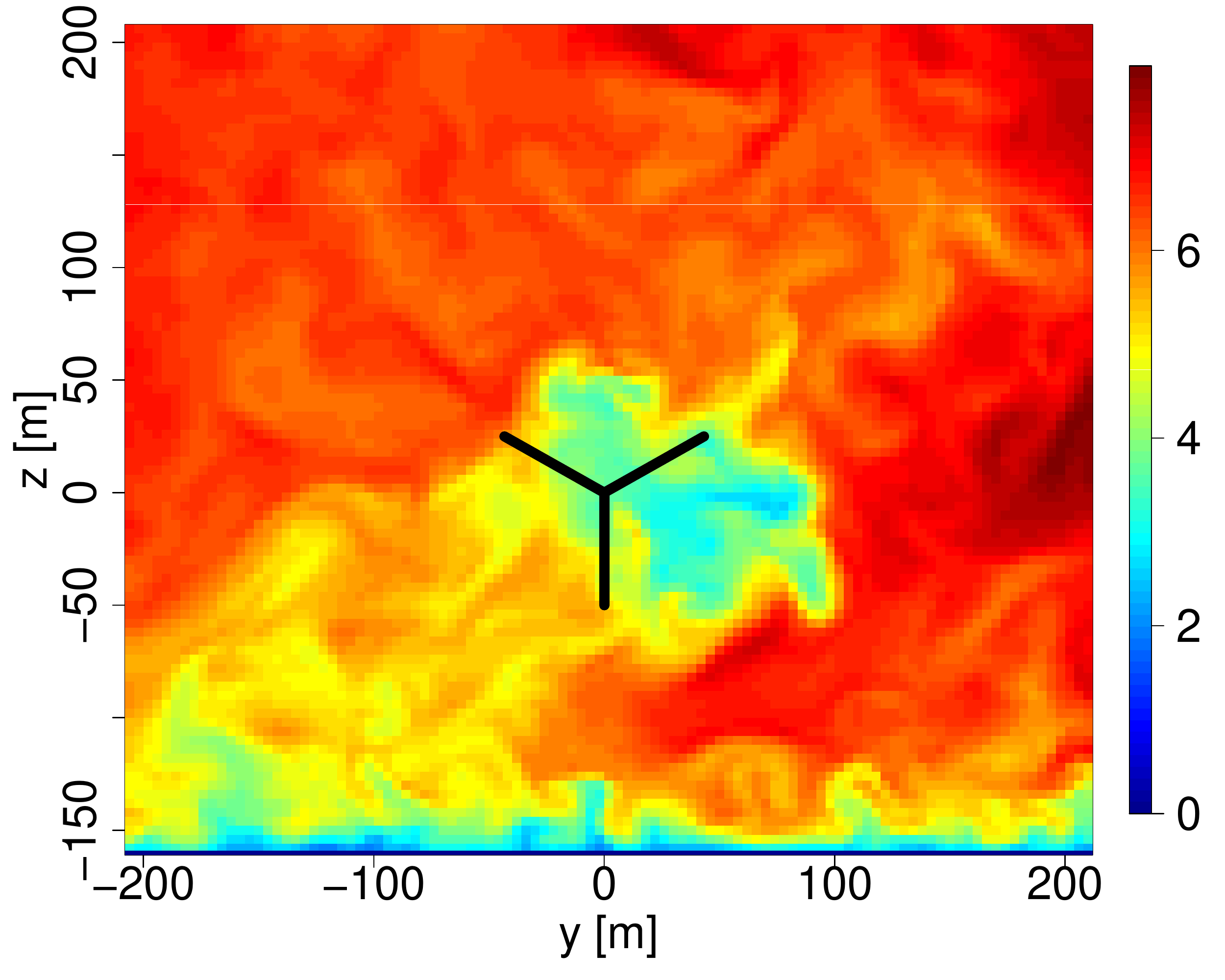}
 \subcaption{}
 \label{fig:qsnapblade}
\end{subfigure}
\caption{Illustrations corresponding to the different measures: (\textbf{a}) Measures $u_{\mbox{eff}}$, $P$ and $\tau_z$ are
obtained through averaging over a disk in the wake flow; (\textbf{b}) $\tau_z$ is related to the torque $t_z$ on a disk in the flow. 
The dashed line represents the rotational axis; (\textbf{c}) The measure $T$ results from averaging over three rotor blades in the flow.
The blades rotate at 10 \mbox{rpm}.}
\label{Fig:measures}
\end{figure}

We now define four different measures as $M(t)$, which are related to the power output or forces
on a sequential turbine. For this purpose, we take the turbine as a disk in the wake, as shown in
Figure \ref{fig:qsnapcirc}. 
The~first quantity that we analyze is the effective velocity, which we define as the average velocity
over the disk area:
\begin{equation}
 u_{\mbox{eff}}(t):=\frac{1}{A}\integrate{disk}{} \hspace{-.3cm}\integrate{}{}dydz~ u (y,z,t)=~\langle u(y,z,t)\rangle_{disk}~
\end{equation}

\noindent
where $\langle ... \rangle_{disk}$ now plays the role of the operator $\fat{\hat{M}}$ in Equation \eqref{eq:mdef}.
In the spirit of actuator disk theory, this quantity is related to the power output by $u_{\mbox{eff}}^3$
and the thrust force on the turbine by $u_{\mbox{eff}}^2$.

As a second measure, we use the energy flux through the disk given by:
\begin{equation}
 P(t):= \frac{1}{2} \rho A \langle u^3(y,z,t)\rangle_{disk}~
\end{equation}

\noindent
where $A$ is the area of the disk. $P$ is obviously related to the potential power output of the turbine.

The third measure defined here is related to the torque along the z-axis through the disk (Figure \ref{fig:qsnapcircyaw}) and is defined by:
\begin{equation}
\tau_{z}(t) := \langle u^2(y,z,t)\cdot y \rangle_{disk}~
\end{equation}

\noindent
where $y$ is the signed distance to the rotational axis.
In the case of a wind turbine, the torque in $z$-direction is also called the tower top yaw moment,
which is particularly important for
the yaw drive of the turbine. In contrast to $u_{\mbox{eff}}$ and $P$, the averaging over the disk
is now weighted by $y$, yielding a dependence on the spatial distribution of the deficit over the disk.

The last measure we define is slightly more complex, taking into account that the rotor consists of three rotating
blades. We define:
\begin{equation}
 T(t):= \integrate{blades}{} \hspace{-.3cm}\integrate{}{}dydz~ u^2(y,z,t)\propto \langle u^2(y,z,t)\rangle_{blades}~
\end{equation}

\noindent
where $blades$ denotes the area of the three blades illustrated in Figure \ref{fig:qsnapblade}. The blades are
rotating with $\omega=10$~\mbox{rpm}, which is a typical value for modern wind turbines. The
results presented in this work do not change qualitatively when varying $\omega$ in a realistic range.
$T$ is obviously related to the thrust force on a~rotor. Since, here, we do not only average
over an entire disk, smaller structures of the wake flow become more important.

\subsection{Results and Discussion}
\label{sec:results}

\begin{figure}[H]
\centering
\begin{subfigure}[H]{.45\textwidth}
 \centering
 \includegraphics[width=.95\linewidth]{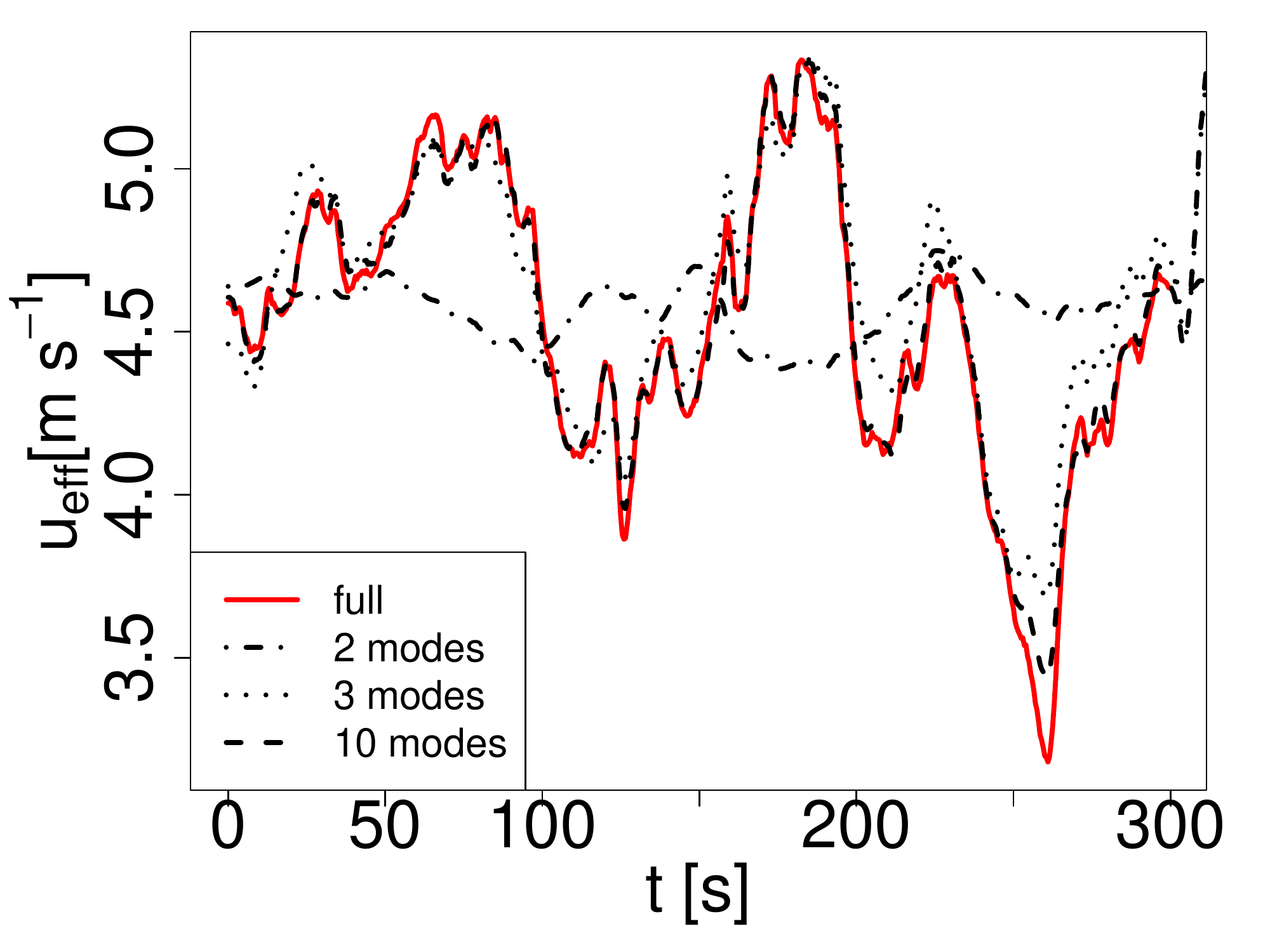}
 \subcaption{}
 \label{fig:ueff-recons-d_1}
\end{subfigure}%
\vspace {-12pt}
\begin{subfigure}[H]{.45\textwidth}
 \centering
 \includegraphics[width=.95\linewidth]{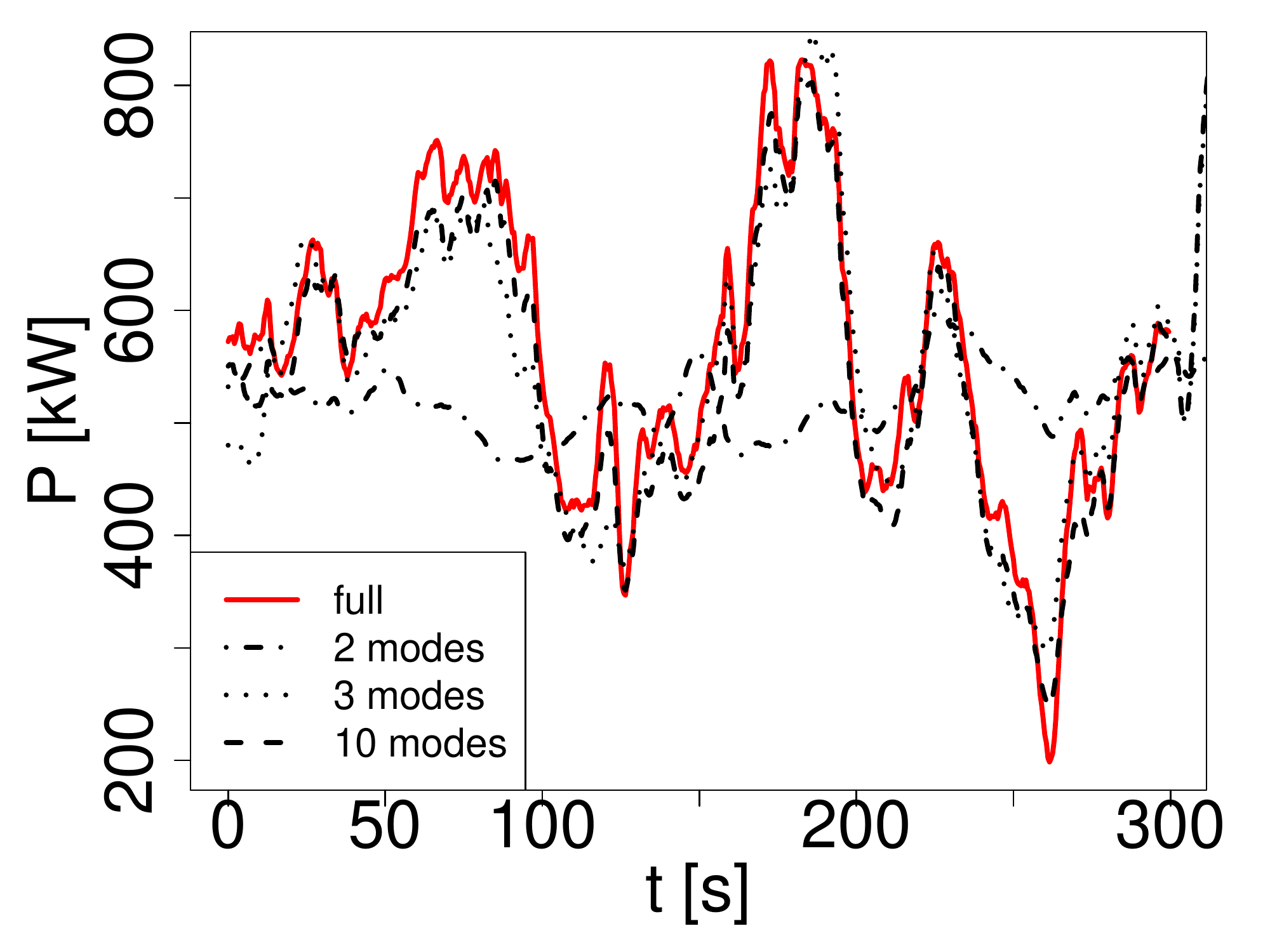}
 \subcaption{}
 \label{fig:P-recons-d_1}
\end{subfigure}
\caption{The measures (\textbf{a}) $u_{\mbox{eff}}$ and (\textbf{b}) $P$ for different numbers of
 modes used for the POD~reconstructions.}
\label{Fig:ueff-recons-d_1}
\end{figure}

\begin{figure}[H]
\centering
\begin{subfigure}[H]{.45\textwidth}
 \centering
 \includegraphics[width=.95\linewidth]{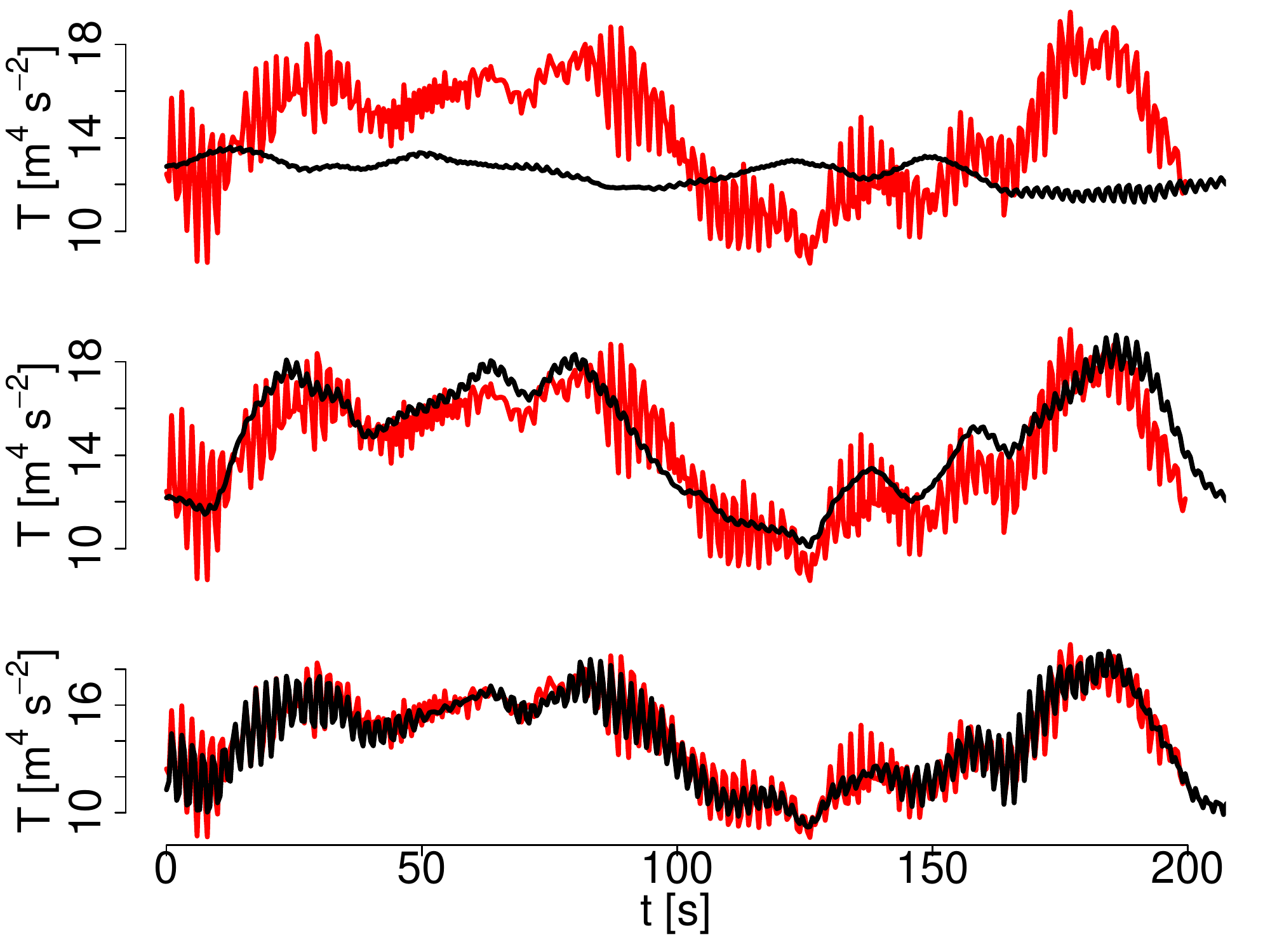}
 \subcaption{}
 \label{fig:althrustrecons}
\end{subfigure}%
\vspace {-12pt}
\begin{subfigure}[H]{.45\textwidth}
 \centering
 \includegraphics[width=.95\linewidth]{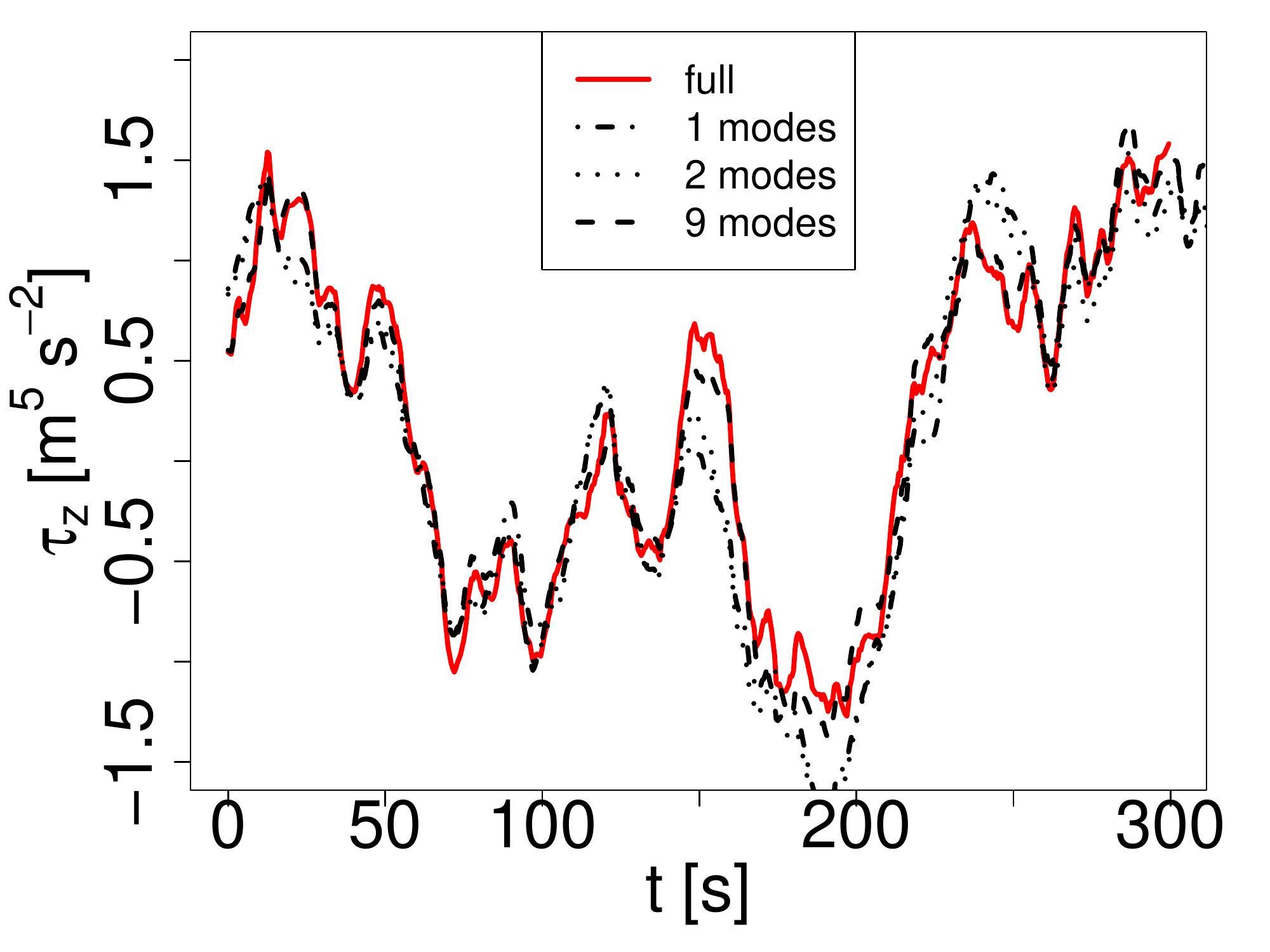}
 \subcaption{}
 \label{fig:zyaw-recons-d_1}
\end{subfigure}
\caption{(\textbf{a}) The measure $T$ for different numbers of modes used
 for the reconstruction. The~red lines show $T$ for the original field, while the black line
 shows $T$ of a reconstructed field ((\textbf{top}) two modes; (\textbf{middle}) three modes; (\textbf{bottom}) $10$ modes);
 (\textbf{b}) Dynamics of the measure $\tau_z$ for different numbers of modes used
 for the reconstruction. Note that since the reconstruction is already quite good with one mode, the lines lie almost on 
 top of each other. However, there is still a visible improvement in the case of nine modes.\\}
\label{Fig:althrustrecons}
\end{figure}

The measures can now be determined for our data and the corresponding reconstructions described in Section \ref{sec:podrecon}.
In Figures \ref{Fig:ueff-recons-d_1} and \ref{Fig:althrustrecons}, sections of the time series 
for $u_{\mbox{eff}}$, $P$, $\tau_z$ and $T$ are shown, which we obtained for the original field and reconstructions with
different numbers of modes. From these time series, one can already see that the dynamics of $u_{\mbox{eff}}$, $P$ and $T$
cannot be captured using two or less modes, but already, three modes grasp the basic dynamic features. 
For $\tau_z$, basic dynamical aspects are already captured when including only the first mode
(Figure \ref{fig:zyaw-recons-d_1}).

\vspace{-32pt}
\subsubsection{Behavior of the Dynamical and Standard Error}
\label{sec:errorbehavior}
\begin{figure}[H]
\centering
\begin{subfigure}[t]{.45\textwidth}
 \centering
 \includegraphics[width=.95\linewidth]{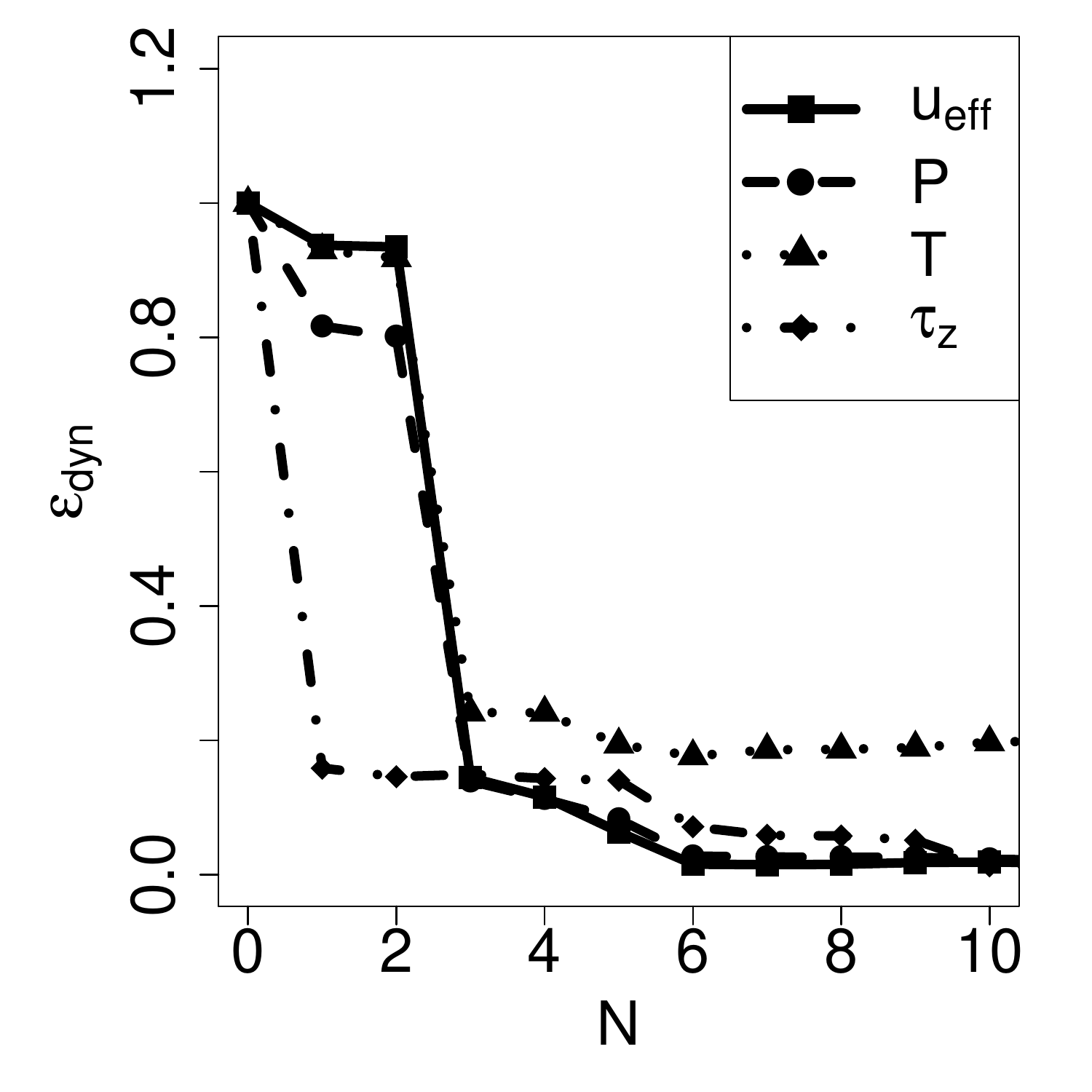}
 \subcaption{}
 \label{fig:edyn_allm}
\end{subfigure}%
\vspace {-12pt}
\begin{subfigure}[t]{.45\textwidth}
 \centering
 \includegraphics[width=.95\linewidth]{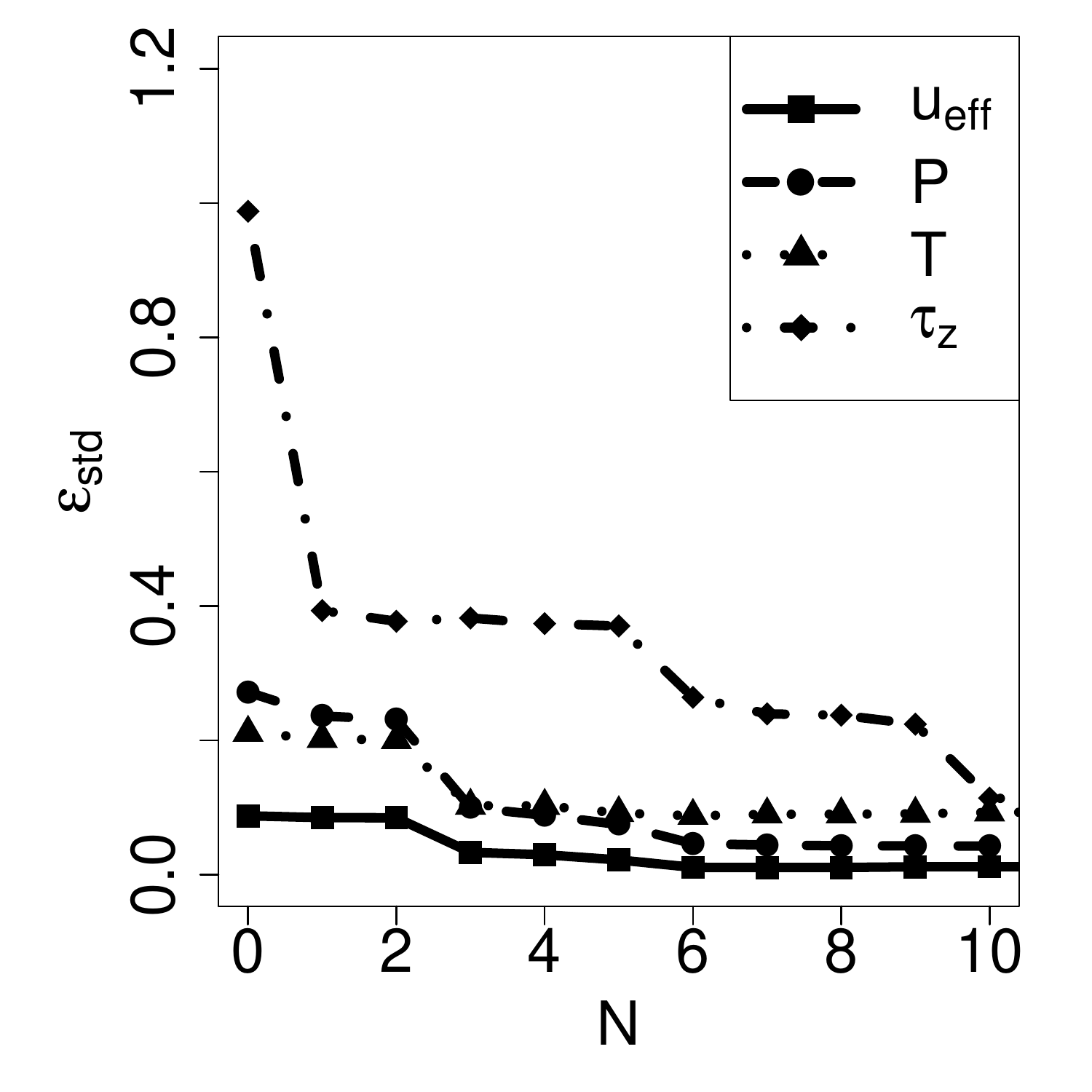}
 \subcaption{}
 \label{fig:estd_allm}
\end{subfigure}
\caption{(\textbf{a}) Dynamical and (\textbf{b}) standard error for different
 measures \textit{versus} the number $N$ of modes used for reconstruction.
 $N=0$ corresponds to a reconstruction with the mean field~only.
 }
\label{Fig:edyn_allm}
\end{figure}

The differences between the time series of the original field and the reconstructions can now be quantified
by the errors $\varepsilon_{\mbox{dyn}}$ and $\varepsilon_{\mbox{std}}$ shown in Figure \ref{Fig:edyn_allm}.
We start with a discussion of the dynamical error $\varepsilon_{\mbox{dyn}}$, followed
by a comparison with $\varepsilon_{\mbox{std}}$.

As expected from the inspection of the time series of $u_{\mbox{eff}}$, $P$ and $T$,
$\varepsilon_{\mbox{dyn}}$ shows a sudden decline from above
80\% to below 20\% when including the third mode (Figure \ref{fig:edyn_allm}). For $\tau_z$,
$\varepsilon_{\mbox{dyn}}$ goes down to below 20\% when only the first mode is included. 
Increasing to six modes results in a further reduction of the dynamical error to less than 10\% for $u_{\mbox{eff}}$ and $P$,
about 10\% for $\tau_z$ and around 20\% for $T$. 

Another remarkable result, revealed in Figure \ref{fig:edyn_allm}, is the discontinuous decay of $\varepsilon_{\mbox{dyn}}$.
The most pronounced jumps occur for $u_{\mbox{eff}}$, $P$ and $T$ when including Mode $3$ and for $\tau_z$ when including the first
mode. The improvement of the error when including a mode might be seen as an indicator
of its importance. The jump with Mode $3$ therefore indicates that the third mode plays a
special role for the description of $u_{\mbox{eff}}$, $P$ and $T$, while the first mode is of major importance for
$\tau_z$. This will be discussed further in Section \ref{sec:singlerecon}.

The similar performance of $u_{\mbox{eff}}$, $P$ and $T$, particularly the sudden jump with the third mode,
is probably caused by the fact that all three measures are integrated over the whole rotor or blade area
without any specific weighting. It seems to be of minor importance whether we integrate over $u$, $u^2$ or $u^3$.
The slightly weaker performance of $T$, especially with more modes, probably results from the integration
over the smaller blade area. Therefore, smaller structures are important for the dynamics of $T$, which
cannot be resolved by only a few modes. This can also be seen in the time series in Figure \ref{fig:althrustrecons},
where three modes nicely recover the macroscopic behavior of $T$, but $10$ modes still fail to grasp
some of the finer structures of the signal. 
The distinct behavior of $\varepsilon_{\mbox{dyn}}$ for $\tau_z$ is likely to be caused by the weighting
with the signed distance to the rotational axis. Due to this weighting, modes symmetric to the rotational axis
become less relevant than antisymmetric modes.

The standard error shows the same discontinuous behavior as the dynamical one with jumps
at identical numbers of modes
as identified for $\varepsilon_{\mbox{dyn}}$ (Figure \ref{fig:estd_allm}). However, two major differences
can be found.
The first difference is that for $u_{\mbox{eff}}$, $P$ and $T$, 
the standard error is already very low when only using the mean field of the flow, 
with around 10\% for $u_{\mbox{eff}}$ and 20\% for $P$. The second is the generally weaker performance of $\tau_z$. 
This different behavior stems from the fact that
$\varepsilon_{\mbox{dyn}}$ neglects the mean values of the measures in contrast to $\varepsilon_{\mbox{std}}$ (compare Equations (\ref{eq:errorstd}) and (\ref{eq:errordyn})).
For $u_{\mbox{eff}}$, $P$ and $T$
the steady mean field without any modes already yields a good description of the mean values of $u_{\mbox{eff}}$, $P$ and $T$.
Since the mean values of these measures are larger than the fluctuating parts, 
this yields the small standard error in Figure \ref{fig:estd_allm}.
Further details, also on the behavior of $\tau_z$, can be found in Appendix \ref{app:standarderror}.

\subsubsection{Selection of Modes}
\label{sec:singlerecon}

So far, we included the extracted modes ordered from high
to low energy content, as usual for POD reconstructions.
In the previous section, Section \ref{sec:results}, we already noted that the discontinuous
behavior of the reconstruction error indicates
that different modes are relevant for different measures.
This~naturally suggests choosing only modes that correspond to jumps in 
the errors when focusing on a specific~measure.

\begin{figure}[H]
\centering
\begin{subfigure}[t]{.45\textwidth}
\centering
\includegraphics[width=.98\linewidth]{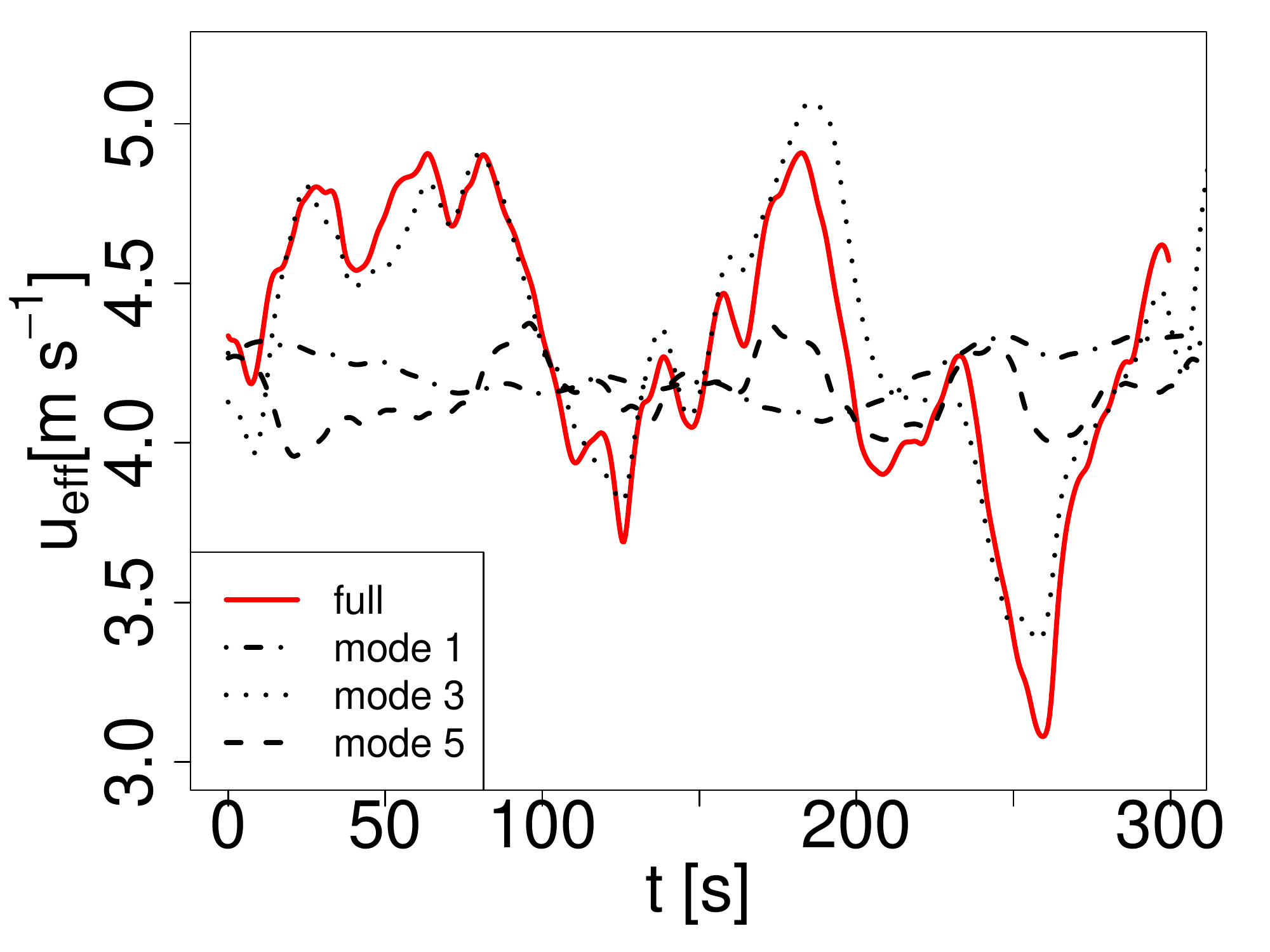}
\subcaption{}
\label{fig:ueff-ncl-recons}
\end{subfigure}%
\begin{subfigure}[t]{.45\textwidth}
\centering
\includegraphics[width=.98\linewidth]{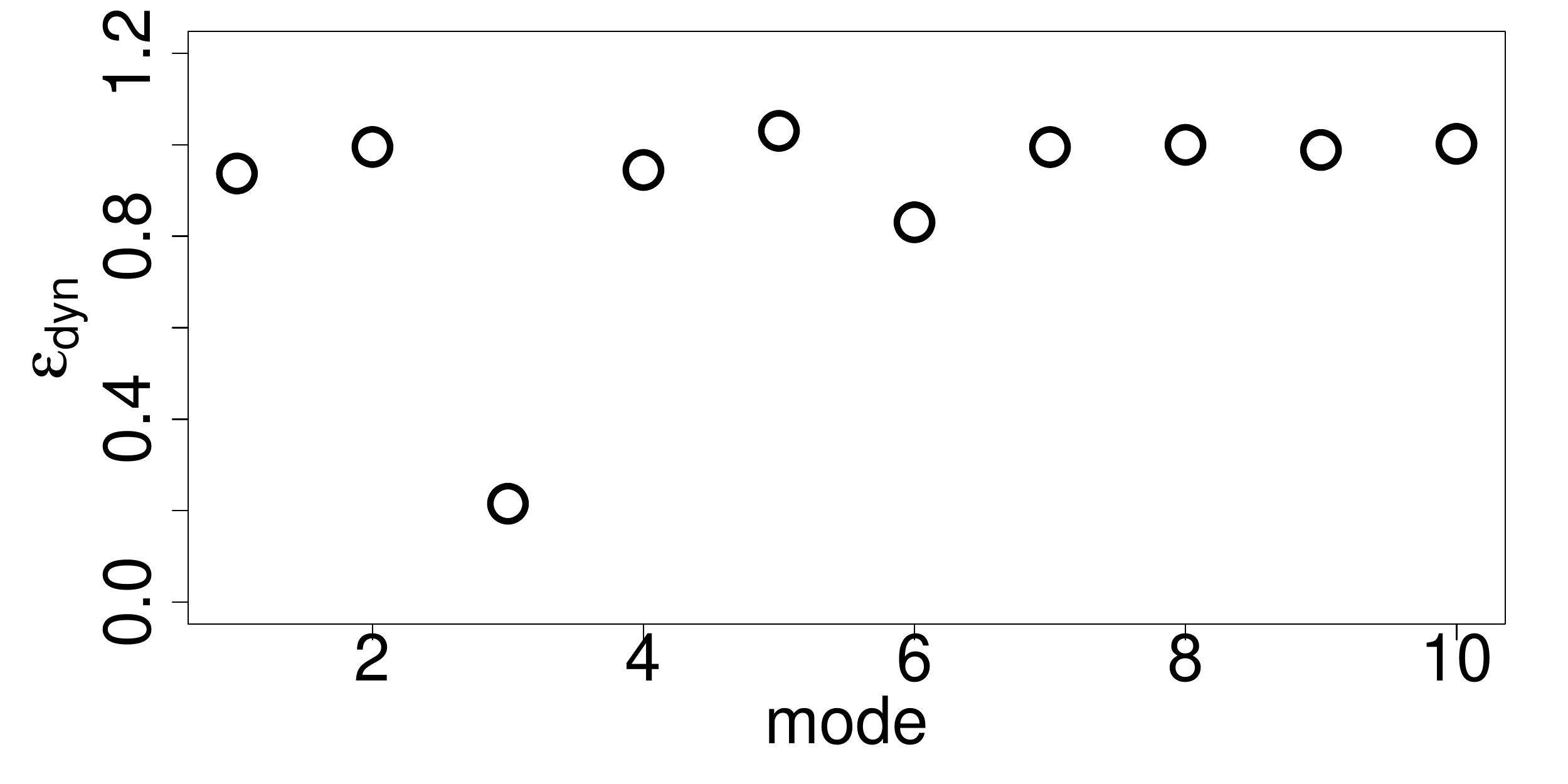}
\subcaption{}
\label{fig:ueff-ncl-ekin}
\end{subfigure}
\caption{(\textbf{a}) Effective velocity $u_{\mbox{eff}}$ for different reconstructions of $u(y,z,t)$
using only one single mode (plus mean field); (\textbf{b}) dynamical error for reconstructions using a single mode only. The $x$-axis denotes
the number of the mode used for the reconstruction.\vspace{-6pt}}
\label{Fig:ueff-ncl-recons}
\end{figure}

\begin{figure}[H]
\centering
\includegraphics[width=.45\linewidth]{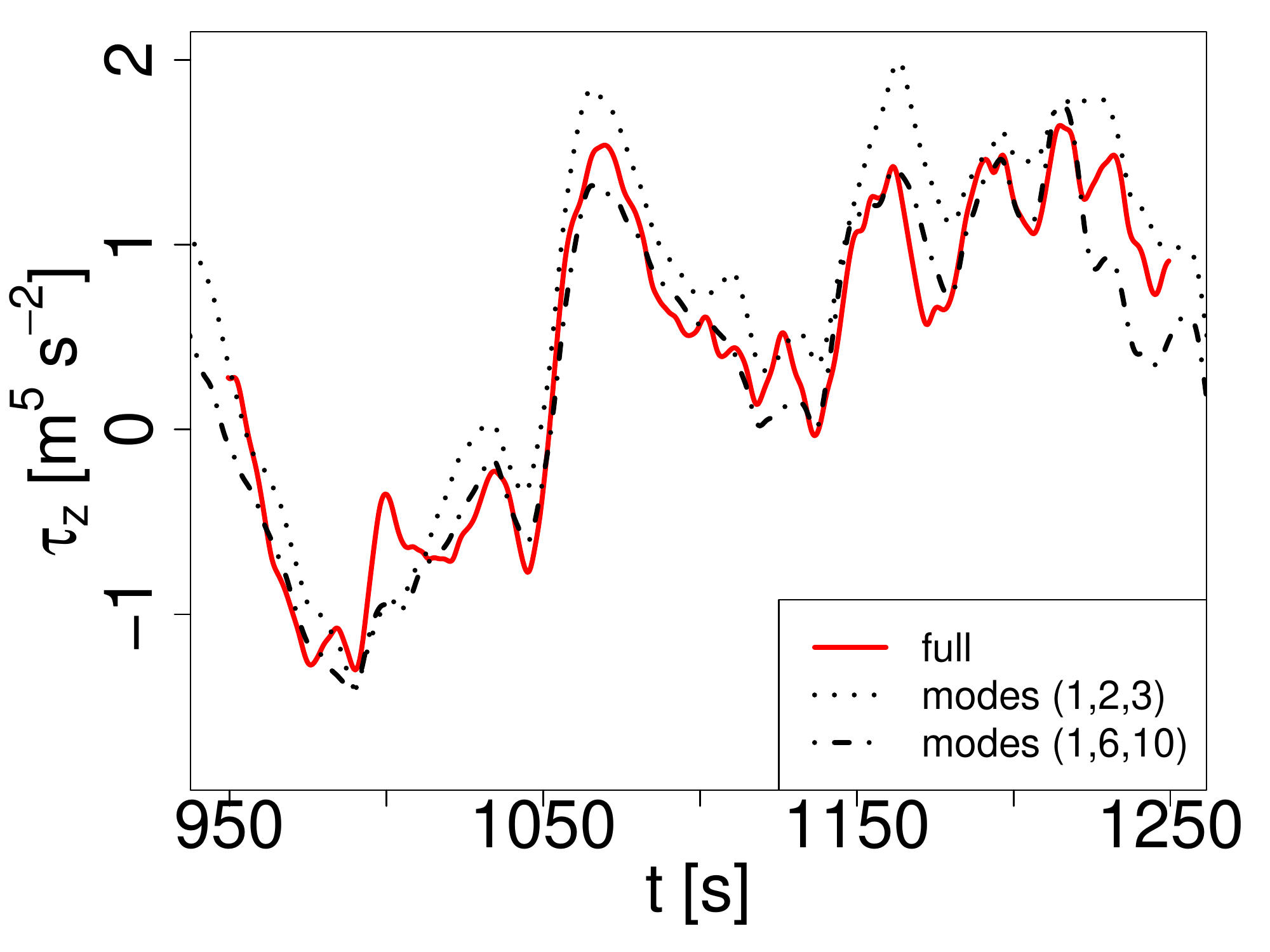}
\caption{Torque-related measure $\tau_z$ for the original field
and two reconstructions using a~different set of three modes.
}
\label{fig:tauz-ncl-recons}
\end{figure}

The first example for this approach can be seen in Figure \ref{Fig:ueff-ncl-recons}.
The third mode, which corresponds to a sharp decrease in the error of $u_{\mbox{eff}}$,
yields a better reconstruction of $u_{\mbox{eff}}$ than the first mode. 
In~the second example (Figure \ref{fig:tauz-ncl-recons}), the measure $\tau_z$
is reconstructed choosing two different sets of modes. The~performance of the modes $(1,6,10)$,
which correspond to sharp decreases in both errors (see Figure~\ref{Fig:edyn_allm}),
yields a better approximation of $\tau_z$ than the most energetic modes $(1,2,3)$.

From these results, we can conclude that the usual energetic order is not necessarily the
best choice for other measures than the turbulent kinetic energy.
Finding the optimal modes can therefore improve the results
or reduce the necessary dimension when aiming for the description of a specific quantity.
It~is important to note that the sharp decreases in the errors
can, in general, only be a first indicator for the importance of the modes,
since most measures of interest are nonlinear. An alternative idea to find the
optimal modes could be to take symmetry considerations
into account, as indicated above for $\tau_z$ in Section~\ref{sec:errorbehavior} and proposed by Saranyasoontorn and Manuel \cite{Saranyasoontorn2006}. 

\section{Interpretation of the Modes}
\label{sec:interpretation}

For a better understanding of the results in the last section, we now relate the first three POD modes
to the dynamical properties of the wake deficit.

\subsection{Meandering}

The first dynamical property that we analyze is the meandering
or large-scale movement of the velocity deficit.
For this purpose, we determine the center of the deficit as:

\begin{equation}
\left( \begin{array}{c}
y_c(t) \\
z_c(t)
\end{array}
\right)
:=\frac{\integrate{}{}dy dz ~ \tilde{u}^2(y,z,t) \left( \begin{array}{c}
y \\
z 
\end{array}
\right)}{\integrate{}{}dy dz~ \tilde{u}^2 (y,z,t)}~
\label{eq:center}
\end{equation}

\noindent
which we evaluate for every time step. 
The definition can be interpreted as the center of energy of the deficit.
Other approaches have been used in the literature \cite{Espana2011,Trujillo2011,Muller2013},
but for these data, Equation (\ref{eq:center}) yielded the most robust results. 

\begin{figure}[H]
\centering
\begin{subfigure}[t]{.45\textwidth}
 \centering
 \includegraphics[width=.95\linewidth]{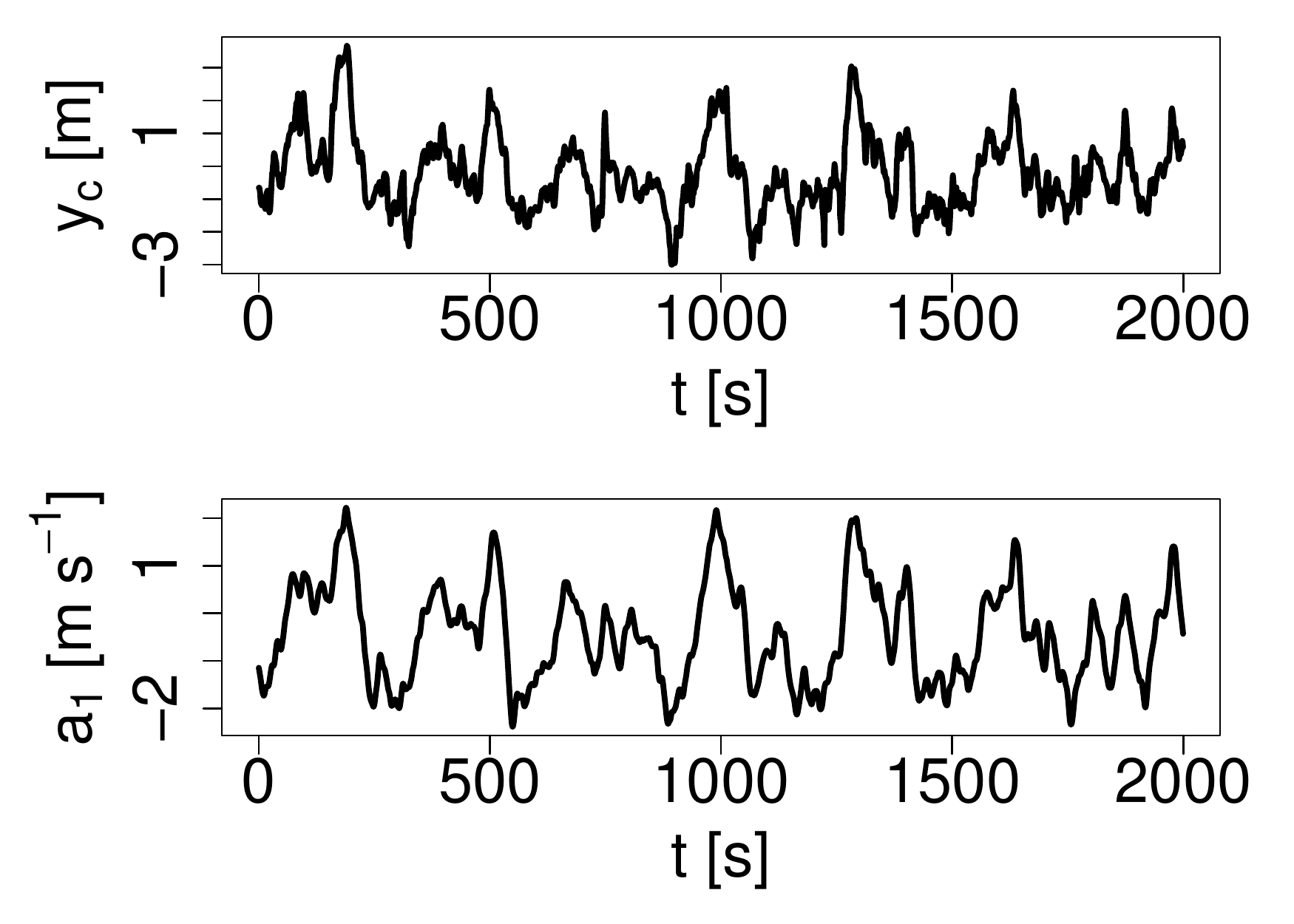}
 \subcaption{}
 \label{fig:ym_vs_a1}
\end{subfigure}%
\vspace {-12pt}
\begin{subfigure}[t]{.45\textwidth}
 \centering
 \includegraphics[width=.91\linewidth]{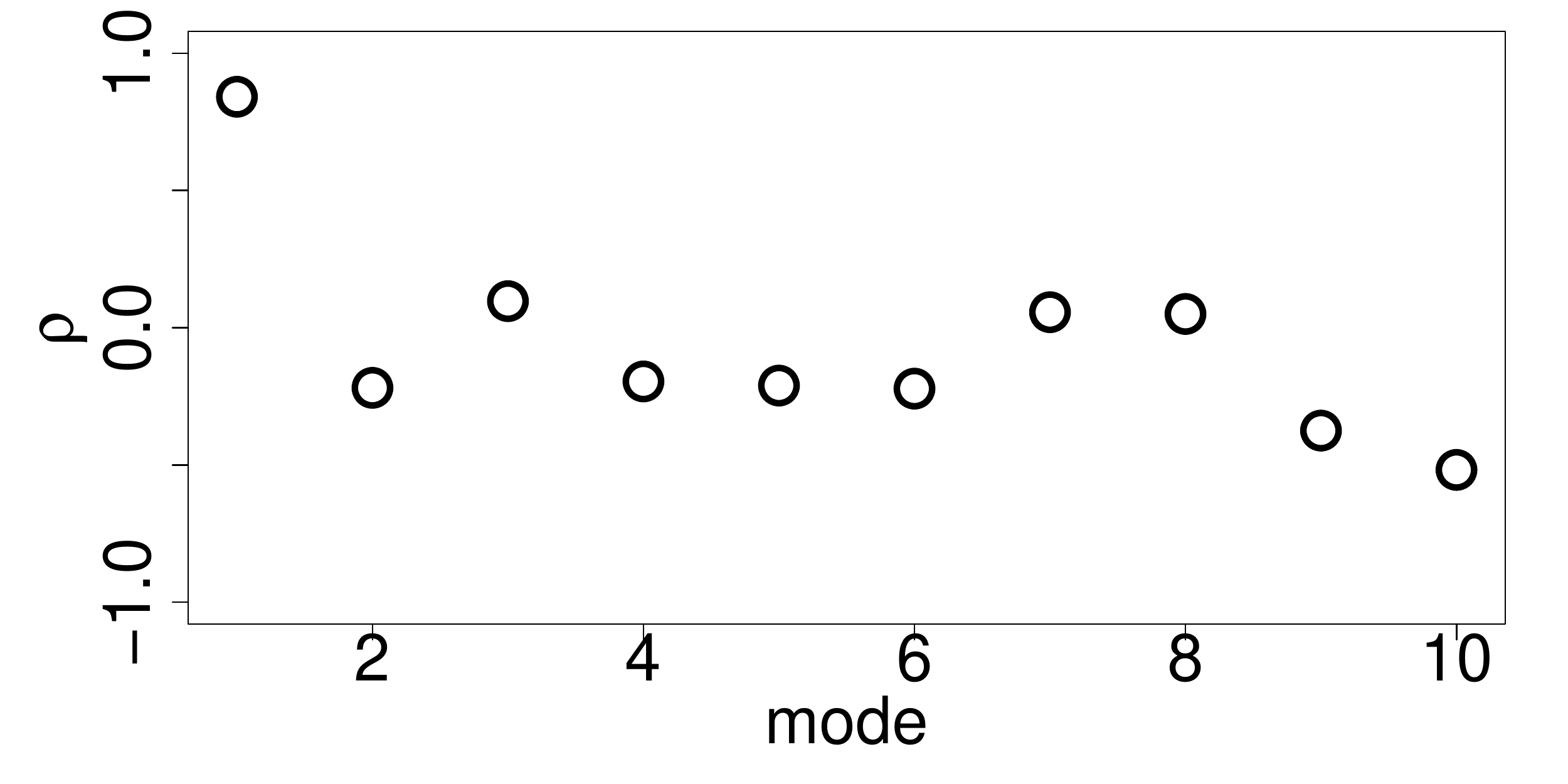}
 \subcaption{}
 \label{fig:ym_corr}
\end{subfigure}
\caption{(\textbf{a}) Time series of the first weighting coefficient $a_1$ and the
horizontal position of the deficit $y_c$; (\textbf{b}) correlation coefficients $\rho$ of the 
horizontal deficit position $y_c$ with the weighting coefficients of the first 10 modes.}
\label{Fig:ym_vs_a1}
\end{figure}

\begin{figure}[H]
\centering
\begin{subfigure}[t]{.5\textwidth}
 \centering
 \includegraphics[width=.95\linewidth]{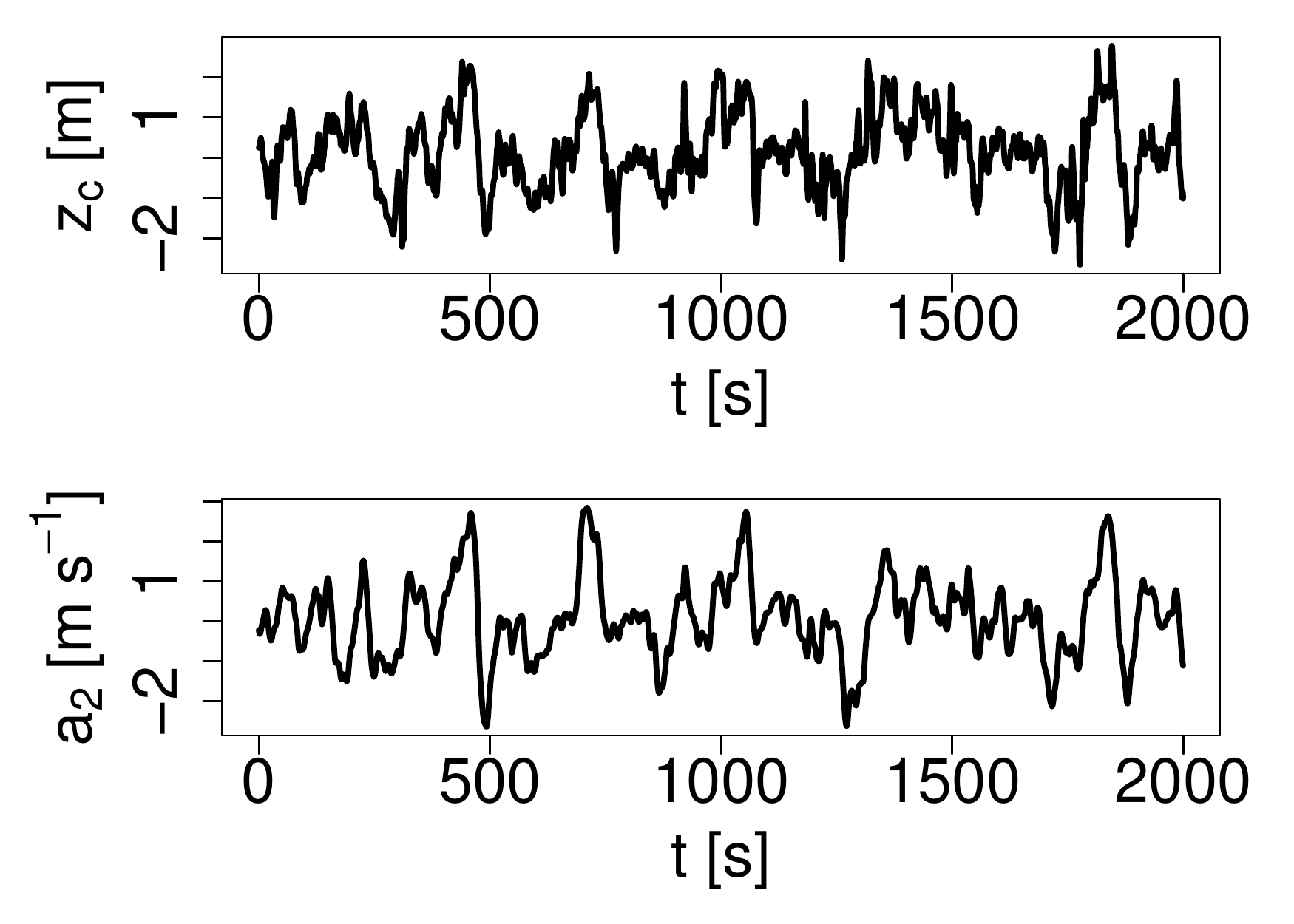}
 \subcaption{}
 \label{fig:zm_vs_a2}
\end{subfigure}%
\begin{subfigure}[t]{.5\textwidth}
 \centering
 \includegraphics[width=.91\linewidth]{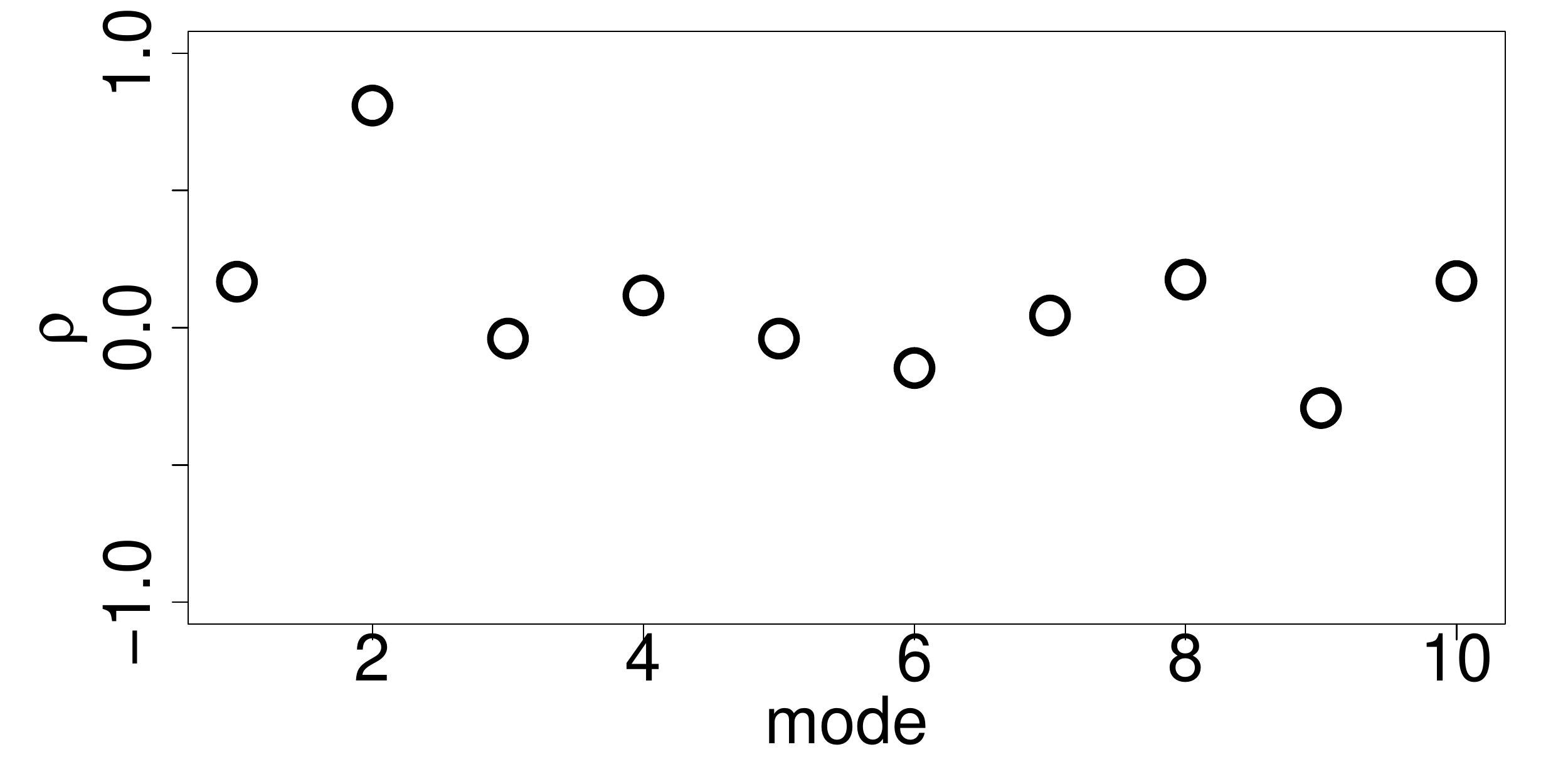}
 \subcaption{}
 \label{fig:zm_corr}
\end{subfigure} 
\caption{(\textbf{a}) Time series of the second weighting coefficient $a_2$ and the
vertical position of the deficit $z_c$; (\textbf{b}) correlation coefficients $\rho$ of the 
horizontal deficit position $z_c$ with the weighting coefficients of the first 10 modes.}
\label{Fig:zm_vs_a2} \vspace{-12pt}
\end{figure}

\begin{figure}[H] 
\centering
\includegraphics[width=.5\linewidth]{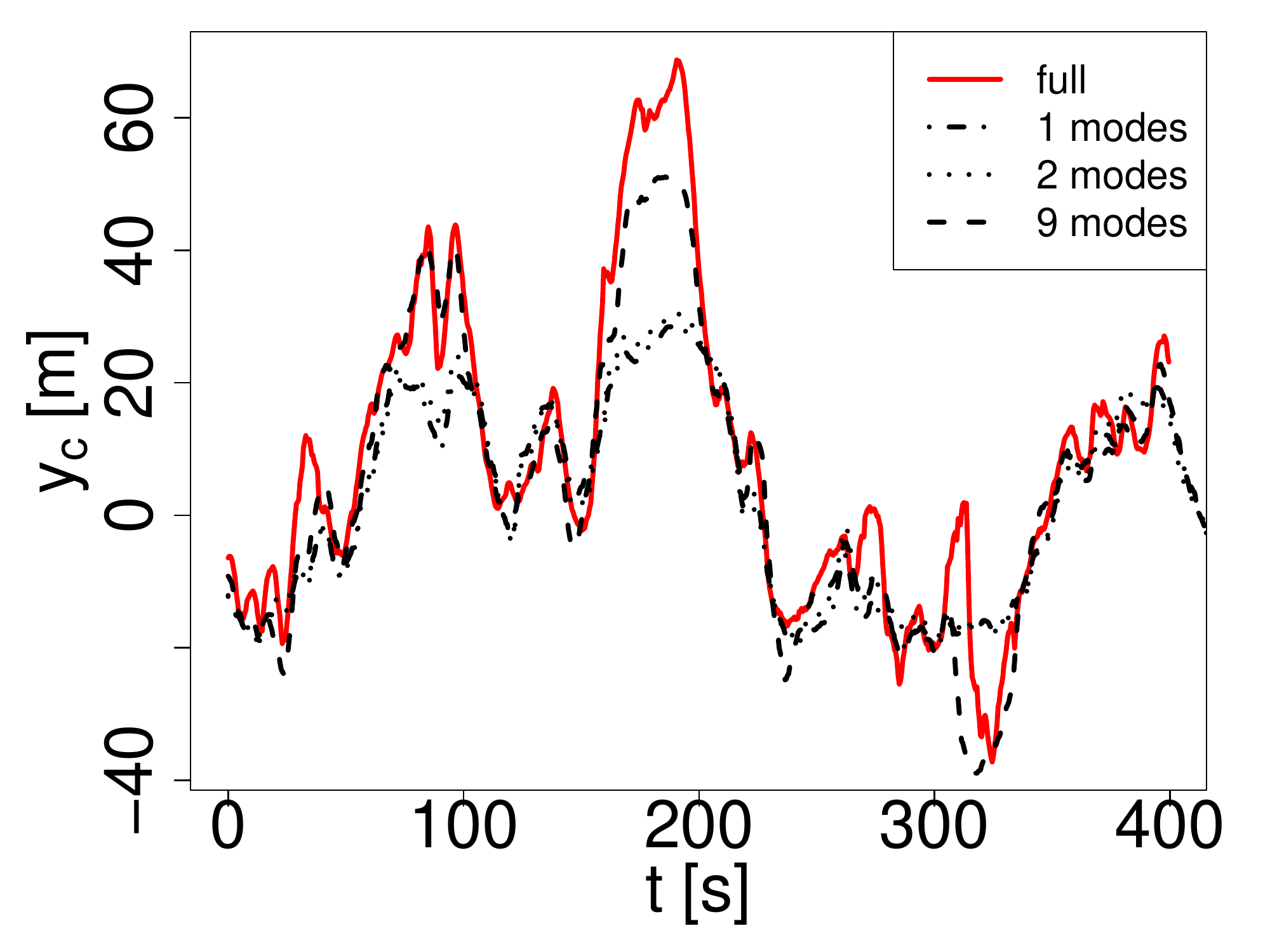} 
\caption{The horizontal position of the deficit $y_{c}$ for different reconstructions of $u(y,z,t)$
using different numbers of modes.}
\label{fig:ym-recons-d_1}
\end{figure}

To investigate the relation to the POD modes, we compare the trajectories of the center to the 
weighting coefficients $a_j(t)$ of the POD modes.
The horizontal trajectory is strongly correlated with the amplitude of the first mode (Figure \ref{fig:ym_vs_a1}),
yielding a linear correlation coefficient of $\rho(y_c,a_1)=0.84$ (Figure \ref{fig:ym_corr}).
The correlations to other weighting coefficients are almost negligible.
Analogously, we find $\rho(z_c,a_2)=0.81$ for the vertical movement and the second amplitude
(Figure \ref{Fig:zm_vs_a2}). Thus, the first and second mode are related to the horizontal and vertical
movement, respectively. To support the former arguments, Figure \ref{fig:ym-recons-d_1}
shows the horizontal position for the original field and different reconstructions. One mode actually
grasps the horizontal low amplitude movement of the deficit. To recover the high amplitudes,
more modes need to be included.

The connection of the first mode to the horizontal meandering also yields a deeper understanding
of the behavior of the measure $\tau_z$. The torque in the z-direction and, thus, $\tau_z$ are
obviously strongly influenced by the horizontal position of the deficit. A large movement to the right,
for example, yields a strongly asymmetric force on the disk, inducing a positive torque in the $z$-direction.
Since one mode is enough to capture the basic dynamics of the horizontal movement, it is also sufficient
for an approximate description of $\tau_z$. Analogous arguments can also be made for the second
mode and the vertical meandering, which induces torque in the $y$-direction.

\vspace{-12pt}

\subsection{Amplitude}

\begin{figure}[H]
\centering
\begin{subfigure}[t]{.45\textwidth}
 \centering
 \includegraphics[width=.95\linewidth]{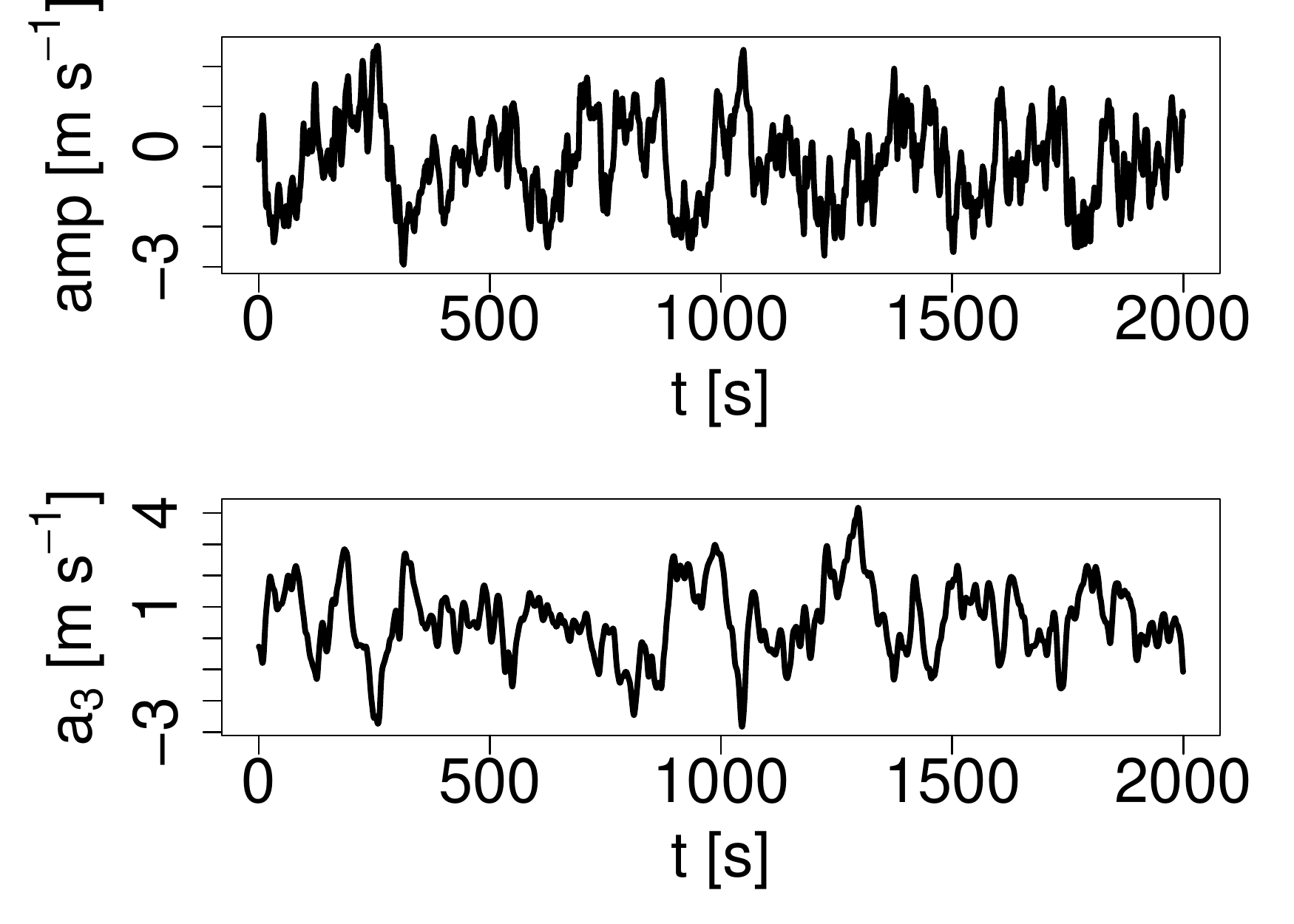}
 \subcaption{}
 \label{fig:amp_vs_a3}
\end{subfigure}%
\vspace {-12pt}
\begin{subfigure}[t]{.45\textwidth}
 \centering
 \includegraphics[width=.91\linewidth]{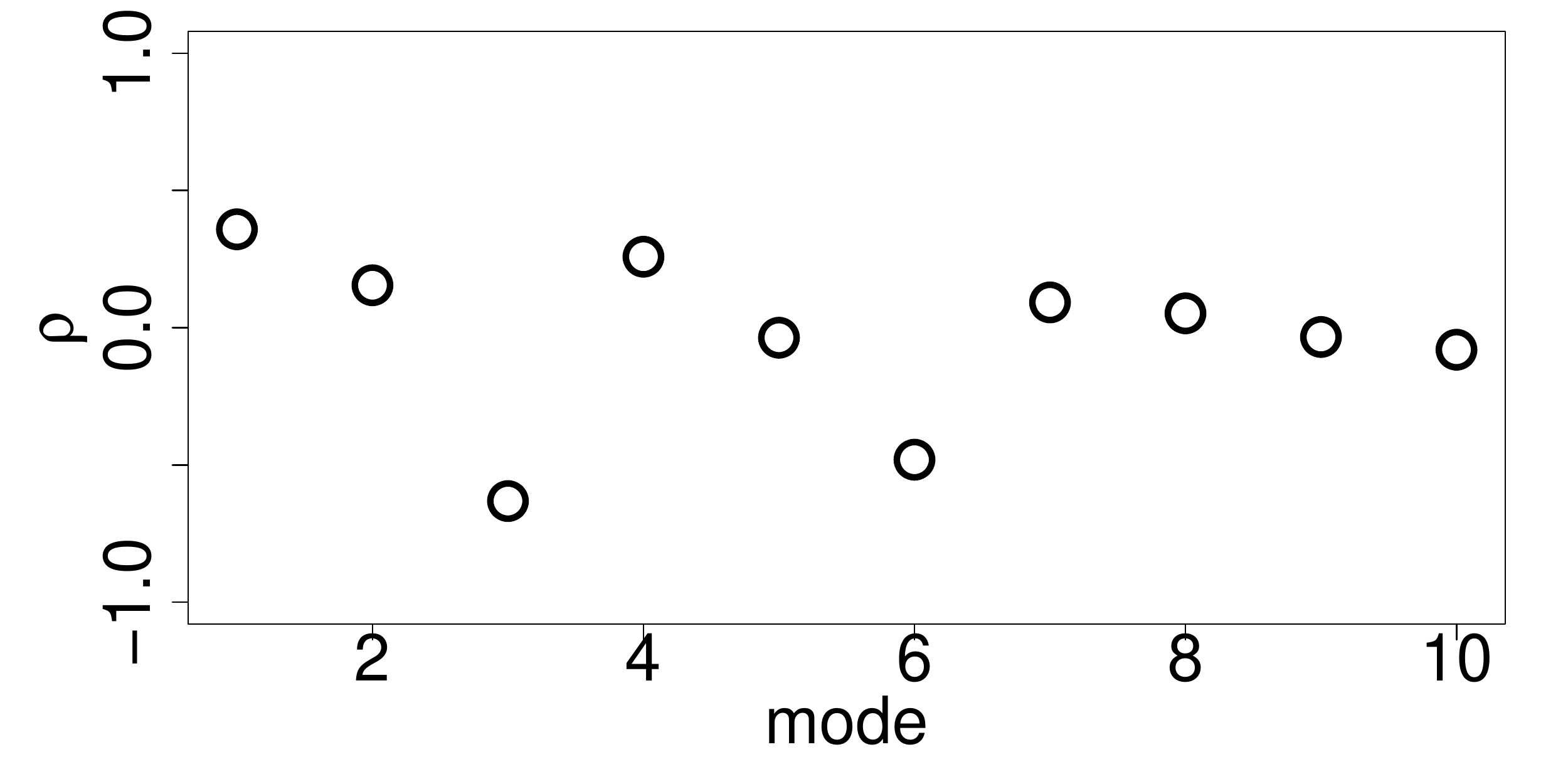}
 \subcaption{}
 \label{fig:amp_corr}
\end{subfigure}
\caption{(\textbf{a}) Time series of the third weighting coefficient $a_3$ and the
 average amplitude of the deficit (called amp in the figure); (\textbf{b}) correlation coefficients $\rho$ of the 
 average amplitude of the deficit with the weighting coefficients of the first 10 modes.}
\label{Fig:amp_vs_a3}
\end{figure}

As the third dynamical property of the deficit, we investigate the average amplitude of the deficit,
which we define as the velocity spatially averaged over the complete deficit
(extracted through the threshold application). As before, we compare
the time series of this amplitude to the weighting coefficients of the modes and calculate the
corresponding correlation coefficients. Figure \ref{Fig:amp_vs_a3} shows that this amplitude is strongly
(anti-) correlated to the third weighting coefficient with $\rho=-0.64$. Since
the third mode is very important for recovering the measures $u_{\mbox{eff}}$, $P$ and $T$, the average
amplitude probably plays a major role for the power output and thrust on a turbine. Therefore,
only considering the meandering is not enough to describe these quantities.

\section{Conclusions}
\label{sec:conclusions}

A modified POD analysis has been applied to the large eddy simulation of an actuator disk in a turbulent atmospheric boundary layer.
The main goal of this work was to identify modes that yield a strong dimensional reduction of the wake flow.
Such a reduction would 
facilitate the application of low dimensional modeling methods
to the temporal dynamics of the weighting coefficients of the modes.

Based on our results, we propose that the quality of modal reconstructions of the field should not only
be assessed by considering the recovery of turbulent kinetic energy, but should be based on quantities relevant to a
sequential turbine in the wake. The basic dynamics of such relevant quantities (e.g., the energy flux through a disk) could
already be captured using the three most energetic modes of our analysis. This strong dimensional
reduction of the flow indicates that developing wake models of very low order is possible.
Therefore, these results strongly support the idea of building simplified dynamic wake models
based on a superposition of spatial modes. 
Such models could be a useful alternative to dynamic models mainly based on the meandering process.
They could therefore play an important role for the layout optimization and controlling of wind farms.
Furthermore, our results support the idea of simplified dynamic wake models
in general, since we show that relevant information is contained in only a few degrees of freedom.

A further dimensional reduction could possibly be achieved when a wake model is used to describe a very specific
impact on a sequential turbine (e.g., the tower top yaw moment), since our results suggest that the necessary
level of complexity
depends on the quantity that we aim to describe. Furthermore, we showed that including modes in the usual energetic order is not generally
the best choice. Thus, modes could be chosen with respect to the application, neglecting irrelevant modes, despite a possible 
high energy~content.

We showed that relating spatial modes to specific properties of the wake can yield a deeper understanding
of models based on modal decompositions. This interpretation can reveal relations to features
of other dynamic wake models. In our case, a description with the first two POD modes is related to a dynamic wake model,
which includes meandering as its only dynamic feature. The importance of Mode $3$ and, thus, the
average amplitude of the wake indicates that meandering alone might not be
sufficient to capture the basic dynamics of, e.g., the thrust on the turbine. Therefore, a fluctuating amplitude
might be a useful extension to a pure meandering model of the wake.

The modes that we obtained after applying a threshold to the field had features corresponding to a
statistically rotationally symmetric field. Since these modes performed well, this slightly 
supports an approach also used in dynamic wake meandering models, where an axial-symmetric wake is assumed before including the ABL. 
It might therefore be possible to model the wake and ABL separately and include possible interactions at a different level. 

To show the robustness of our results, this study could obviously benefit from the analysis of additional datasets,
such as more detailed simulations or high-speed PIV
 data from wind tunnel experiments.
The quality measures introduced, even though related to the performance of a turbine in the wake, are simple.
Therefore, an aeroelastic code could be used, as a next step, to obtain better measures of quality.
Another possible improvement of models based on decompositions could be achieved by including all three velocity
components. However, it is not clear whether a sufficient dimensional reduction is possible in this case.

So far, we have mainly investigated the quality of the reduced descriptions of the wake dynamics. 
Building a reduced order model based on the extracted POD modes is a very challenging task; since we need
to model the temporal dynamics of the weighting coefficients of the modes.
To do this, we plan to consider the weighting coefficients as stochastic processes
and to estimate the corresponding model equations in the spirit
 of, e.g., \cite{kleinhans2007, Friedrich2011, Milan2014}.


\acknowledgments{Acknowledgments}
We would like to acknowledge discussions with Philip Rinn, Patrick Milan, Pedro Lind,
 Davide Trabucchi, Hauke Beck, Marc Bromm, Lukas Vollmer, Bernd Kuhnle, Juan Trujillo,
Felix~Gadeberg and Mohammed Reza Rahimi Tabar. 
For a fruitful discussion at the Making Torque Conference
in Copenhagen, we would like to thank S{\o}ren Juhl Andersen
who did similar work during his Ph.D.

This work has been funded by the {Bundesministerium f\"ur Wirtschaft und Energie} (BMWi)
due to a decision of the German {Bundestag} (FKZ
 0325397A). 

\authorcontributions{Author Contributions}
David Bastine performed the analysis and wrote the major part of this article. 
 Joachim Peinke and Matthias W\"achter are supervisor and co-supervisor of
David Bastine and his current Ph.D. work. Therefore, they helped with the ideas and discussions and
proofread the manuscript. Bj\"orn Witha performed the LES simulations
and mainly wrote the part describing the data.

\vspace {12pt}
\appendix\noindent{\bf {Appendix}}

\section{POD Modes without a Threshold}
\label{app:classicpod}

\setcounter{figure}{0}
\renewcommand\thefigure{A\arabic{figure}}

In this Appendix, we describe the POD modes obtained without applying any kind
of preprocessing and discuss the corresponding results for the different quality
measures used in Section \ref{sec:alternative}. 

When not using the threshold as introduced in Section \ref{sec:preprocessing}, the POD modes
and values strongly depend on the spatial region used for the analysis. 
Here, we chose the region $[-120, 120]$ m $\times$ $[-120, 120]$ m
centered at the hub height to apply the POD. It should be noted that some of the modes were sensitive even to relatively
small changes of the size of this window.

\begin{figure}[H]
\begin{center}
\begin{subfigure}[t]{.32\textwidth}
 \includegraphics[width=.98\linewidth]{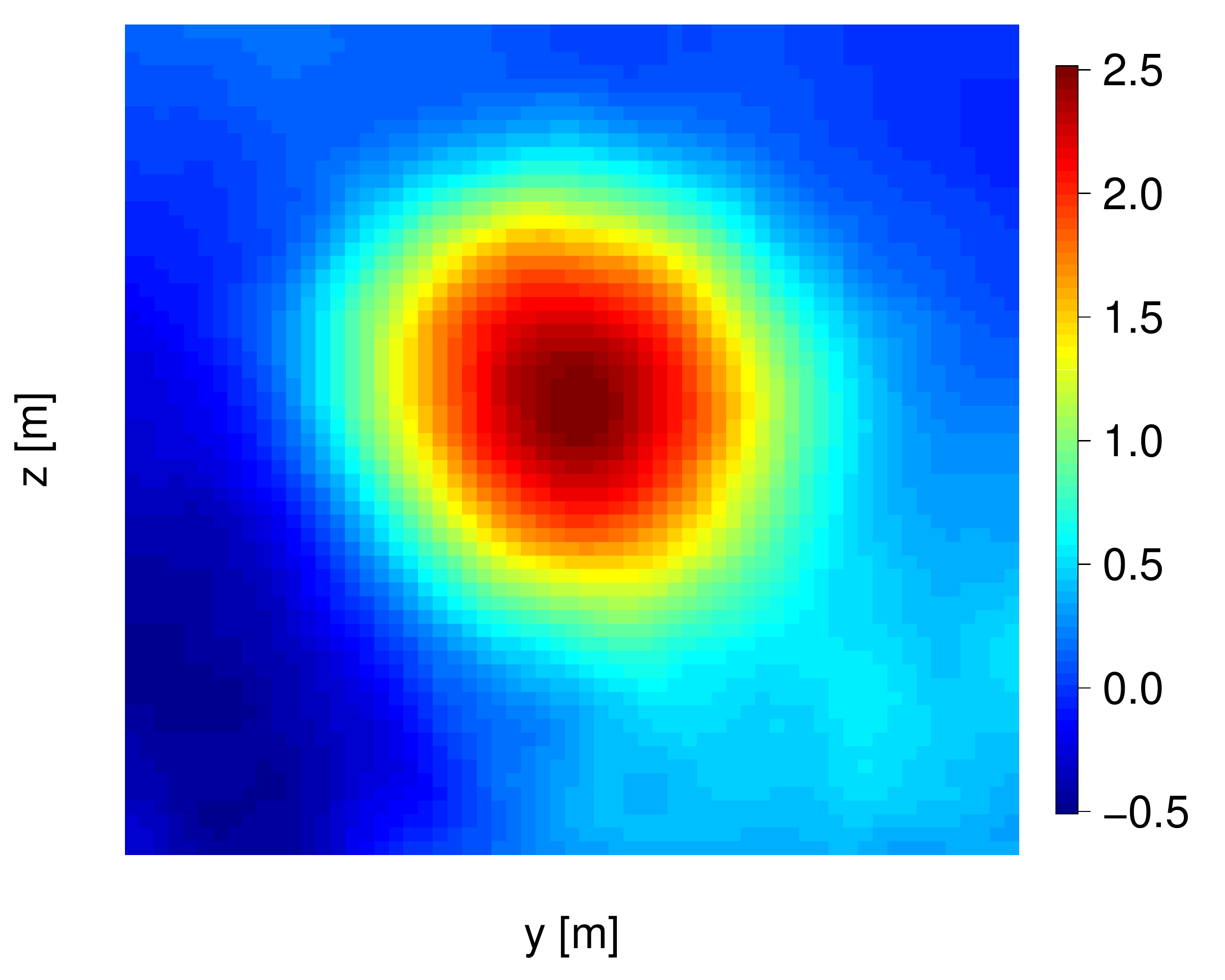}
 \subcaption{}
 \label{fig:qcutmeanb}
\end{subfigure}%
\vspace {-12pt}
\begin{subfigure}[t]{.32\textwidth}
 \includegraphics[width=.98\linewidth]{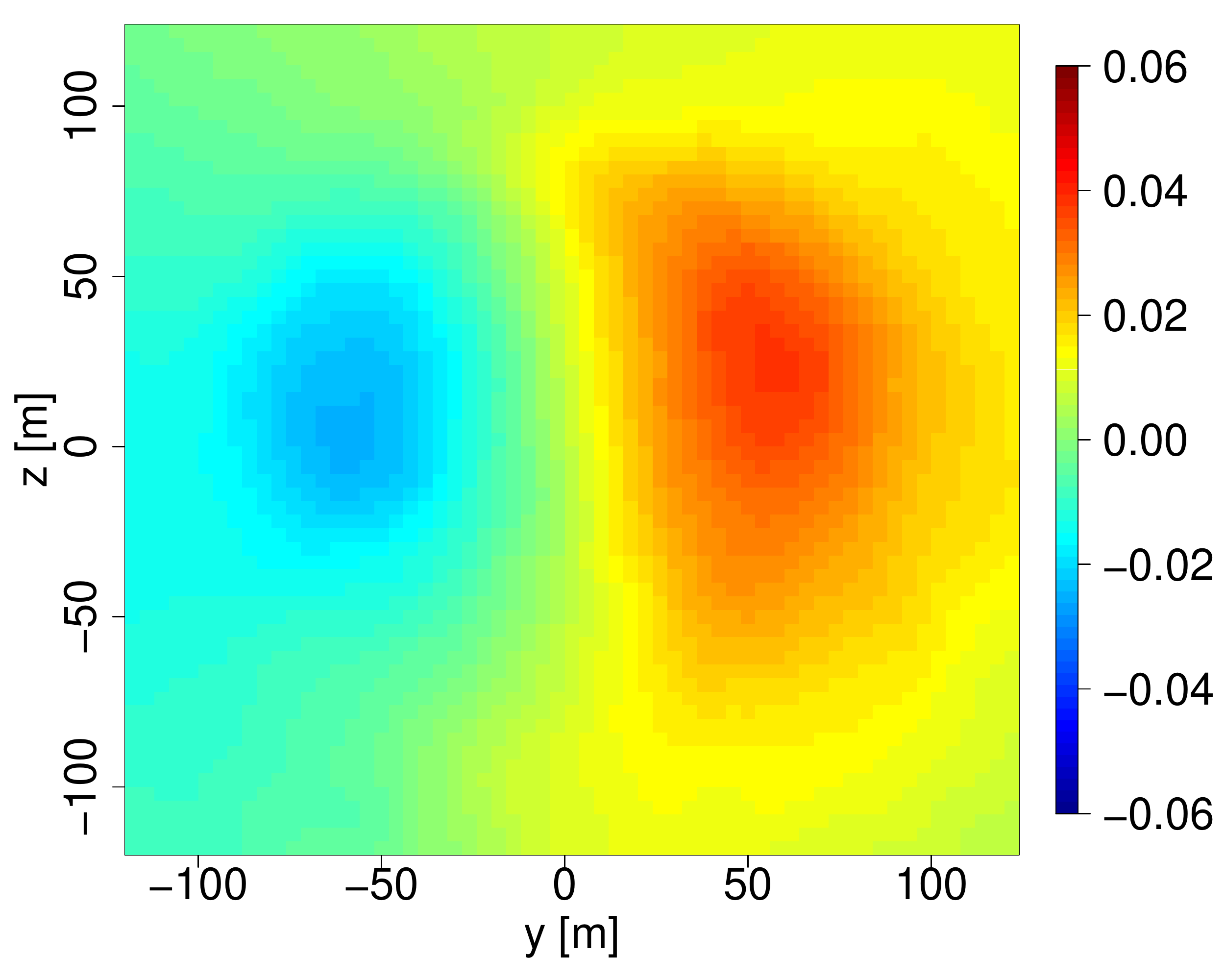}
 \subcaption{}
 \label{fig:modeb1}
\end{subfigure}
\begin{subfigure}[t]{.32\textwidth}
 \includegraphics[width=.98\linewidth]{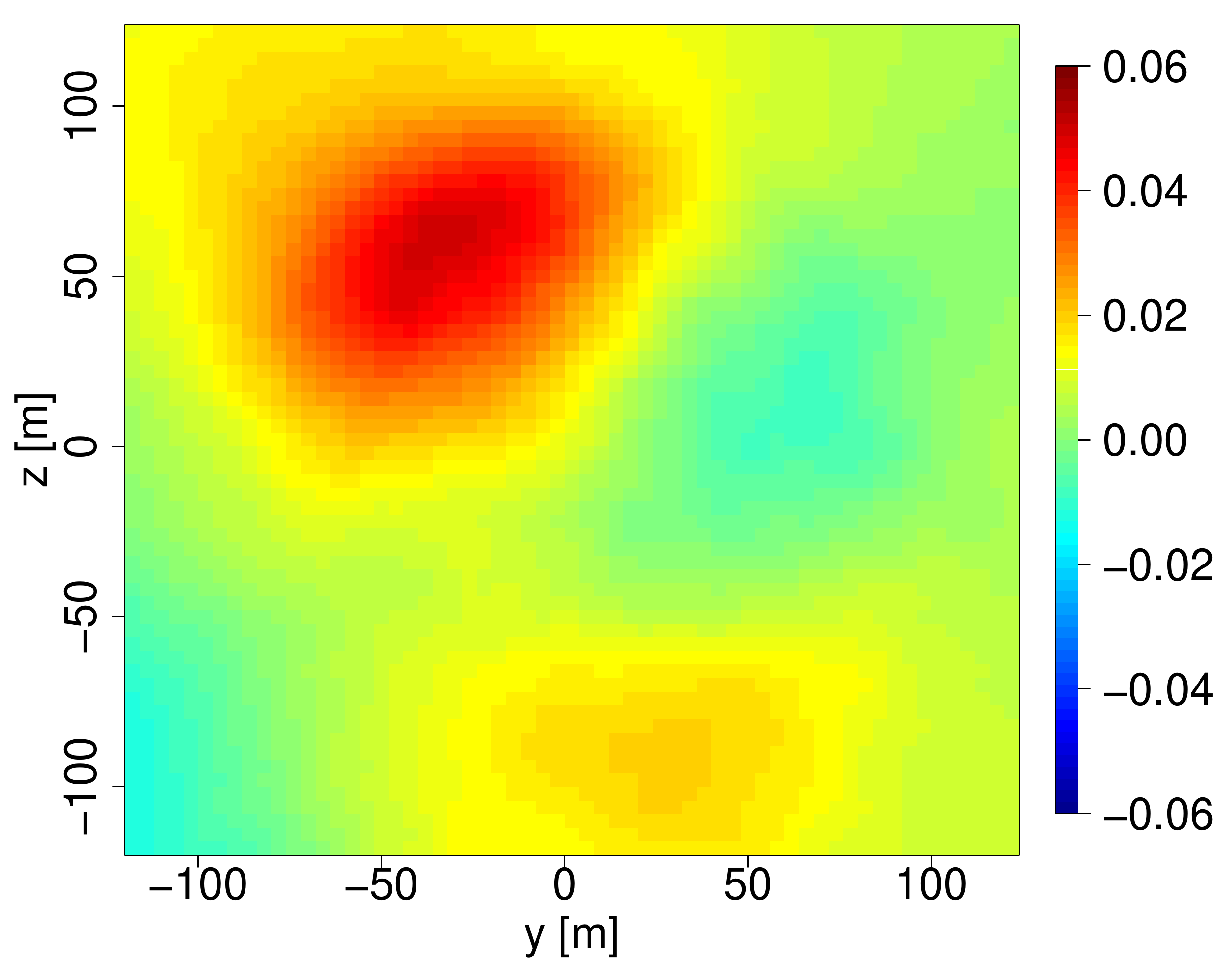}
 \subcaption{}
 \label{fig:modeb2}
\end{subfigure}
\end{center}
\begin{center}
 \begin{subfigure}[t]{.32\textwidth}
 \includegraphics[width=.98\linewidth]{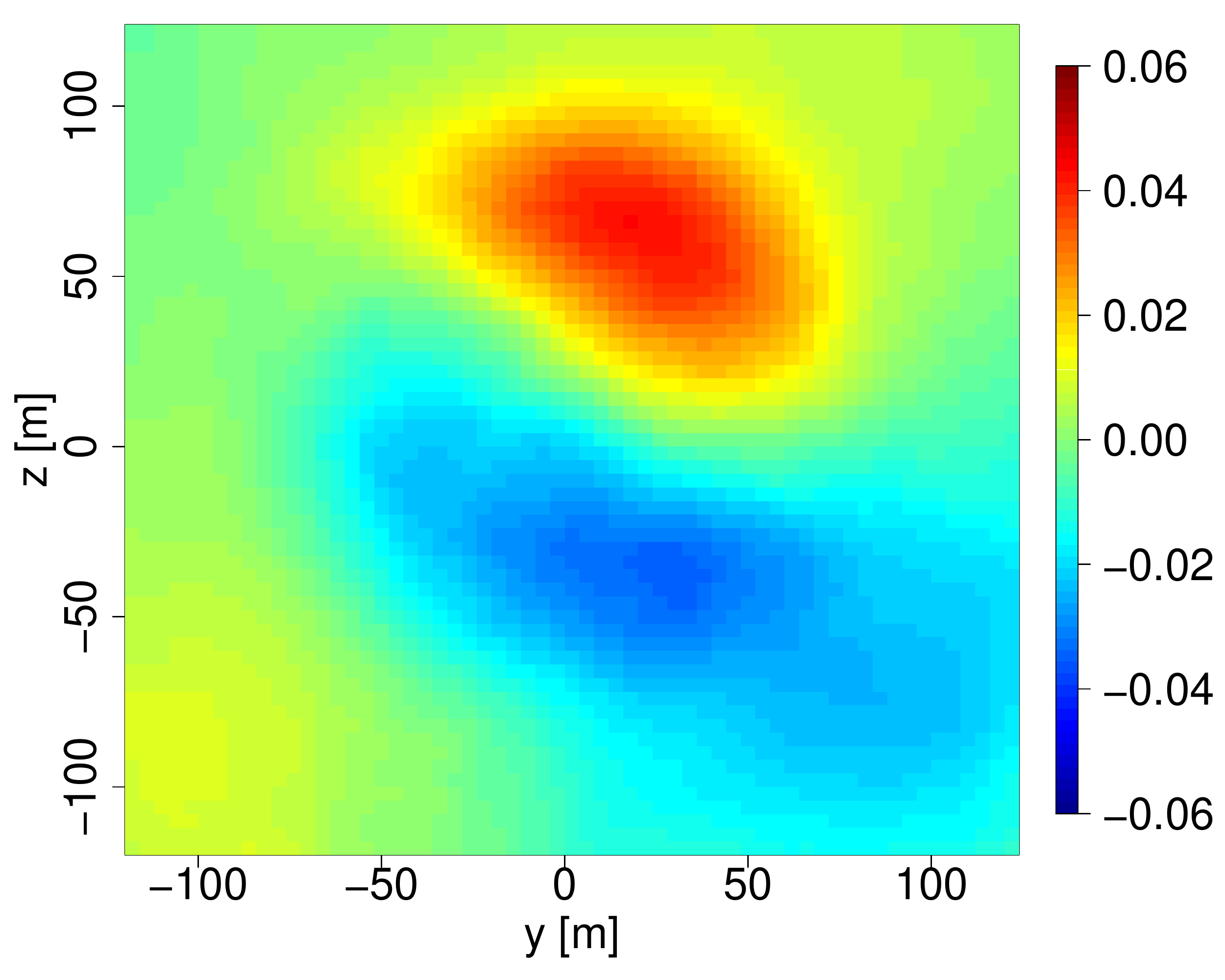}
 \subcaption{}
 \label{fig:modeb3}
\end{subfigure}%
\vspace {-12pt}
\begin{subfigure}[t]{.32\textwidth}
 \includegraphics[width=.98\linewidth]{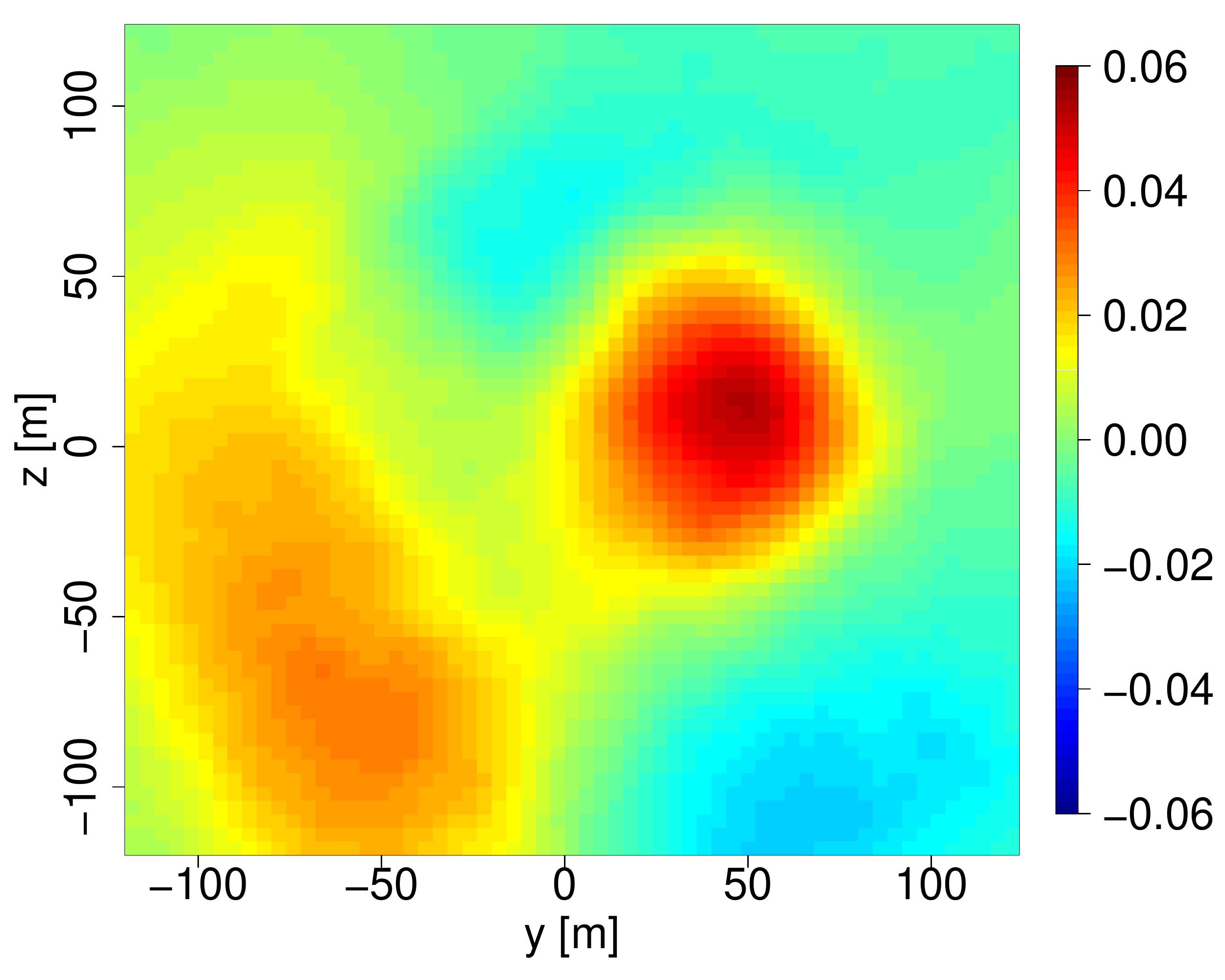}
 \subcaption{}
 \label{fig:modeb4}
\end{subfigure}
\begin{subfigure}[t]{.32\textwidth}
 \includegraphics[width=.98\linewidth]{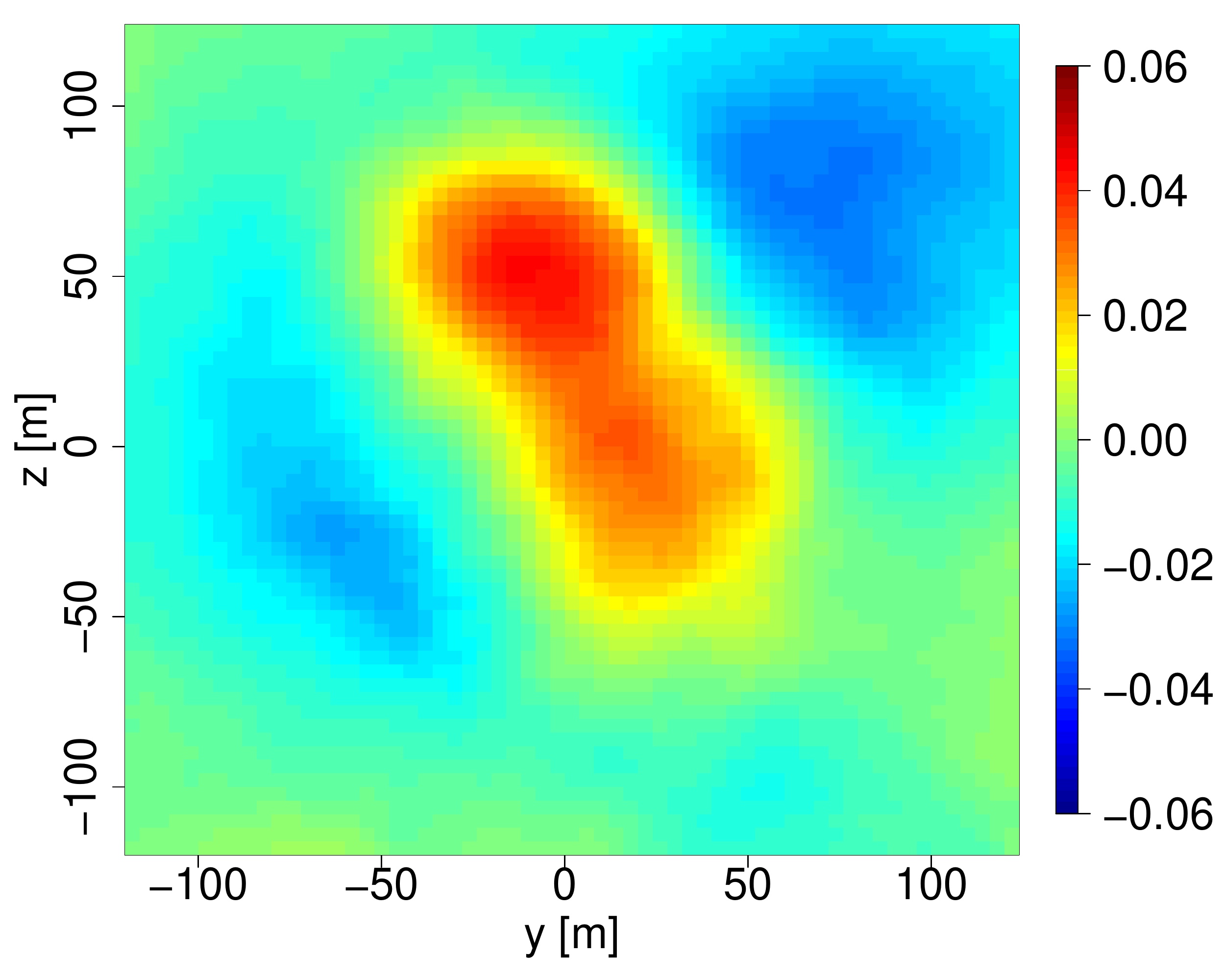}
 \subcaption{}
 \label{fig:modeb5}
\end{subfigure}%
\end{center}
\begin{center}
\begin{subfigure}[t]{.32\textwidth}
 \centering
 \includegraphics[width=.98\linewidth]{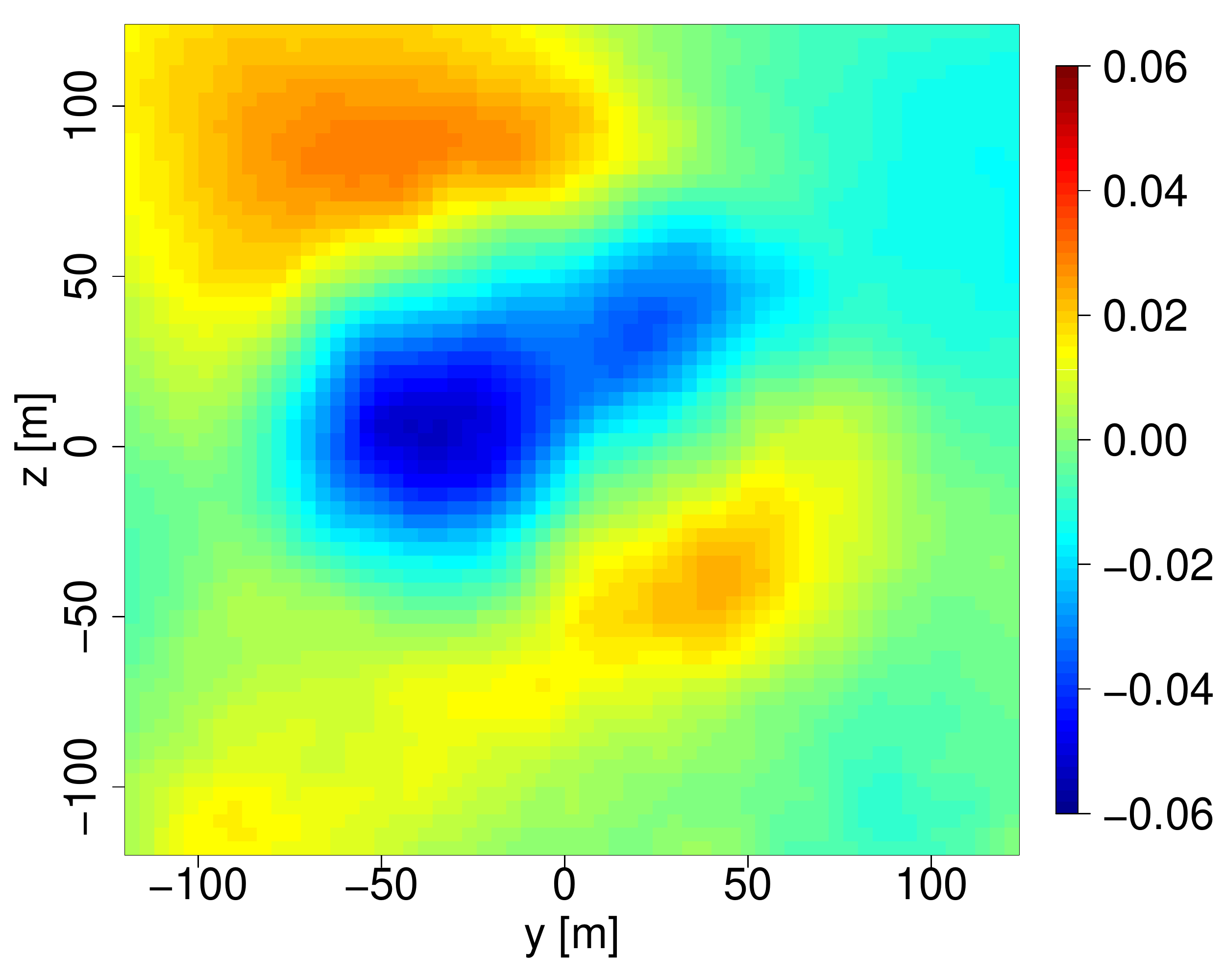}
 \subcaption{}
 \label{fig:modeb6}
\end{subfigure}%
\begin{subfigure}[t]{.32\textwidth}
 \centering
 \includegraphics[width=.98\linewidth]{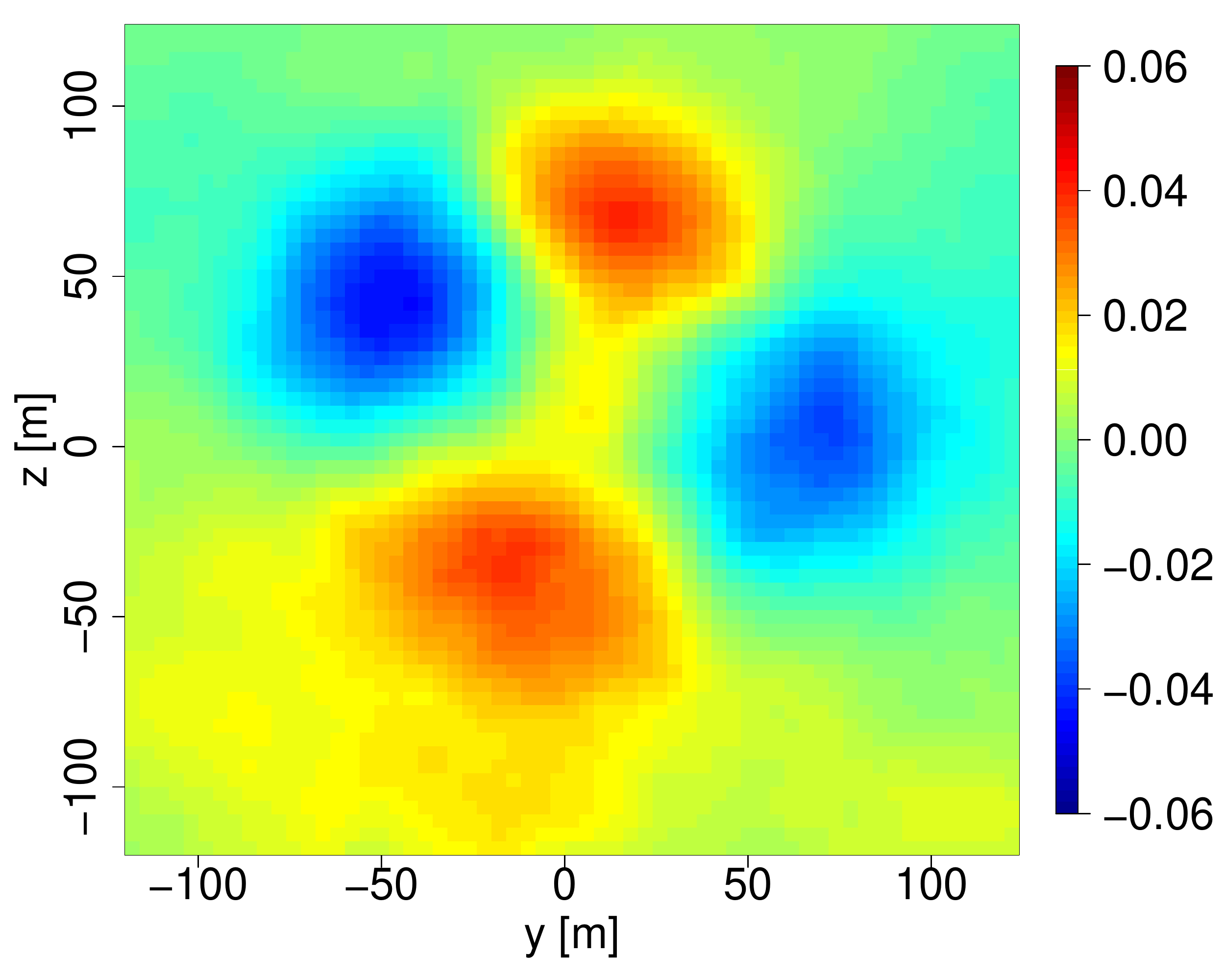}
 \subcaption{}
 \label{fig:modeb7}
\end{subfigure}
\begin{subfigure}[t]{.32\textwidth}
 \centering
 \includegraphics[width=.98\linewidth]{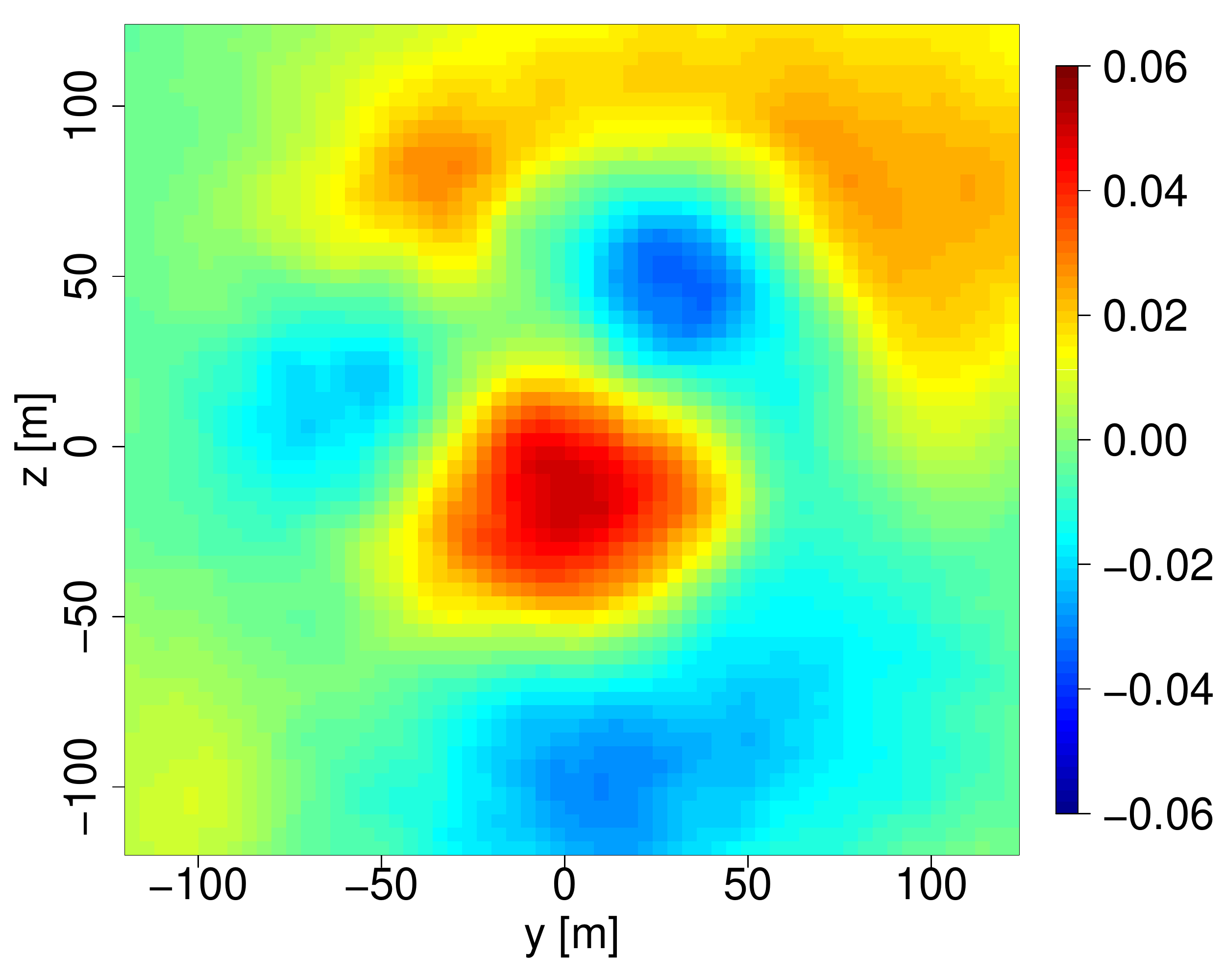}
 \subcaption{}
 \label{fig:modeb8}
\end{subfigure}%
\end{center}\vspace{-12pt}
\caption{Mean field and POD modes. (\textbf{a}) Mean field; (\textbf{b}) Mode 1; (\textbf{c}) Mode 2; (\textbf{d})~Mode~3; (\textbf{e}) Mode 4; (\textbf{f}) Mode 5; (\textbf{g}) Mode 6; (\textbf{h}) Mode 7; (\textbf{i}) Mode 8.}
\label{Fig:qcutmeanb}
\end{figure}

The obtained modes are shown in Figure \ref{Fig:qcutmeanb}. 
Some modes are quite similar to the modes obtained after preprocessing (compare to Figure \ref{Fig:qcutmean}).
For example, Mode $1$ stays almost the same, and Mode $7$ resembles the former Mode $4$.
Other modes, like Mode $4$ or Mode $8$, appear slightly more complex. 
Generally, we can say that the modes appear less clearly structured, even though some of the
azimuthal structures are~retained.

\begin{figure}[H]
\centering
\begin{subfigure}[H]{.45\textwidth}
 \centering
 \includegraphics[width=.95\linewidth]{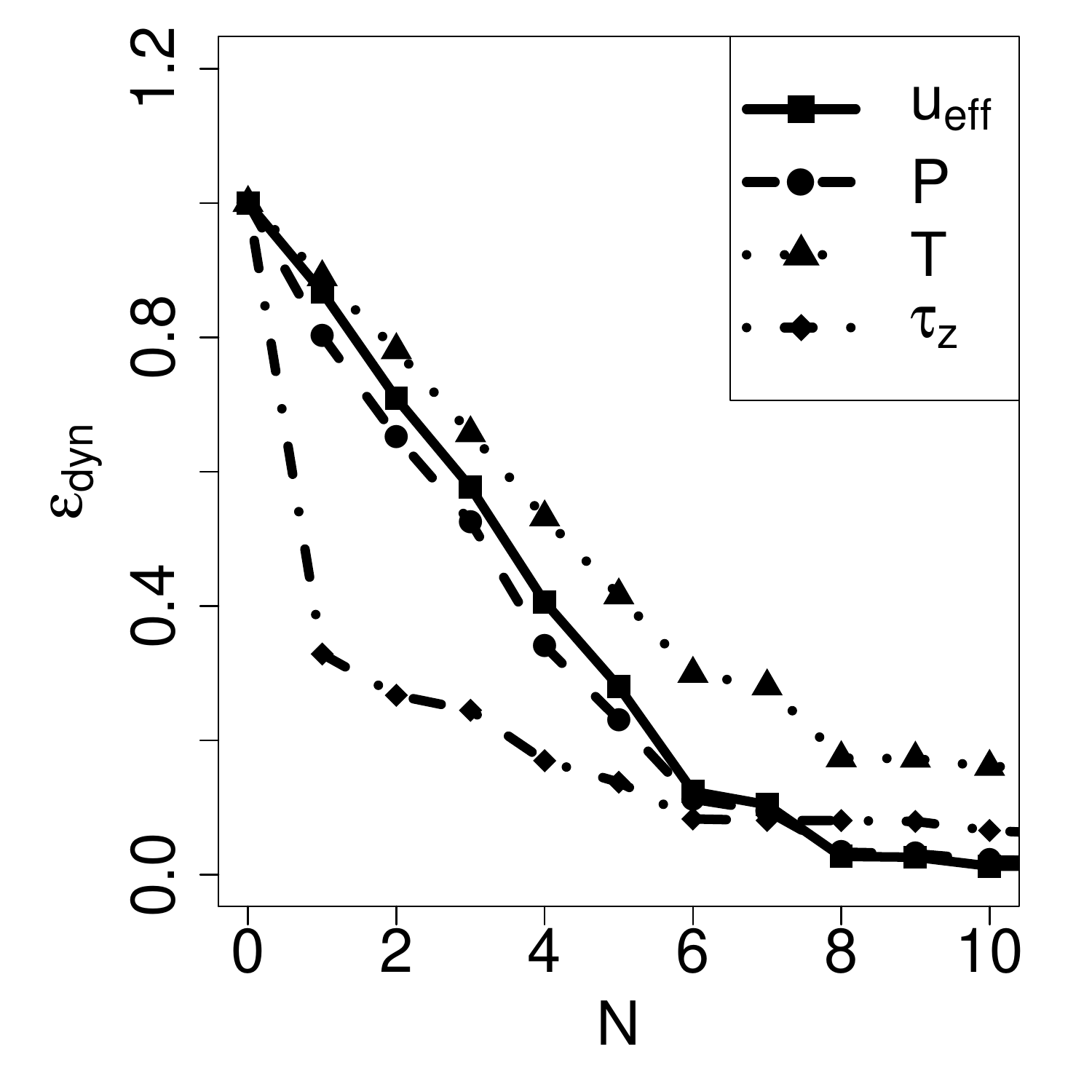}
 \subcaption{}
 \label{fig:edyn_allmb}
\end{subfigure}%
\begin{subfigure}[H]{.45\textwidth}
 \centering
 \includegraphics[width=.91\linewidth]{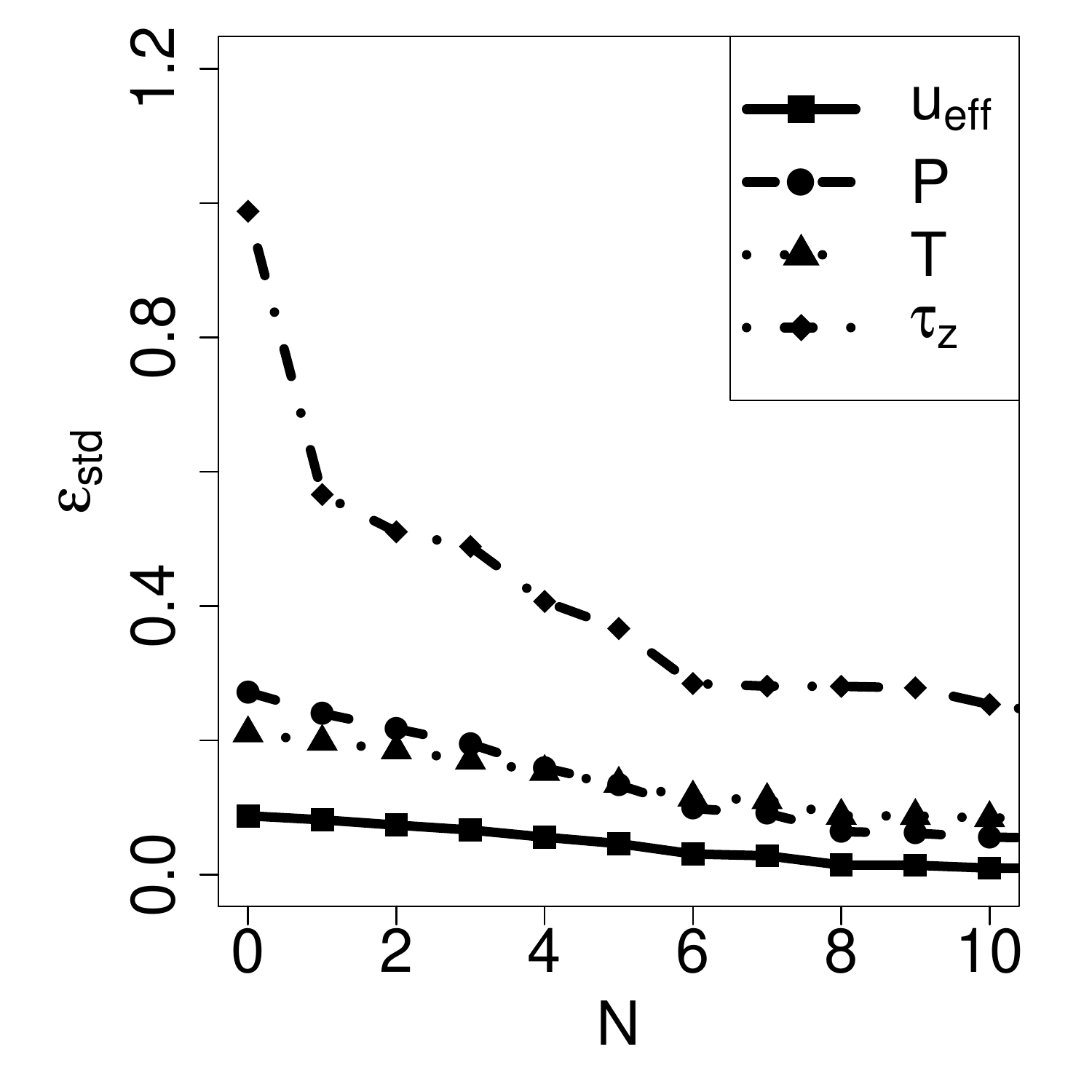}
 \subcaption{}
 \label{fig:estd_allmb}
\end{subfigure}
\caption{(\textbf{a}) Dynamical and (\textbf{b}) standard error for different
 measures \textit{versus} the number $N$ of modes used for the reconstruction.
 $N=0$ corresponds to a reconstruction with the mean field only.}
\label{Fig:edyn_allmb}
\end{figure}

As the next point, we investigate the influence of these new modes 
on the quality of reconstructions with the measures defined in Section \ref{sec:alternative}.
In Figure \ref{Fig:edyn_allmb}, the corresponding error graphs are shown.
The major difference is the steady decrease of the dynamical error for $u_{\mbox{eff}}$, $P$ and $T$.
The sudden jumps are missing. Particularly, the large decrease with the third mode has vanished
(compare with Figure \ref{fig:edyn_allm}).
Therefore, a model based on less than six modes
would probably perform worse when using the modes obtained without a threshold.

\label{sec:classicpod}

\section{Convergence Study of POD Modes and Values}
\label{app:convergence}

\setcounter{figure}{0}
\renewcommand\thefigure{B\arabic{figure}}

\begin{figure}[H]
\centering
\begin{subfigure}[H]{.45\textwidth}
 \centering
 \includegraphics[width=.95\linewidth]{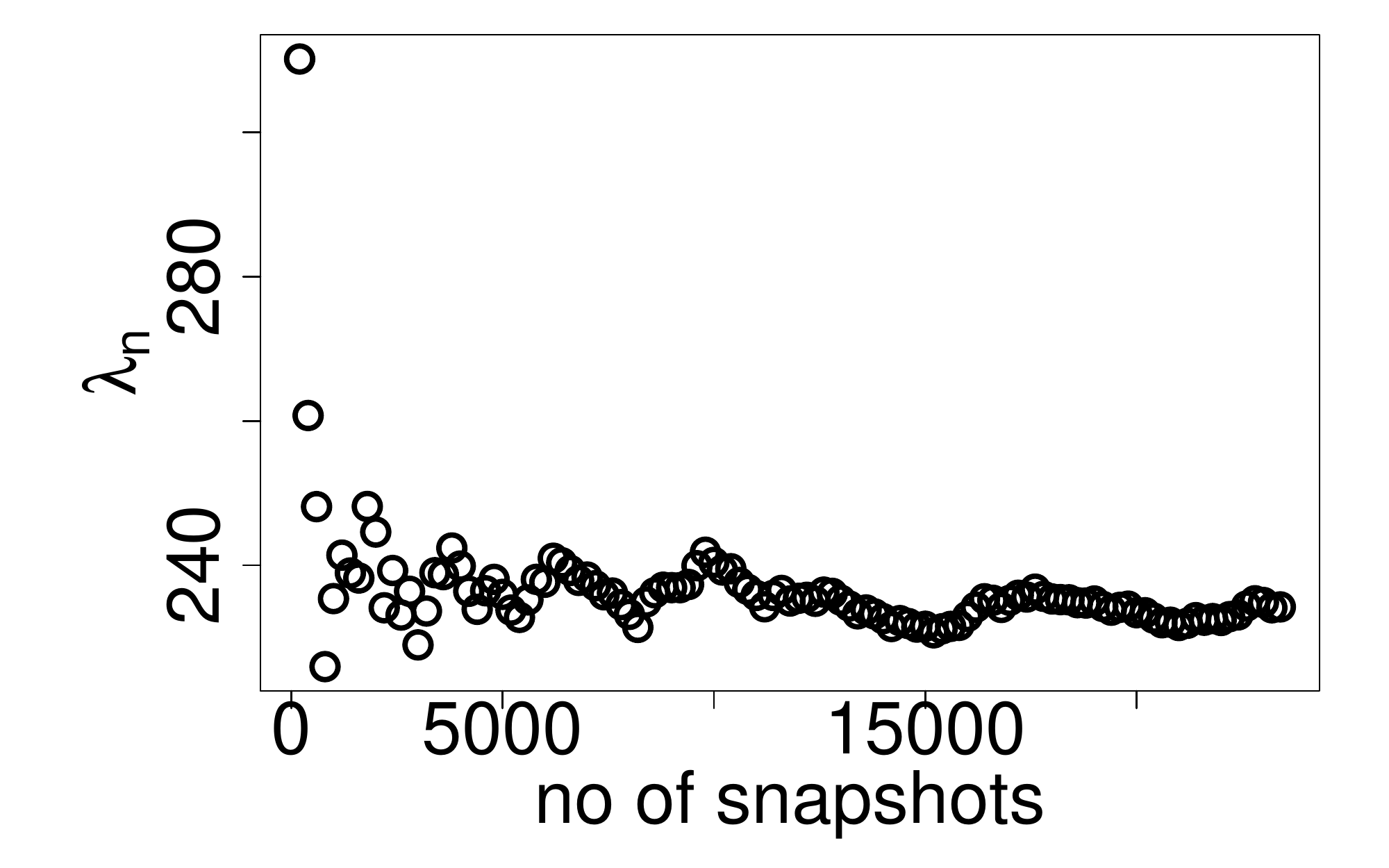}
 \subcaption{}
 \label{fig:lambda1convNT}
\end{subfigure}%
\begin{subfigure}[H]{.45\textwidth}
 \centering
 \includegraphics[width=.91\linewidth]{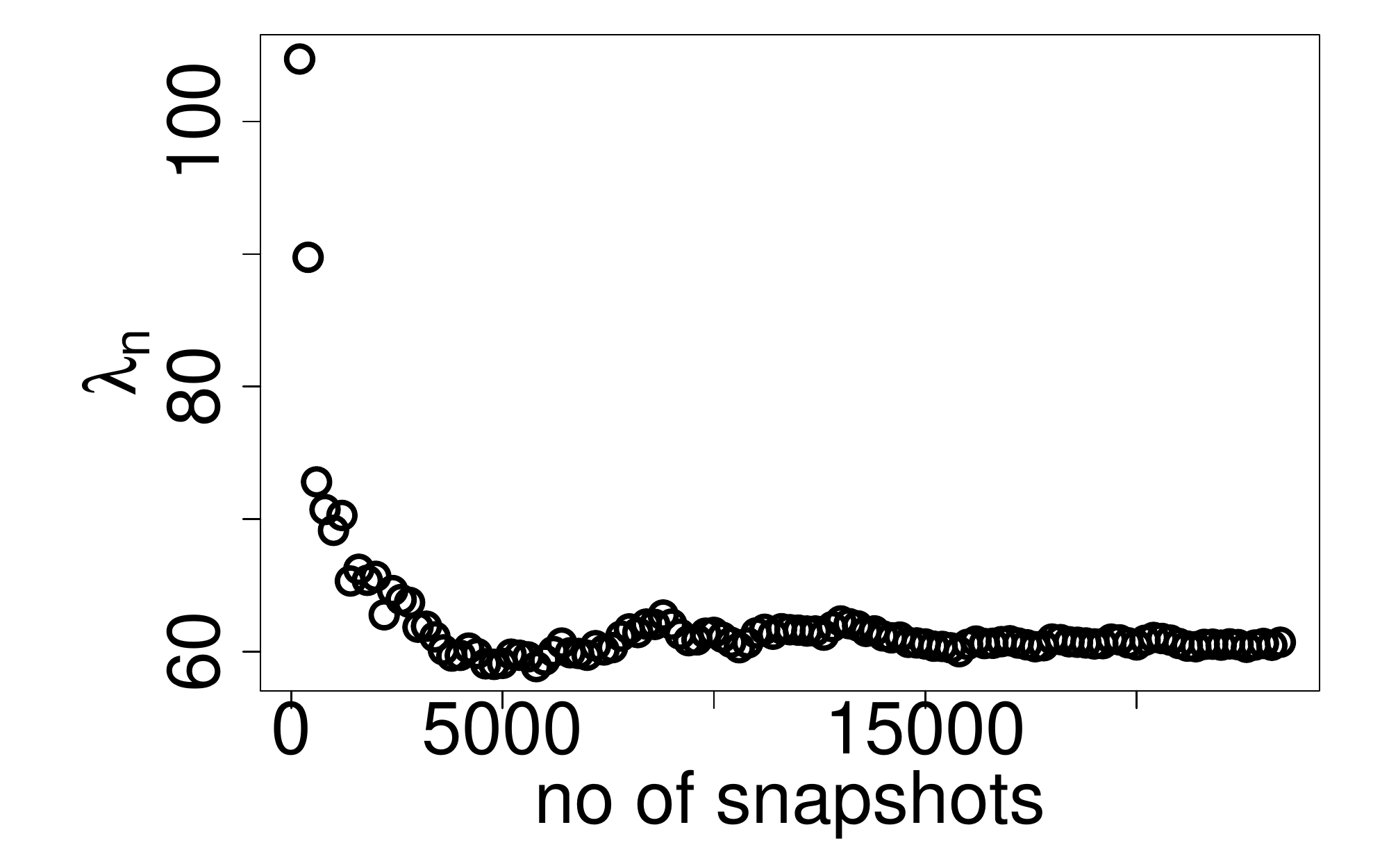}
 \subcaption{}
 \label{fig:lambda4convNT}
\end{subfigure}
\caption{Eigenvalue \textit{versus} the number of snapshots used for its estimation. 
(\textbf{a}) First eigenvalue (\textit{n} = 1); (\textbf{b}) second eigenvalue (\textit{n} = 2).}
\label{Fig:lambda1convNT}
\end{figure}

In this appendix, we briefly discuss the convergence behavior of the POD modes and the values presented
in Section \ref{sec:podmodes}. For this purpose, we investigate the sensitivity of these results to the
number of snapshots used for their estimation. 
In Figure \ref{Fig:lambda1convNT}, the behavior of two different eigenvalues is illustrated.
After~around 11,000 snapshots, the change when including more snapshots stays below 5\%.
Thus, the values are roughly converged. Note that less independent snapshots would be necessary due to
the temporal correlations in the field.

\begin{figure}[H]
\centering
\begin{subfigure}[t]{.24\textwidth}
 \centering
 \includegraphics[width=.98\linewidth]{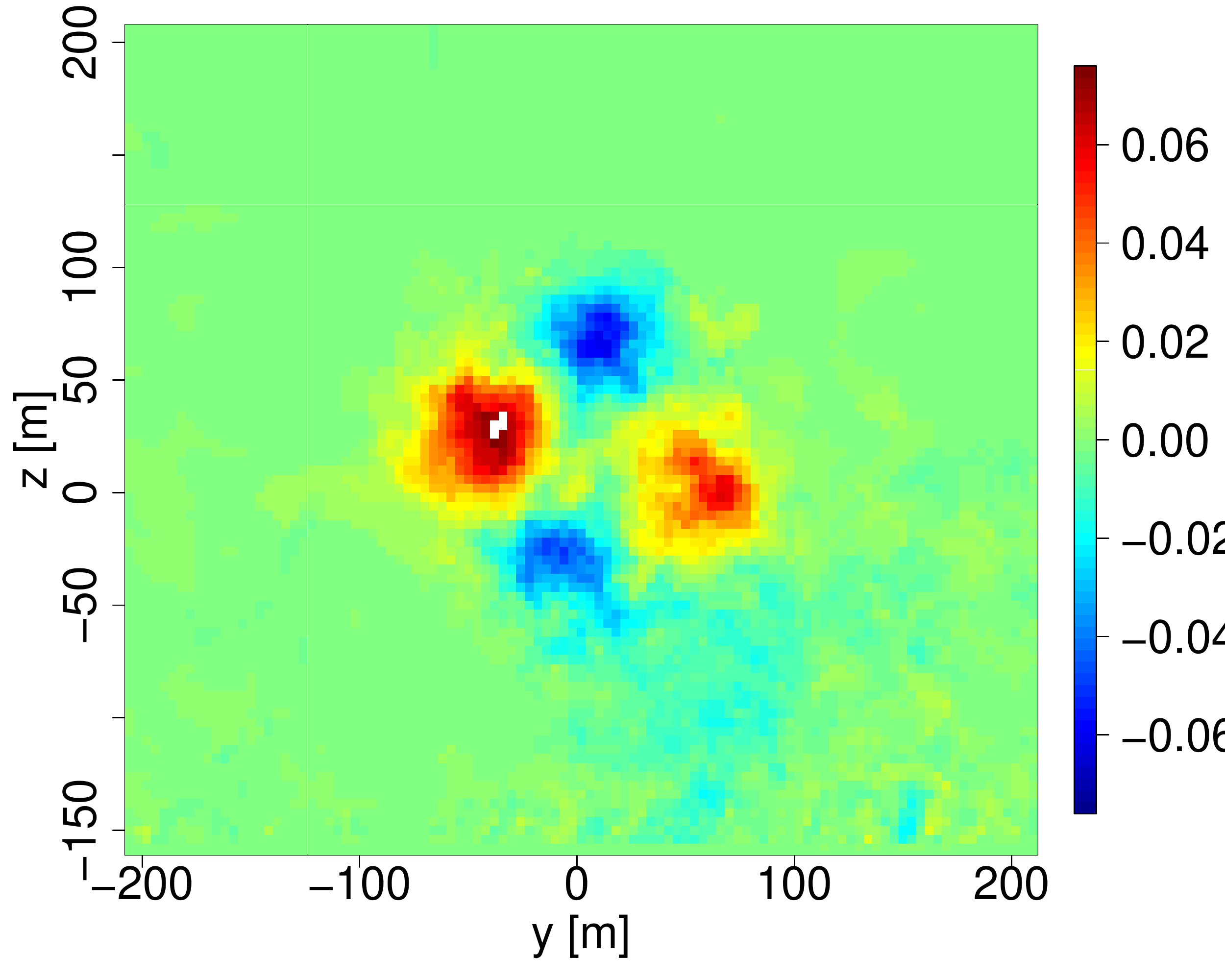}
 \subcaption{}
 \label{fig:pu1}
\end{subfigure}%
\begin{subfigure}[t]{.24\textwidth}
 \centering
 \includegraphics[width=.98\linewidth]{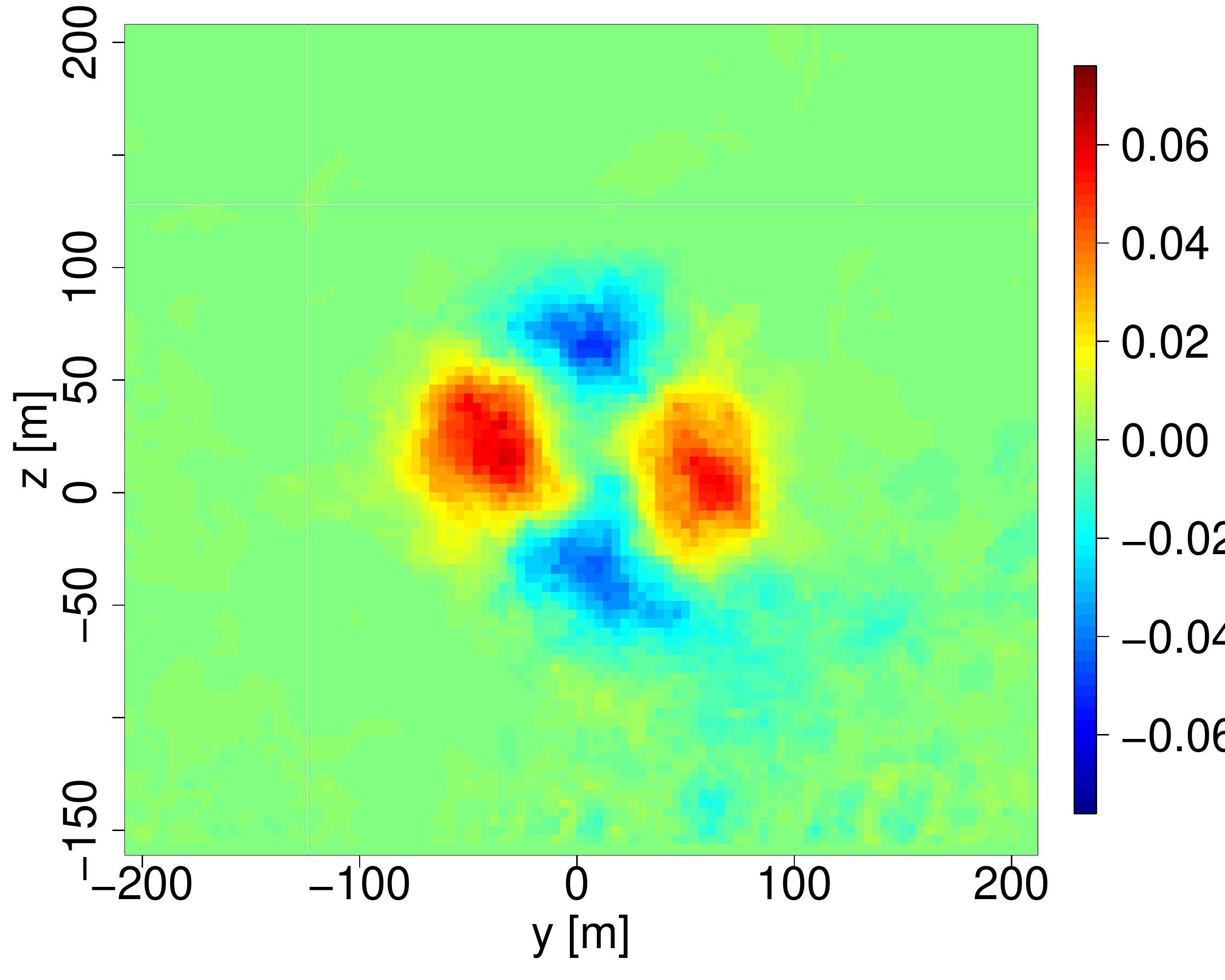}
 \subcaption{}
 \label{fig:pu2}
\end{subfigure}
\begin{subfigure}[t]{.24\textwidth}
 \centering
 \includegraphics[width=.98\linewidth]{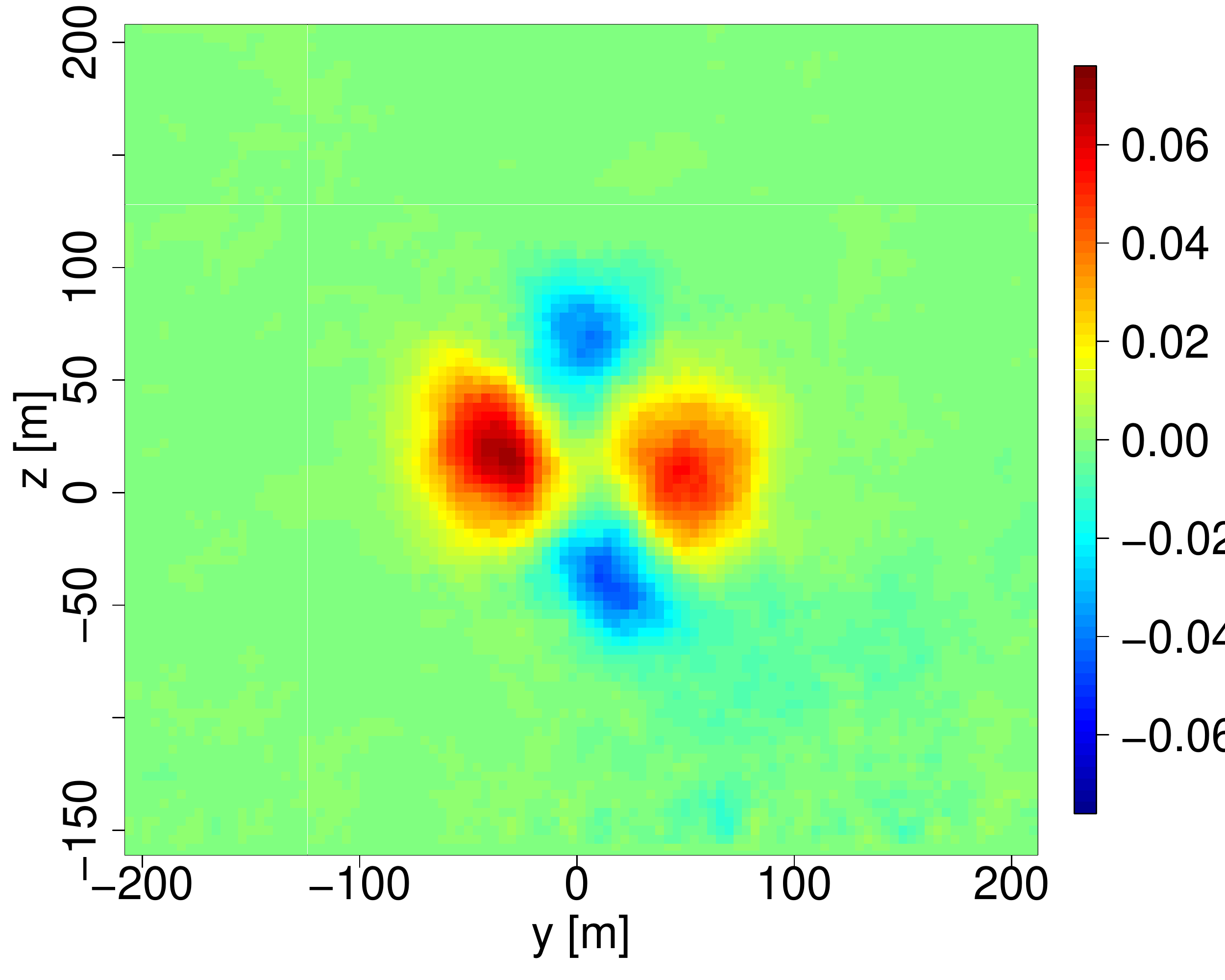}
 \subcaption{}
 \label{fig:pu3}
\end{subfigure}%
\begin{subfigure}[t]{.24\textwidth}
 \centering
 \includegraphics[width=.98\linewidth]{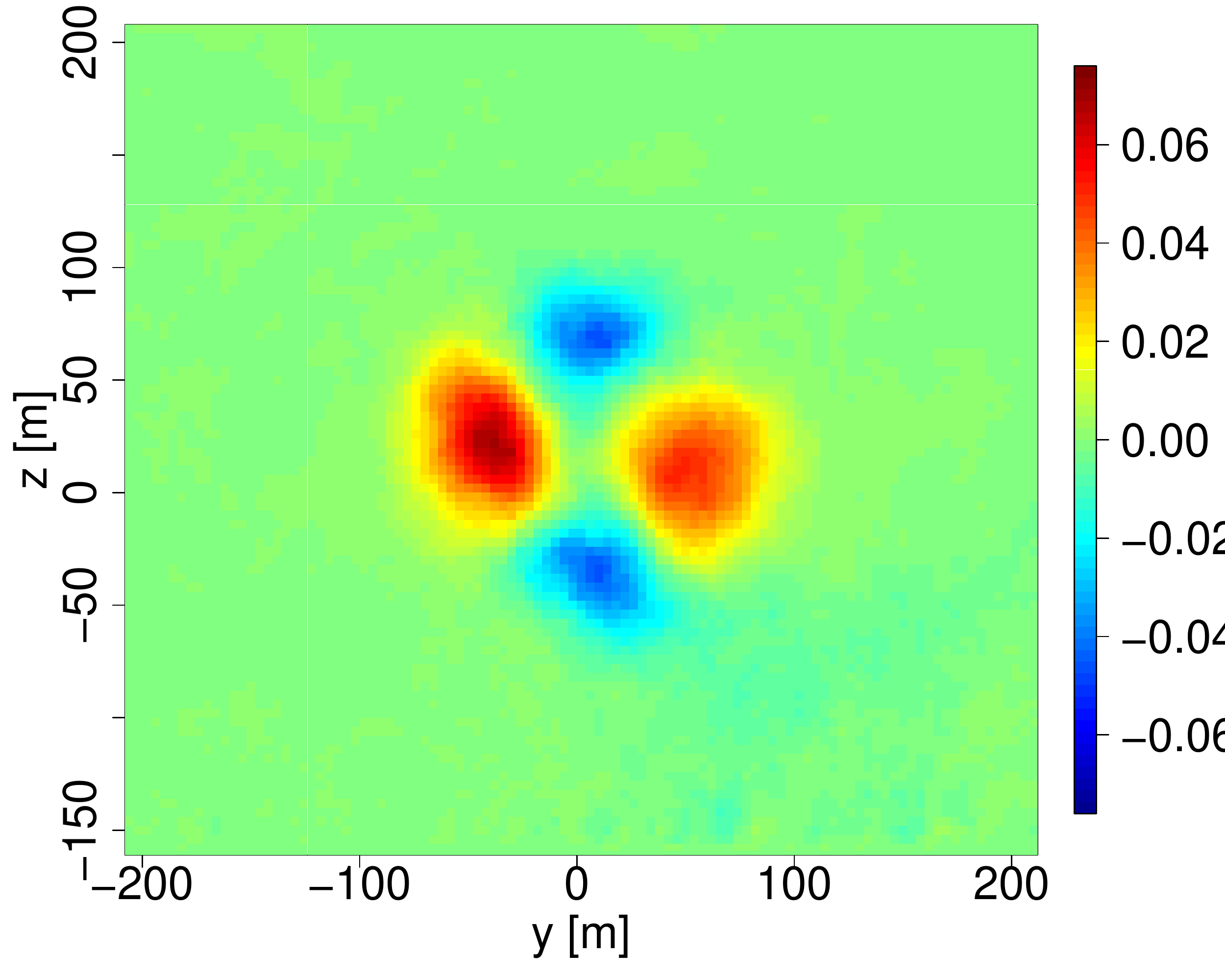}
 \subcaption{}
 \label{fig:pu4}
\end{subfigure}
\caption{Fourth POD mode estimated using different numbers of snapshots: (\textbf{a}) 1000; (\textbf{b})~2000; (\textbf{c}) 6000; (\textbf{d}) 10,000.}
\label{Fig:mode_4_N}
\end{figure}

Figure \ref{Fig:mode_4_N} illustrates the change of the estimate of Mode $4$ when increasing
the number of snapshots. The basic structure is already visible after around 1000 snapshots.
After around 6000, no strong changes in the mode are visible anymore.
Similar results have been found for the other modes, but the convergence of the
modes gets weaker with higher mode number.
These results indicate that the modes used for the reconstructions in this paper
are converged and will not change strongly when using longer simulations. 

\section{Further Explanations on the Standard Error}
\label{app:standarderror}

\setcounter{equation}{0}
\renewcommand\theequation{C\arabic{equation}}

In this Appendix, the discussion in Section \ref{sec:results} about the results for the standard error
(Figure \ref{fig:estd_allm}) is continued.  
We start by illustrating why the standard error is already low for $u_{\mbox{eff}}$ and $P$ 
when using only the mean field followed by a comparison with $\tau_z$.

For $u_{\mbox{eff}}$,
using only the steady mean field already yields a perfect reconstruction of $\langle u_{\mbox{eff}} \rangle_t$:
\begin{equation}
\langle \langle u(y,z,t) \rangle_t \rangle_{disk} = \langle \langle u(y,z,t) \rangle_{disk} \rangle_t=
\langle u_{\mbox{eff}}(t) \rangle_t~
\label{eq:ueffmean}
\end{equation}
Since the fluctuations of $u_{\mbox{eff}}$ are much smaller than the mean value, the correct description of the mean
obviously leads to a low standard error (see Equation (\ref{eq:errorstd})), due to the normalization.
For the dynamical error (Equation (\ref{eq:errordyn})), the recovery of the mean
obviously does not play any role. For $P$, $\langle P \rangle_t$ cannot be fully recovered using just
$\langle u(y,z,t) \rangle_t$ due to:
\begin{equation}
\langle \langle u \rangle_t^3 \rangle_{disk} \neq \langle \langle u^3 \rangle_t \rangle_{disk}=\langle \langle u^3 \rangle_{disk} \rangle_{t}
\propto \langle P \rangle_t
\end{equation}
Consequently, the standard error is always higher for $P$ than for $u_{\mbox{eff}}$. 
Similarly, we have for $\tau_z$:
\begin{equation}
\langle \langle u \rangle_t^2 y \rangle_{disk} \neq \langle \langle u^2 \rangle_t y\rangle_{disk}=\langle \langle u^2 y \rangle_{disk} \rangle_{t}
= \langle \tau_z \rangle_t~
\end{equation}
For $\tau_z$, the fluctuations are much higher than for $P$. This yields a large difference between
$\langle u \rangle_t^2$ and $\langle u^2 \rangle_t$, resulting in the large standard error compared with
$P$ and $u_{\mbox{eff}}$. Similar arguments can be found for $T$.

The short discussion above does not only explain the different behavior of $\varepsilon_{\mbox{dyn}}$ and
$\varepsilon_{\mbox{std}}$, but also nicely illustrates a major challenge for steady mean field models.
Since the conversion from wind to relevant quantities, such as power or loads, can be a nonlinear process,
the mean velocity field does not necessarily offer a good description of other mean quantities.

\conflictofinterests{Conflicts of Interest}

The authors declare no conflict of interest. 

%
%
%
%



%


\end{document}